\documentclass[11pt,twoside,a4paper]{article}

\usepackage{arxiv}

\usepackage[utf8]{inputenc} 
\usepackage[T1]{fontenc}    
\usepackage{hyperref}       
\usepackage{url}            
\usepackage{booktabs}       
\usepackage{amsfonts}       
\usepackage{nicefrac}       
\usepackage{microtype}      
\usepackage{lipsum}		
\usepackage{graphicx}
\usepackage{natbib}
\usepackage{doi}
\usepackage{caption}
\usepackage{subcaption}

\usepackage{mathtools,upgreek,eucal,mathrsfs,enumitem,amsthm,amsmath}
\usepackage{times,epsfig,makeidx,amsfonts,amsmath,dcolumn,multirow,amssymb,colortbl,bm,url}
\usepackage{algorithm}
\usepackage[]{algorithmic}
 \usepackage[flushleft]{threeparttable}
 
\usepackage{xr}
\newtheorem{lemma}{Lemma}

\newcommand{\bxi}{\bm{\bxi}}

\newcommand{\bgamma}{\bm{\gamma}}
\newcommand{\btheta}{\bm{\theta}}
\newcommand{\bdelta}{\bm{\delta}}

\newcommand{\bmeta}{\bm{\eta}}
\newcommand{\bbeta}{\bm{\beta}}
\newcommand{\balpha}{\bm{\alpha}}

\newcommand{\td}{\tilde{d}}

\newcommand{\bh}{{\bf h}}

\newcommand{\bt}{{\bf t}}

\newcommand{\bX}{{\bf X}}
\newcommand{\bx}{{\bf x}}
\newcommand{\by}{{\bf y}}
\newcommand{\bY}{{\bf Y}}

\newtheorem{proposition}{Proposition}

\usepackage{xcolor}

\usepackage{soul}

\title{Hazard-based distributional regression via ordinary differential equations}

\author{
	\href{https://orcid.org/0000-0000-0000-0000}{\includegraphics[scale=0.06]{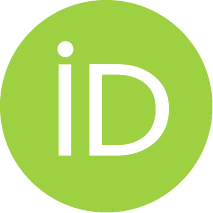}\hspace{1mm} J. Andres Christen} \\
	Department of Statistics\\
	Centre for Research in Mathematics (CIMAT) \\
	Guanajuato, M{\' e}xico\\
	\texttt{jac@cimat.mx} 
 \And	
	\href{https://orcid.org/0000-0001-7183-8407}{\includegraphics[scale=0.06]{orcid.pdf}\hspace{1mm} F. Javier Rubio} \\
	Department of Statistical Science\\
	University College London \\
	London, UK\\
	\texttt{f.j.rubio@ucl.ac.uk} 
	}

\renewcommand{\shorttitle}{Hazard-based distributional regression}

\hypersetup{
pdftitle={XXX},
pdfsubject={stat.ME, stat.AP},
pdfauthor={XXX, Rubio},
}

\begin{document}
\maketitle
\date{}
\begin{abstract}
The hazard function is central to the formulation of commonly used survival regression models such as the proportional hazards and accelerated failure time models. However, these models rely on a shared baseline hazard, which, when specified parametrically, can only capture limited shapes. To overcome this limitation, we propose a general class of parametric survival regression models obtained by modelling the hazard function using autonomous systems of ordinary differential equations (ODEs). Covariate information is incorporated via transformed linear predictors on the parameters of the ODE system.
Our framework capitalises on the interpretability of parameters in common ODE systems, enabling the identification of covariate values that produce qualitatively distinct hazard shapes associated with different attractors of the system of ODEs. This provides deeper insights into how covariates influence survival dynamics. We develop efficient Bayesian computational tools, including parallelised evaluation of the log-posterior, which facilitates integration with general-purpose Markov Chain Monte Carlo samplers. We also derive conditions for posterior asymptotic normality, enabling fast approximations of the posterior.
A central contribution of our work lies in the case studies. We demonstrate the methodology using clinical trial data with crossing survival curves, and a study of cancer recurrence times where our approach reveals how the efficacy of interventions (treatments) on hazard and survival are influenced by patient characteristics.
\end{abstract}


\section{Introduction}\label{sec:intro}

Modern statistical applications require advanced models and methods capable of capturing complex data features. However, since the ultimate goal is to apply these models and inform decision-makers, it is crucial to develop interpretable models that allow for translating their results into quantities that can be effectively communicated to multidisciplinary groups or general audiences. One key area where this balance between complexity and interpretability is essential is the analysis of survival data, a longstanding area of core interest in statistics.
Survival data naturally appears in many applied areas such as medicine, biology, reliability engineering, advanced manufacturing, and industrial process monitoring, to name but a few, where the aim is to model a sample of times-to-event using the information in the available covariates (\textit{i.e.}~individual characteristics). For example, suppose that we are interested in understanding how individual characteristics -- such as age, sex, sociodemographic characteristics, and others -- predict survival times in a population. This question can be framed in statistical terms as a regression model for survival data. This leads to the need for developing survival regression models that capture covariate effects over time and facilitate their interpretation. 
Survival regression models are usually formulated using the hazard function as their foundation \citep{chen:2000,rubio:2019}, since this function provides information about the probability of the event (death) happening at each time point.  The hazard function of a random variable $T>0$, with probability density function $f(t)$ and cumulative distribution function $F(t) = \Pr(T \leq t)$, is defined as \citep{rinne:2014}
$$h(t) = \lim_{dt \downarrow 0} \dfrac{\Pr[t \leq  T < t+dt \mid  T \geq t]}{dt}= \dfrac{f(t)}{1-F(t)}.$$
Several strategies to incorporate the information of the available covariates $\bx_i$, $i=1,\dots,n$, into the hazard function $h(t \mid \bx_i)$, and define a survival regression model, have been proposed. Most survival regression models incorporate covariates into the hazard function by assuming a common ``baseline hazard'' $h_0(t)$. The most popular model in practice is the proportional hazards model \citep{cox:1972}, which assumes the structure $h(t \mid \bx_i,\bbeta) = h_0(t)\exp\left\{\bx_i^\top\bbeta\right\}$. The assumption of proportional hazards is often unrealistic in practice, as the effect of some of the covariates may vary over time. Consequently, a number of hazard-based regression models that can capture time-level covariate effects have been proposed, such as the accelerated failure time (AFT), accelerated hazards (AH), general or extended hazards (GH), and additive models \citep{chen:2001,rubio:2019,eletti:2022,basak:2025}. However, those models still assume a common baseline hazard, which limits the understanding of the effect of individual characteristics over time. For example, a group of patients with characteristics $\bx^*$ might exhibit an increasing baseline hazard, while another group of patients with characteristics $\bx^{**}$ could exhibit a decreasing baseline hazard. 
A potential solution to this limitation involves assuming a ``baseline parametric model'' and introducing covariates by modelling the (suitably transformed) distributional parameters using linear or additive predictors \citep{stasinopoulos:2018,kneib:2021}. This approach has the advantage of enabling the modelling of individual distributional features such as the mean, variance, and skewness, as functions of covariates. However, a key limitation is the reliance on a specific baseline distribution. In survival analysis, this is often the Weibull, Power Generalised Weibull, or another generalisation of the Weibull distribution \citep{burke:2020}, each of which imposes structural restrictions on the shape of the hazard function.


This work develops a general class of parametric survival regression models based on systems of ordinary differential equations (ODEs), accompanied by practical guidelines for model building, interpretation, and numerical and software implementations. These models offer a more flexible structure than classical alternatives and provide additional insights into the evolution of mortality (or failure) risk for individuals of interest. Specifically, we build on the framework proposed by \citet{christen:2024}, which models the dynamics of the hazard function using a parametric system of first-order ODEs.
\citet{christen:2024} proposed modelling the hazard function with a system of ODEs, where the hazard function is treated as a state variable within the system.
They introduce two examples of such systems: the Logistic model and the ``Hazard-Response'' model, the latter representing a species competition system of ODEs. We now extend this approach to incorporate covariates.
We introduce covariate dependence directly into the parameters governing the ODE system that defines the hazard function, allowing the hazard's evolution to vary with individual characteristics $\bx_i$. To capitalise on the interpretability of ODE parameters \citep{christen:2024}, covariates are introduced via suitably transformed linear predictors. While closely related to distributional regression \citep{kneib:2021} and machine learning methods for modelling cumulative hazard functions \citep{tang:2022a}, our approach differs in that covariates enter through the ODEs defining the hazard function rather than the distributional parameters or the weights of a neural network.
This distinction allows for a more direct interpretation of covariate effects on the shape and dynamics of the hazard function. Notably, our method enables identification of covariate values that lead the hazard function to exhibit different qualitative behaviours, corresponding to distinct attractors in the ODE system, which helps understanding the underlying hazard shape and evolution over time.
We illustrate the proposed methodology with two real-world applications: one in clinical trials and another concerning the recurrence of cancer in patients undergoing treatment. 

The paper is organised as follows. 
Section~\ref{sec:model} introduces the formulation of the hazard function using systems of ODEs, along with our specific proposal for incorporating covariates. This section also presents the Logistic and Hazard-Response ODE models, which will be used in our simulation studies and case studies. 
Section~\ref{sec:postinf} discusses the likelihood function and posterior distribution, and presents a general Bernstein–von Mises result that justifies a normal approximation to the posterior. This facilitates posterior inference and avoids the need for Markov Chain Monte Carlo methods, which may become prohibitive in large samples, specifically for the Hazard-Response model.
Further details on prior specification and variable selection are presented in 
Section~\ref{sec:modelbuild}.
A comprehensive simulation study is presented in Section~\ref{sec:simulation} and our two case studies are presented in Section~\ref{sec:applications}.  Finally, a discussion of the paper is presented in Section~\ref{sec:discusssion}. 
The R and Julia code, along with the data used in our real-data applications, are available at \url{https://github.com/FJRubio67/SurvMODE}.

\section{Survival regression models via ODEs}\label{sec:model}

Let ${\bf o} = \{o_1,\dots,o_n\}$ denote a sequence of survival times, and let $c_i \in \mathbb{R}_+$ be the corresponding right-censoring times. Define the observed times as $t_i = \min\{o_i, c_i\}$ and the censoring indicator as $\delta_i = \mathrm{I}(o_i \leq c_i)$, for $i = 1, \dots, n$. Let $\bx_i \in \mathbb{R}^p$ denote the vector of individual covariates for observation $i$. Throughout, we denote the probability density function (pdf), cumulative distribution function (cdf), survival function, and hazard function by $f(t \mid \bx_i)$, $F(t\mid \bx_i)$, $S(t\mid \bx_i)$, and $h(t\mid \bx_i)$, respectively, for $t > 0$ and each covariate vector $\bx_i$. We also assume that $h(t\mid \bx_i)$ is a differentiable function with respect to $t$ 
for each covariate vector $\bx_i$.

{As already mentioned, we generalize \cite{christen:2024} to handle covariates.}
We define a class of hazard-based parametric survival regression models through a system of first-order ODEs with an initial condition, as follows. Define the system of ODEs 
\begin{equation}
\begin{cases}
\bY'(t \mid \btheta, \bx_i) =  \psi_{\btheta}\left(\bY(t\mid \btheta,\bx_i), t, \bx_i \right), \\
H'(t \mid \btheta, \bx_i) = h(t \mid \btheta, \bx_i),
\end{cases}
\label{eq:general_ode}
\end{equation}
{where $\bY(t\mid \btheta, \bx_i) = \left(h(t \mid \btheta, \bx_i), q_1(t \mid \btheta, \bx_i), \ldots, q_m(t \mid \btheta, \bx_i)\right)^{\top}$, $t > 0$, and $q_j: \mathbb{R}^+ \to \mathbb{R}$, $j = 1, \dots, m$, are differentiable functions with respect to $t$ for each covariate vector $\bx_i$ and parameter $\btheta \in \Theta \subset \mathbb{R}^d$. The right-hand side, $\psi_{\btheta}: D \to R^{m+1}$, is parametrised by $\btheta$, and determines the dynamic behaviour of $\bY(t\mid \btheta,\bx_i)$, including the dynamic behaviour of the hazard function $h(t \mid \btheta, \bx_i)$ and the state variables $q_1(t \mid \btheta, \bx_i), \ldots, q_m(t \mid \btheta, \bx_i)$. The initial conditions are given by $\bY(t_0\mid \btheta, \bx_i) = \bY_0(\bx_i)$ and $H(t_0 \mid \btheta, \bx_i) = H_0(\bx_i)$, taking $t_0 = 0$ and $H_0(\bx_i) = 0$ (corresponding to all individuals being alive at the initial time) as the typical choice, without loss of generality. The idea behind formulation \eqref{eq:general_ode} is that we are interested in modelling the dynamic behaviour of the hazard function $h(t \mid \btheta, \bx_i)$. However, in some cases the behaviour of $h(t \mid \btheta, \bx_i)$ cannot be adequately described without accounting for other quantities that help predict its evolution over time \citep{hirsch:2013}. These quantities are represented through state variables $q_j(t \mid \btheta, \bx_i)$, $j=1,\dots,m$ \citep[see][for details]{christen:2024}.}

{As in \cite{christen:2024}}, throughout, we restrict our attention to autonomous systems of ODEs:
\begin{equation}
\begin{cases}
\bY'(t \mid \btheta, \bx_i) =  \psi_{\btheta}\left(\bY(t\mid \btheta,\bx_i), \bx_i \right), \\
H'(t \mid \btheta, \bx_i) = h(t \mid \btheta, \bx_i).
\end{cases}
\label{eq:general_ode_auto}
\end{equation}
{In the ODE system \eqref{eq:general_ode_auto}, the right-hand side $\psi_{\btheta}: D \to \mathbb{R}^{m+1}$ no longer depends explicitly on $t$, but it remains parametrised by $\btheta \in \Theta \subset \mathbb{R}^d$.
The existence of solutions for autonomous systems of ODEs has been extensively studied, and general results are available that guarantee the existence and uniqueness of solutions for broad classes of such systems. In particular, if the vector field $\psi_{\btheta}: D \to \mathbb{R}^{m+1}$ is Lipschitz continuous in $\by$ for each covariate vector $\bx_i$ (\textit{i.e.}, $\psi_{\btheta}(\by, \bx_i)$), and the initial condition $\bY_0(\bx_i)$ lies in the interior of $D$, then a unique solution to \eqref{eq:general_ode_auto} exists in a neighbourhood of the initial value (see, \textit{e.g.}, \cite{po-fang1999}).
Since hazard functions are positive, \textit{i.e.}~it should be the case that $h(t\mid \btheta,\bx_i) > 0$, the main conditions on \eqref{eq:general_ode_auto}, specifically on the vector field $\psi_{\btheta}$, its domain $D$, and the initial condition $\bY_0(\bx_i)$, should be such that the solution for $h(t\mid \btheta,\bx_i)$ remains positive for all $t > 0$, and for any parameter value $\btheta \in \Theta \subset \mathbb{R}^d$ and covariate vector $\bx_i$. That is, \eqref{eq:general_ode_auto} represents a family of ODE systems in which one state variable corresponds to the hazard function $h$, leading to a family of hazard functions defined by $\btheta$ and $\bY_0(\bx_i)$. 
The cumulative hazard function $H$ is also included in the formulation \eqref{eq:general_ode_auto}, and is therefore obtained as part of the system's solution. The initial condition must satisfy $h(0\mid \btheta,\bx_i) \geq 0$, but is otherwise arbitrary. For instance, the initial condition in \eqref{eq:general_ode_auto} could specify $h(0\mid \btheta,\bx_i) = h_0 > 0$ and $H(0 \mid \btheta, \bx_i) = 0$, indicating that the hazard function takes a non-negative, finite value at $t = 0$, and that all individuals are alive at the the initial time.}

Autonomous systems \citep{boyce:2021} are fundamental in dynamical systems theory because their behaviour is determined solely by the state variables, making them ideal for studying long-term system behaviour. 
Autonomous systems often allow for qualitative analysis (\emph{e.g.}~equilibrium points, stability, bifurcations) without requiring explicit solutions. 
Qualitative analyses of systems of ODEs help gain understanding of the evolution of the solution over time, and extending their interpretation across individuals with the same or similar characteristics.
This feature is particularly useful in survival analysis or epidemiology, as autonomous systems of ODEs can help explain why certain covariate patterns lead to increasing, decreasing, or plateauing hazard functions, even when the exact functional form of the solution is complex or not available analytically.
In the following sections, we present specific examples of systems of ODEs that can be used in \eqref{eq:general_ode_auto}, along with interpretations of their parameters.
{This general formulation allows for the use of, for example, population growth or species competition type models, or more general families of dynamical systems in \eqref{eq:general_ode_auto} \citep{hirsch:2013}. Some specific choices for the right-hand side of \eqref{eq:general_ode_auto} are presented below, and later applied in the real-data examples.}

To capitalise on the interpretability of the parameters of \eqref{eq:general_ode_auto}, we propose modelling the parameters $\btheta$ in \eqref{eq:general_ode_auto} using a transformed linear predictor. As previously discussed, this idea is similar to the modelling strategy used in generalised additive models for location, scale and shape (GAMLSS), although at the level of the hazard function obtained as a solution to \eqref{eq:general_ode_auto}. More specifically, we incorporate information about the covariates on each of the parameters through linear predictors:
\begin{eqnarray}
\varphi_k(\theta_{k,i}) = \beta_{k,0} + \bx_{k,i}^{\top}\bbeta_k, \quad k = 1, \dots d, \quad i=1,\dots,n, 
\label{eq:paramreg}
\end{eqnarray}
where $\varphi_k$ is an appropriate monotone, differentiable, mapping of the parameter $\theta_k$ into ${\mathbb R}$, for each value of the covariates $\bx_{k,i} \subseteq \bx_i$, $\bx_{k,i} \in {\mathbb R}^{p_k}$, used to model the parameter $\theta_k$, and  $\bbeta_k = (\beta_{k,1},\dots,\beta_{k,p_k})^{\top}$.
For example, for parameters taking values in the entire real line, $\varphi_k$ can be the identity function, while for positive parameters a possible choice is $\varphi_k(\theta_k) = \log(\theta_k)$.
Note that this formulation includes an intercept, which represents the baseline level of each parameter and account for potential sparsity. Let $\bmeta \in \tilde{\Theta} \subseteq {\mathbb R}^{\td}$ denote the full vector of parameters (including all regression coefficients), and $\td = \sum_{k=1}^d p_k$ be the corresponding dimension. 

Depending on the system of ODEs in \eqref{eq:general_ode_auto}, closed-form solutions for $h$ and $H$ may be available. For models with analytical solutions, this formulation resembles the modelling strategy used in GAMLSS, particularly in terms of computational cost and implementation. However, many ODE systems of practical interest lack closed-form solutions and therefore require numerical methods (ODE solvers). Consequently, implementing these models involves solving a system of ODEs for each individual, with distinct parameter values and time intervals $[0, t_i]$. Specifically, this requires solving
\begin{equation*}
\begin{cases}
\bY'(t\mid \bmeta, \bx_i) =  \psi_{\bmeta}\left(\bY(t \mid \bmeta,\bx_i), \bx_i \right), \\
H'(t \mid \bmeta, \bx_i) = h(t \mid \bmeta, \bx_i) ,
\end{cases}
\end{equation*}
at $t = t_i$, for $i = 1, \dots, n$, with initial conditions $\bY(0\mid \bmeta,\bx_i) = \bY_0(\bx_i)$ and $H(0\mid \bmeta,\bx_i) = 0$ (indicating that all individuals are alive at time $t = 0$).
Solving this system yields the values $h(t_i \mid \bmeta, \bx_i)$ and $H(t_i \mid \bmeta, \bx_i)$ for $i = 1, \dots, n$, which define the likelihood function of $\bmeta$, as discussed in Section \ref{sec:postinf}. The computational tools required for this implementation are described in detail in the following sections.

\subsection*{Logistic growth regression model}
A basic but useful example of a system of ODEs of type \eqref{eq:general_ode_auto} is the Logistic growth hazard model \citep{christen:2024}, which is defined as,
\begin{eqnarray}
\begin{cases}
h'(t)  =  \lambda h(t) \left(1 - \dfrac{h(t)}{\kappa}\right), & h(0) = h_0\\
H'(t)  =  h(t), & H(0) = 0.
\end{cases}
\label{eq:logisODE}
\end{eqnarray}
where $\lambda > 0$ represents the intrinsic growth rate of the hazard function, $\kappa > 0$ (carrying capacity) represents the upper or lower bound of the hazard function, and $h_0 > 0$ is the value of the hazard function at $t=0$. This ODE has the following analytic solution
\begin{eqnarray*}
h(t \mid \lambda, \kappa, h_0) &=& \frac{\kappa h_0 e^{\lambda t}}{\kappa + h_0 (e^{\lambda t} - 1)},\\
H(t \mid \lambda, \kappa, h_0) &=& \dfrac{\kappa}{\lambda }  \log \left(\dfrac{\kappa + h_0 \left(e^{\lambda  t}-1\right)}{\kappa} \right).
\end{eqnarray*}
This model can capture sigmoidal shapes (increasing, or decreasing) similar to the Weibull distribution, but with the important difference that the hazard function is upper bounded in contrast to the Weibull distribution. The upper-bound is arguably more realistic for modelling real data. For instance, when modelling human mortality or device failure, there is a physiological or physical limit to how high the instantaneous risk of failure can be. Some hazard models (\textit{e.g.}~Weibull with shape parameter $> 1$) allow the hazard function to grow unboundedly, which may lead to unrealistic predictions for extreme covariate patterns. An upper-bounded hazard mitigates this problem and reduces sensitivity to extrapolation.

We incorporate information about the covariates on the parameters of \eqref{eq:logisODE} through the log-linear predictors $\log(\lambda_i) = \beta_{1,0} + \bx_{1i}^{\top}\bbeta_1$, and $\log(\kappa_i) = \beta_{2,0} + \bx_{2i}^{\top}\bbeta_2$. We recommend assuming a fix value of $h_0$ (not dependent on covariates), and either estimating this parameter or fixing it using previous information (see \citealp{christen:2024} for a discussion on how to fix this value using contextual information). This strategy allows for modelling varying growth rates and carrying capacities, thus allowing for capturing crossing hazard and survival functions for groups of individuals, as illustrated in Section \ref{sec:applications}. 

\subsection*{Hazard-response regression model}
We now describe a more flexible hazard regression model that accommodates a variety of hazard shapes and enables interpretation of the time-varying effects of interventions (\textit{e.g.}~treatments) on the hazard function. The hazard-response model is defined through the system of ODEs \citep{christen:2024}:
\begin{eqnarray}
\begin{cases}
h'(t)  =  \lambda h(t) \left(1 - \dfrac{h(t)}{\kappa}\right) - \alpha q(t) h(t), & h(0) = h_0 \\
q'(t) =  \mu q(t) \left( 1- \dfrac{q(t)}{\kappa} \right) -\alpha q(t) h(t)  ,  &  q(0) = q_0 \\ 
H'(t)  =  h(t), & H(0) = 0,
\end{cases}
\label{eq:hazardresponse}
\end{eqnarray}
with $\lambda>0$, $\alpha \geq 0$, $\mu>0$, $\kappa>0$, $h_0>0$, and $q_0>0$. {This model can be viewed as a competitive Lotka–Volterra system \citep{murray:2002,christen:2024}, in which the hazard function $h(t)$ and the response $q(t)$ describe the magnitudes of two interacting (or competing) agents. The quantity $q(t)$ can be viewed as an unobserved or latent process. Its scale is therefore arbitrary, and, without loss of generality, its (potential) carrying capacity can be set equal to that of $h(t)$, namely $\kappa$. While it is possible to assign distinct competition coefficients to each equation, we adopt a parsimonious formulation using a common coefficient $\alpha$. The fully parametrised version of this system is discussed in \cite{murray:2002} in the context of ecological modelling, as a generalisation of the Lotka–Volterra predator–prey model.

A more intuitive description of \eqref{eq:hazardresponse} seems appropriate \citep[see also][]{christen:2024}. In the absence of a response or intervention, that is, when $q(t)=0$ for $t>0$, the hazard function evolves according to a logistic growth pattern, as described in \eqref{eq:logisODE}.
When the response function $q(t)$ is incorporated into \eqref{eq:hazardresponse}, it interacts with the hazard through the competitive term $\alpha q(t)h(t)$. The function $q(t)$ acts as a mitigating mechanism that counterbalances the hazard, representing effects such as immune response, therapeutic action, or other responses aimed at reducing the hazard level.
If there is no competition ($\alpha=0$), then $h(t)$ follows a logistic growth, as in \eqref{eq:logisODE}, and converges to its carrying capacity $\kappa$ as $t \to \infty$. 
For $\alpha>0$, the interaction term $\alpha q(t)h(t)$ introduces competition between the two processes, reducing both since the term appears with a negative sign in each equation.
We therefore interpret the system of ODEs \eqref{eq:hazardresponse} as representing a hazard function that captures the combined forces of mortality over time, including, for instance, those associated with a specific disease such as cancer, and a competing response that acts to counterbalance them (hence the name \textit{hazard-response} model).
}

The solution to \eqref{eq:hazardresponse} does not exist in closed form; that is, there is no {explicit} solution to this system of ODEs. Therefore, evaluating the hazard function at specific time points requires the use of numerical methods (ODE solvers), as we discuss in the next section. However, one of the important features of this model is that the parameter values allow for identifying ``attractors'' of the solution, which define the shape of the hazard function \citep{christen:2024}. Briefly, if we define
$h^* = \kappa \left( \frac{1-\alpha \kappa \lambda^{-1}}{D} \right)$ and $q^* = \kappa \left( \frac{1-\alpha \kappa \mu^{-1}}{D}  \right)$, with $D=1 - \frac{(\alpha \kappa)^2}{\lambda \mu}$, then
\begin{itemize}
    \item    If $q^* < 0$, then the hazard $h(t)$ reaches its carrying capacity as $t\to\infty$ and the response $q(t)$ ``losses'' the competition ($q(t) \to 0$).

    \item If $h^* < 0$, then the response $q(t)$ ``wins'' the competition ($q(t) \to \kappa$) and the hazard ``losses'' the competition ($h(t) \to 0$).

    \item If both $h^* >0$ and $q^* > 0$, the hazard remains positive ($h(t)\to h^* >0$), although not at its maximum (or minimum), but in an equilibrium with the response ($h^* < \kappa$\; and also $q(t)\to q^* < \kappa$).
\end{itemize}

Our strategy for building a regression model consists of modelling the parameters $\btheta = (\lambda,\alpha,\mu,\kappa)^{\top}$ using log-linear predictors $\log(\theta_{k,i}) = \beta_{k,0} + \bx_{k,i}^{\top}\bbeta_k$, $k = 1, \dots 4$, $i=1,\dots,n$, while fixing the initial conditions $h_0$ and $q_0$ using contextual information \citep{christen:2024}, as discussed in Section \ref{sec:applications}. This allows for identifying groups (defined by their covariate values) belonging to different attractors, and thus providing insights about the reasons for the resulting shape of the hazard function in terms of the effectiveness of the response function $q(\cdot)$.

\section{Model building and posterior inference}\label{sec:postinf}
In this section we provide details on the computations required to implement the likelihood function in contexts both with and without an analytical solution to \eqref{eq:general_ode_auto}. We then discuss model building through Bayesian variable selection for the covariates entering each linear predictor. Additionally, we explain how to use general-purpose Markov Chain Monte Carlo (MCMC) samplers to draw from the posterior distribution, and highlight a result on the asymptotic normality of the posterior distribution under certain conditions on the right-hand side of \eqref{eq:general_ode_auto}.
\subsection{Likelihood}
Consider a hazard-based parametric regression model obtained as the solution to \eqref{eq:general_ode_auto}. For the case where the initial conditions $\bY_0(\bx_i)$ are fixed, the log-likelihood function is:
\begin{eqnarray*}
\ell_n(\bmeta) &=&  \sum_{i=1}^n   \delta_i \log h(t_i \mid \bmeta, \bx_i) - \sum_{i=1}^n H(t_i \mid \bmeta, \bx_i)\\
 &=& \sum_{i=1}^n   \delta_i \log f(t_i \mid \bmeta, \bx_i) + \sum_{i=1}^n (1-\delta_i)\log S(t_i \mid \bmeta, \bx_i),
\end{eqnarray*} 
For the logistic growth regression model \eqref{eq:logisODE}, an analytical solution is available, making its implementation cost equivalent to that of any standard parametric regression model. This enables both likelihood-based and Bayesian inference (once priors for all parameters are specified), using general-purpose optimisation routines or MCMC sampling methods. As we will demonstrate in Section~\ref{sec:applications}, the initial condition $h_0$ can also be set as a parameter and included in the likelihood.

For the hazard-response model \eqref{eq:hazardresponse}, implementing the log-likelihood and log-posterior functions requires evaluating the hazard function $h(\cdot)$ and its integral $H(\cdot)$ at time points $t_i$, for $i = 1, \dots, n$, using parameter values determined by the corresponding log-linear predictors $\log(\theta_k) = \beta_{k,0} + \bx_{k,i}^{\top}\bbeta_k$, $k = 1, \dots 4$. Consequently, evaluating the log-posterior involves solving $n$ systems of ODEs defined by \eqref{eq:hazardresponse}, each with different parameter values and over different time intervals $[0, t_i]$. This setup constitutes an ensemble of ODE systems \citep{utkarsh:2024}. A key advantage of such ensembles is their embarrassingly parallelisable structure, which permits efficient multi-threading or the use of advanced techniques such as GPU-accelerated ODE solvers, as available in the Julia programming language \citep{rackauckas:2017,utkarsh:2024}, thereby speeding up log-posterior evaluation. In particular, the \texttt{DifferentialEquations.jl} library in Julia provides a wide range of tools and methods for obtaining numerical solutions to systems of ODEs. As a result, posterior sampling can be carried out using general-purpose MCMC libraries such as \texttt{AdaptiveMCMC.jl} and \texttt{Turing.jl}. Nonetheless, for large sample sizes and long MCMC chains, the computational burden may remain substantial. To mitigate the need for full MCMC sampling, the next section introduces conditions on the right-hand side of \eqref{eq:general_ode_auto} under which a normal approximation to the posterior distribution becomes feasible.

\subsection{Normal approximation of the posterior distribution}

The following result establishes sufficient conditions for the asymptotic normality (Bernstein–von Mises theorem) of the posterior distribution of the parameters in survival regression models defined by an autonomous system of ODEs \eqref{eq:general_ode_auto} with fixed initial conditions $\bY_0$. The key idea is to translate the classical conditions of the Bernstein–von Mises theorem \citep{van:2000} into assumptions on the right-hand side of \eqref{eq:general_ode_auto}, adapted to account for right-censoring.

\begin{proposition}\label{prop:BvM}
Consider a survival regression model defined through an autonomous system of ODEs \eqref{eq:general_ode_auto} with fixed initial conditions $\bY_0$ and true parameter $\bmeta^*$. 
Suppose that the prior distribution on $\bmeta\in \tilde{\Theta} \subseteq {\mathbb R}^{\td}$ is absolutely continuous and satisfies $\pi(\bmeta^*) > 0$.
Define $\Delta_n = \sqrt{n}\left(\bmeta - \bmeta^*\right)$.
Then, under conditions C1-C6 (see Supplementary Material~\ref{app:technical}), it follows that  

\begin{equation*}
\lVert \pi(\Delta_n \mid \bt, \bdelta, \bX) -    N_{\td}\left(\mathbf{0}, Q^{-1}(\bmeta^*) \right) \lVert_{\text{TV}} \, \stackrel{\Pr}{\to} 0, \quad \text{as } n\to\infty,
\end{equation*}
where $Q(\bmeta) = {\mathbb E}\left\{ \left[\nabla_{\bmeta} \ell_i(\bmeta)\right] \left[\nabla_{\bmeta} \ell_i(\bmeta)\right]^{\top} \right\}$, $\ell_i(\bmeta) 
=   \delta_i \log h(t_i \mid \bmeta, \bx_i) - H(t_i \mid \bmeta, \bx_i)$ is the contribution of a single observation to the log-likelihood, and $\lVert \cdot \lVert_{\text{TV}}$ is the total variation norm.
\end{proposition}

Condition C1 restricts the result to the M-closed scenario (correct model specification) and constrains both the parameter space and $t$ to a compact set, as most results on the existence and uniqueness of solutions of systems of ODEs only apply to compact domains.
In practice, this compact set can be assumed to be arbitrarily large, and time-to-event analyses are typically limited to bounded follow-up periods. This condition implies the existence of uniformly consistent hypothesis tests \citep{van:2000,nickl:2013}, which is a fundamental requirement in Bernstein-von Mises theorems.
Condition C2 requires the right-hand side of the ODE system to be three times continuously differentiable with respect to the parameters and $\bY$; this implies twice continuous differentiability and ensures the existence of solutions. 
Condition C3 explicitly assumes identifiability of the hazard function for simplicity. However, more specific conditions on the system of ODEs that guarantee identifiability are available in \cite{miao:2011} and \cite{qiu:2022} for families of linear and nonlinear systems. In addition, state-of-the-art numerical methods for detecting non-identifiability of systems of ODEs are implemented in the Julia package \texttt{StructuralIdentifiability.jl}. The identifiability of many ODE systems of practical interest, including those used in our applications, has already been established.
Condition C4 assumes that the asymptotic variance is non-singular, while Condition C5 prevents collinearity in sufficiently large samples. 
Condition C6 restricts the asymptotic result to the case of non-informative censoring \citep{le:1988}. 

Based on Proposition \ref{prop:BvM}, we consider a normal approximation to the posterior distribution. The idea is to use the approximation 
\begin{equation*}
    \bmeta \mid \bt, \bdelta, \bX \stackrel{\cdot}{\sim} N_{\tilde{d}}\left(\tilde{\bmeta}, H_{\tilde{\bmeta}}^{-1}\right),
\end{equation*}
where $\tilde{\bmeta}$ denotes the maximum a posteriori (MAP), and $H_{\tilde{\bmeta}}^{-1}$ denotes the inverse of the Hessian of $\ell_n(\bmeta) + \log \pi(\bmeta)$ evaluated at $\tilde{\bmeta}$. This kind of approximations have been used in a number of parametric survival models satisfying standard regularity conditions \citep{chen:2014}. 


\subsection{Regression model building}\label{sec:modelbuild}

In principle, one could include all variables in each equation of the regression model defined in \eqref{eq:paramreg}. However, since the parameters control different aspects of the model, it is reasonable to expect that not all variables will play an important role (or any at all) in explaining every parameter. Therefore, it is desirable to construct the regression model by selecting the relevant variables for each linear predictor in \eqref{eq:paramreg}. In the Bayesian framework, two general approaches to achieve this are: (i) shrinking the coefficients toward zero using a shrinkage prior (the typical approach in Bayesian distributional regression), or (ii) performing explicit variable selection \citep{tadesse:2021}. Both approaches are valid, but in this paper we focus on the latter.

To formulate the variable selection problem, let us define the matrix of variable inclusion indicators in each linear predictor $\bgamma = \left(\bgamma_1^{\top},\dots,\bgamma_d^{\top}\right)^{\top}$, where $\bgamma_k = (\gamma_{k,1},\dots, \gamma_{k,p})^{\top}$, $k=1,\dots,d$, and
\begin{eqnarray*}
{\gamma}_{k,j} =
\begin{cases}
0 & \text{if} \quad \beta_{k,j} = 0,\\
1 & \text{if} \quad \beta_{k,j} \neq 0.\\
\end{cases}
\end{eqnarray*}
Note that we do not impose any hierarchy or heredity principle in the variable selection process. However, hierarchical inclusion rules could be incorporated in specific applications where such structure is desirable or when prior information supports a particular hierarchy.
Our specific prior for the $\gamma_{k,j}$'s is discussed below.

We adopt group $g$-Priors \citep{liang:2008,rossell:2023} for the regression coefficients $\bbeta_{\bgamma_k}$ in model $\bgamma$ and diffuse or informative priors for the intercepts, as we will always consider including intercepts. 
More specifically,
\begin{equation}\label{eq:gprior}
\pi(\bmeta_{\bgamma} \mid \bgamma) = \prod_{k=1}^d \pi_{k,0}(\beta_{k,0}) \prod_{k=1, \, p_k>0}^d  N_{p_k}\left(\bbeta_{\bgamma_k} \mid \bm{0}, g_k \left(\bX_{\bgamma_k}^{\top}\bX_{\bgamma_k}\right)^{-1}\right),
\end{equation}
where $\bX_{\bgamma_k}$ denotes the design matrix associated with the linear predictor for parameter $k$ in model $\bgamma_k$, and $p_k = \sum_{j=1}^p \gamma_{k,j}$. 
The motivation for using $g$-priors follows our aim of performing Bayesian variable selection. The $g$-prior scales the prior covariance of regression coefficients by $\left(\bX_{\bgamma_k}^{\top}\bX_{\bgamma_k}\right)^{-1}$, which ensures scale invariance and makes the prior sensitive to the amount of information in the data. 
Several choices of the hyperparameter $g_k$ can be made to control the degree of sparsity, or it can be modelled using a hierarchical prior on $g_k$ \citep{liang:2008}. 
For the intercepts $\beta_{k,0}$, we adopt normal priors with zero mean and large variance. 

For the inclusion indicators $\bgamma$, we adopt a complexity-type prior \citep{castillo:2015}: $\pi(\bgamma) \propto \tilde{d}^{-\tilde{C} \vert \bgamma \vert}$, where $\tilde{C}\geq 0$ is a hyperparameter, and $\vert \bgamma \vert$ is the number of active variables (model size). This prior penalises models with more variables, promoting sparsity and parsimony.

The marginal posterior probability of each model $\bgamma$ can be obtained as \citep{hoeting:1999}:
\begin{equation*}\label{eq:modelposterior}
\pi(\bgamma \mid \bt, \bdelta, \bX) = \frac{p(\bt, \bdelta, \bX \mid \bgamma)\pi(\bgamma)}{\sum_{\bgamma'} p(\bt, \bdelta, \bX \mid \bgamma')\pi(\bgamma')}
\end{equation*}
where $p(\bt, \bdelta, \bX \mid \bgamma)$ denotes the marginal likelihood of the data
\begin{equation}\label{eq:marginallikelihood}
p(\bt, \bdelta, \bX \mid \bgamma)= \int p(\bt, \bdelta, \bX \mid \bm{\eta}_{\bgamma}) \pi(\bm{\eta}_{\bgamma}\mid \bgamma) d\bm{\eta}_{\bgamma} \, .
\end{equation}

We take advantage of the asymptotic normality of the posterior distribution to approximate the marginal likelihood via the Laplace method:
\begin{equation*}
\widehat{p}(\bt, \bdelta, \bX \mid \bgamma)= \exp\{\ell(\tilde{\bmeta}_{\bgamma}) + \log \pi(\tilde{\bmeta}_{\bgamma} \mid \bgamma) \}
(2\pi)^{\td_{\bgamma}/2} \left\vert H(\tilde{\bmeta}_{\bgamma}) + \nabla^2 \log\pi(\tilde{\bmeta}_{\bgamma} \mid \bgamma) \right\vert^{-1/2},
\label{eq:laplace_approx}
\end{equation*}
where $\tilde{\bmeta}_{\bgamma}= \arg\max_{\bmeta_{\bgamma}} \lbrace \ell_n(\bmeta_{\bgamma}) + \log \pi(\bmeta_{\bgamma} \mid \bgamma)\rbrace$ is the maximum a posteriori (MAP) under prior $\pi(\bmeta_{\bgamma} \mid \bgamma)$, $\td_{\bgamma} = \sum_{k,j} \gamma_{k,j}$, and $H$ is the Hessian matrix of the log-likelihood function.

For computation of the MAP estimate and the Hessian matrix, we use the \texttt{Optim.jl} and \texttt{ForwardDiff.jl} libraries in Julia, with evaluations of the log-posterior parallelised via multi-threading. The total number of possible models is $2^{\tilde{d}}$,which may be prohibitively large for computing all model posterior probabilities. To address this, we propose using a Gibbs sampler to explore regions with high posterior probability.
The corresponding algorithm is presented in the Supplementary Material \ref{app:gibbs}. This algorithm updates each of the $p$ variables within a given linear predictor, iterating over all $d$ linear predictors. While reverse moves or more sophisticated proposals could also be considered \citep{liang:2023}, we found that this simple strategy performs well in our real-data application in Section \ref{sec:applications}, quickly eliminating spurious variables from each linear predictor.

\section{Simulation study}\label{sec:simulation}
{We now present a simulation study to assess the inferential properties of the hazard–response model \eqref{eq:hazardresponse} under weakly informative priors, as an illustrative example of the proposed class of survival models \eqref{eq:general_ode_auto}. Specifically, we evaluate the model's ability to recover the true parameter values through Bayesian point and interval estimates, and we compare the posterior samples obtained using an adaptive MCMC sampler with those from the asymptotic normal approximation in Proposition \ref{prop:BvM}. This study highlights the interplay between sample size and censoring, providing simple guidelines on the sample sizes required to reliably fit the proposed models.
In a second simulation study, in section~\ref{subsec:benchmarking}, we compare the ability of the hazard–response model \eqref{eq:hazardresponse} to predict individual hazard functions against that of a flexible model with a rich hazard structure but a common baseline hazard across the population. This study illustrates a scenario in which the assumption of a single baseline hazard leads to poor predictions for certain subgroups that exhibit distinct hazard shapes.}

\subsection{Inferential properties}\label{subsec:validation}
We simulate time-to-event data, $\{o_1,\dots,o_n\}$, from the hazard-response model \eqref{eq:hazardresponse} using Algorithm \ref{alg:simHT} in the Supplementary Material, where each model parameter is associated with a (different) single covariate:
\begin{eqnarray}
\log(\lambda_i) &=& \beta_{1,0} + \beta_{1,1} x_{1i}, \nonumber \\
\log(\kappa_i) &=& \beta_{2,0} + \beta_{2,1} x_{2i}, \nonumber\\
\log(\alpha_i) &=& \beta_{3,0} + \beta_{3,1} x_{3i}, \nonumber\\
\log(\mu_i) &=& \beta_{4,0} + \beta_{4,1} x_{4i}, 
\label{eq:reghr}
\end{eqnarray}
where $x_{1i} \sim \text{Bernoulli}(0.5)$, $x_{2i} \sim \text{Bernoulli}(0.5)$, $x_{3i} \sim \text{N}(0,1)$, $x_{4i} \sim \text{N}(0,1)$, $i=1,\dots,n$, are independent. To evaluate the effect of sample size,  we consider samples sizes $n=500,1000,2000$.
To assess the effect of censoring, we consider administrative censoring values of $C = 2$ and $C = 10$, which induce censoring rates of $20\%$ and $40\%$, respectively, and correspond to markedly different follow-up periods. The observed times are given by $t_i = \min\{o_i, C\}$, and the censoring indicators are defined as $\delta_i = I(o_i \leq C)$.
We assume the true values of the eight regression coefficients as $\bbeta = (\beta_{1,0} , \beta_{1,1},\beta_{2,0} , \beta_{2,1},  \beta_{3,0} , \beta_{3,1}, \beta_{4,0} , \beta_{4,1})^{\top} = (1.5, 0.5, 0.5, -0.5, 1.0, 0.5, 3.0, -0.5)^{\top}$, which resemble the parameter values in the application in Section \ref{sec:applications}. We simulate $M=500$ data sets in each of these scenarios, where the solution obtained with the Julia package \texttt{DifferentialEquations.jl} is saved at steps of length $0.01$. Algorithm \ref{alg:simHT} in the Supplementary Material \ref{app:hrsim} shows the steps to simulate survival times from this model. As discussed in \citep{christen:2024}, the quality of the simulated samples depends on the number of time points and the step length used in the simulation process.
For each data set, we fit the hazard–response regression model defined by equations \eqref{eq:hazardresponse} and \eqref{eq:reghr}, using weakly informative $N(0, 10)$ priors for all parameters. We compute the maximum a posteriori (MAP) estimator and obtain a normal approximation of the posterior distribution to construct the $95\%$ credible intervals for each parameter, obtained as the empirical $2.5\%$ and $97.5\%$ quantiles. The Hessian matrix required to construct the credible intervals is obtained numerically using the \texttt{ForwardDiff.jl} Julia package. We then record whether each parameter is included in its corresponding $95\%$ credible interval. The mean, median, and root mean squared error (RMSE) of the MAPs, along with the coverage proportion and width of the credible intervals is reported in the Supplementary Material \ref{app:simresults}. 

From Table \ref{tab:C20n1000} and Figure \ref{fig:C20n1000} below, as well as \ref{tab:C20n500}-\ref{tab:C40n2000} and Figures \ref{fig:C20n500}-\ref{tab:C40n2000} in the Supplementary Material \ref{app:simresults} for the remaining results, we can observe an overall good coverage (close to the nominal value) in all cases with $20\%$ censoring. However, we can see that administrative censoring can affect the coverage of some parameters as this type of censoring precludes observing survival times beyond a certain point, making it difficult to learn about the parameters in the linear predictor for the shape parameter $\alpha$, which represents the strength of the competition between the hazard and the response functions. The mean length of the credible intervals is reduced by increasing the sample size or reducing the censoring rate, as expected from the asymptotic theory.
The MAPs are relatively close to the true parameter values, which suggests good inferential performance, as well as good choice of the step of length $0.01$ for the simulation study, which controls the quality of the approximation (see \citealp{christen:2024} for more details). The bias and variability of the MAP is clearly reduced by increasing the sample size or the reducing censoring rate.

\begin{figure}[h!]
    \centering
\begin{tabular}{c c c}
\includegraphics[width=0.3\textwidth]{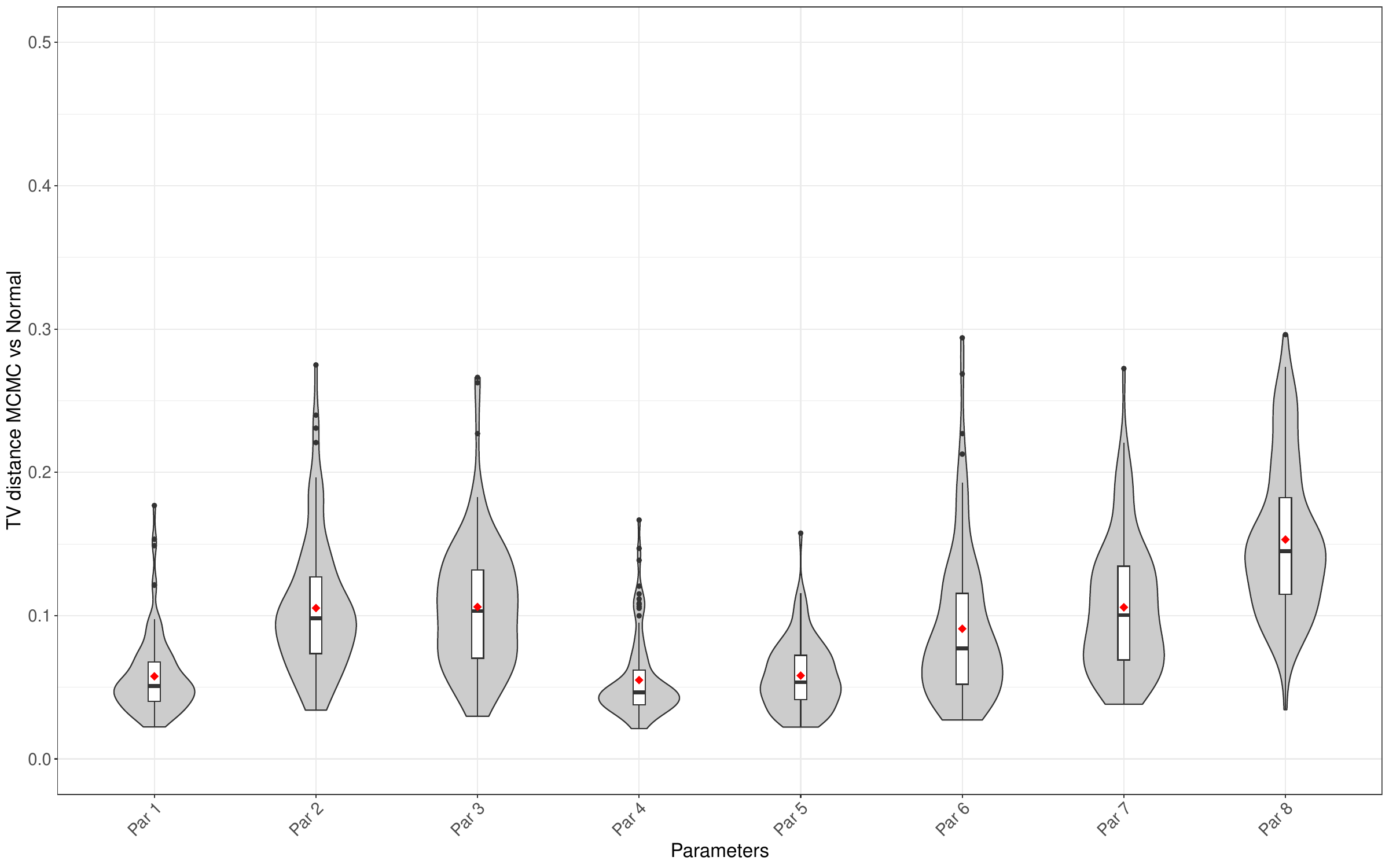} & \includegraphics[width=0.3\textwidth]{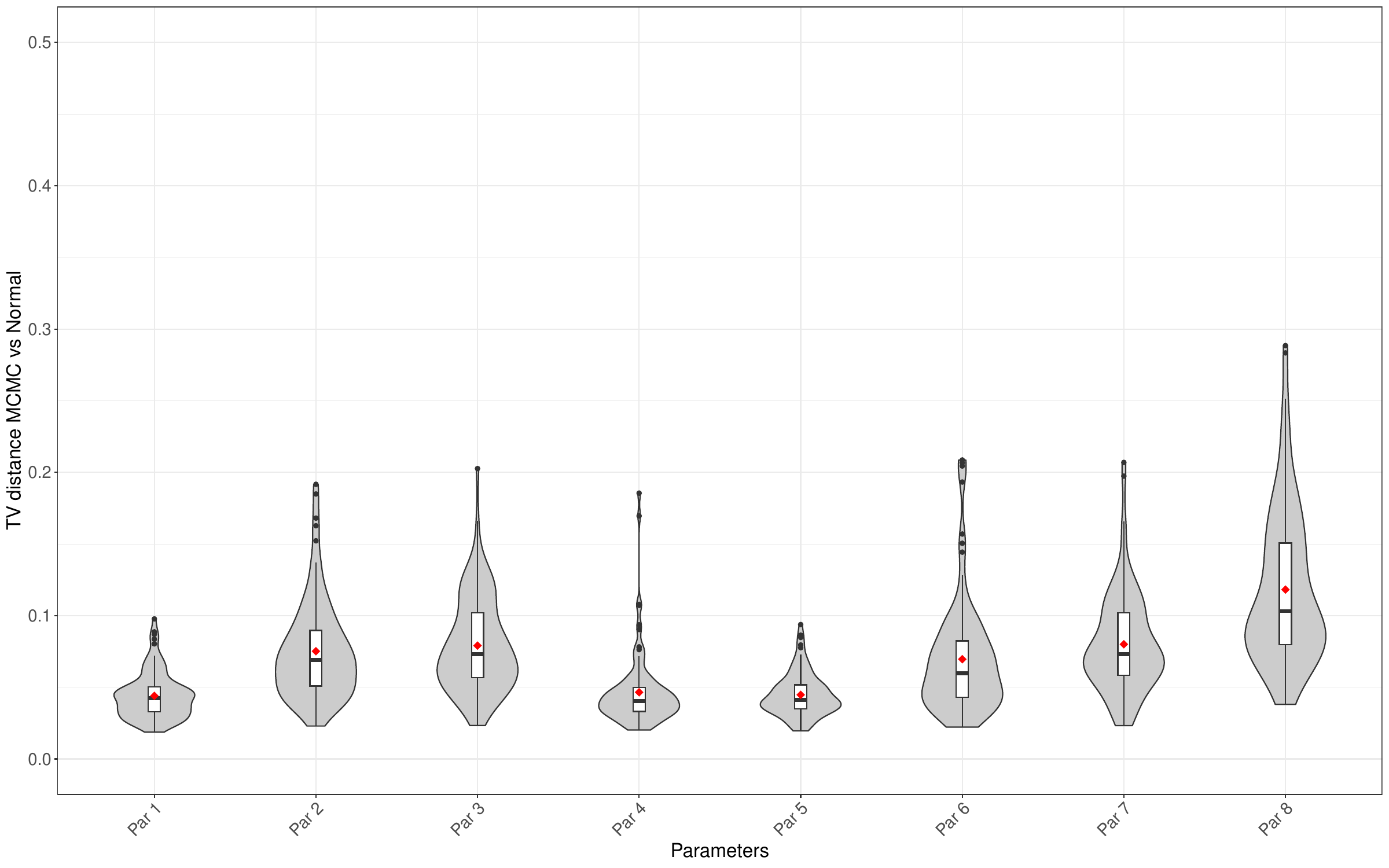} & \includegraphics[width=0.3\textwidth]{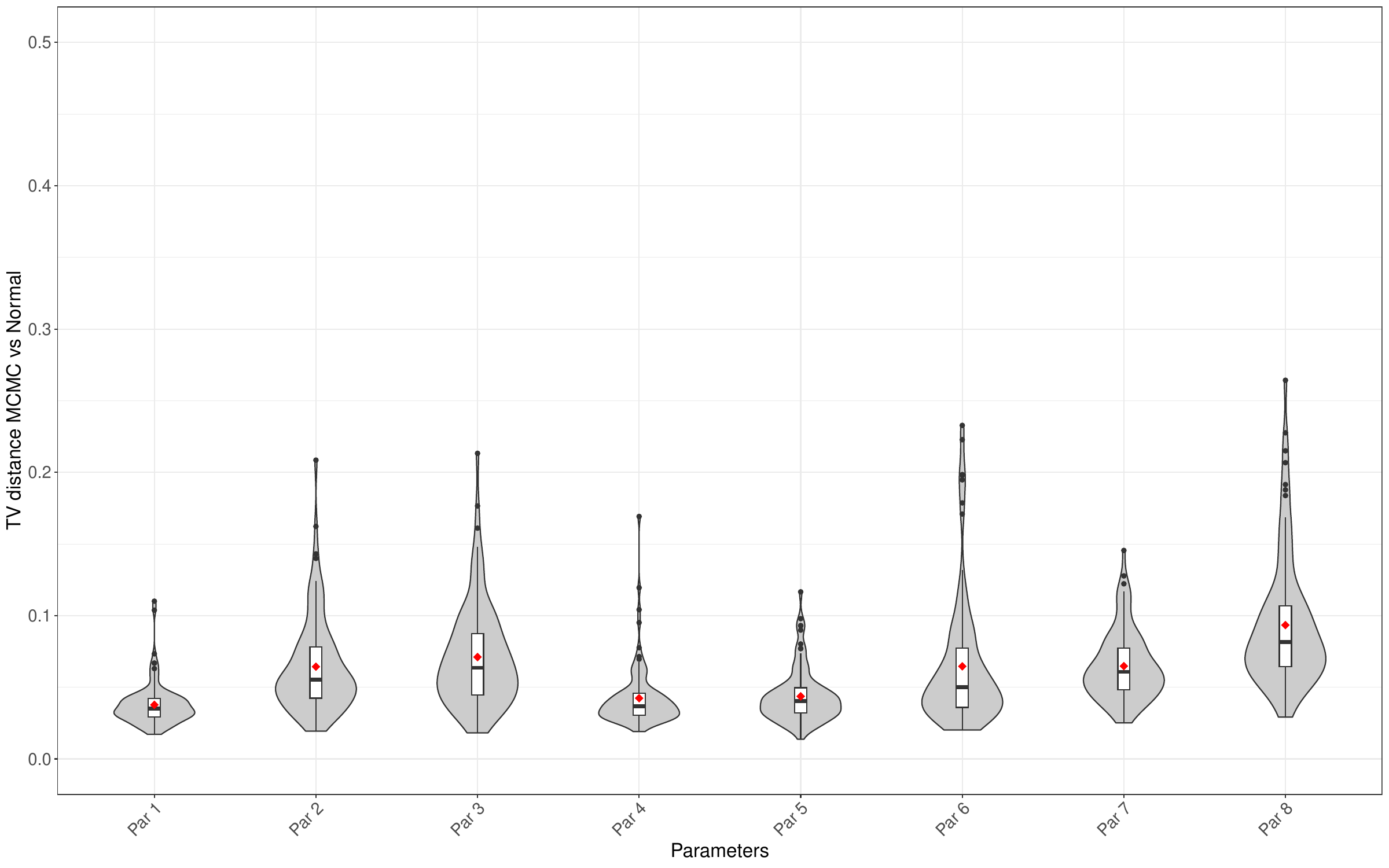} \\
 (a) & (b) & (c) \\
 \includegraphics[width=0.3\textwidth]{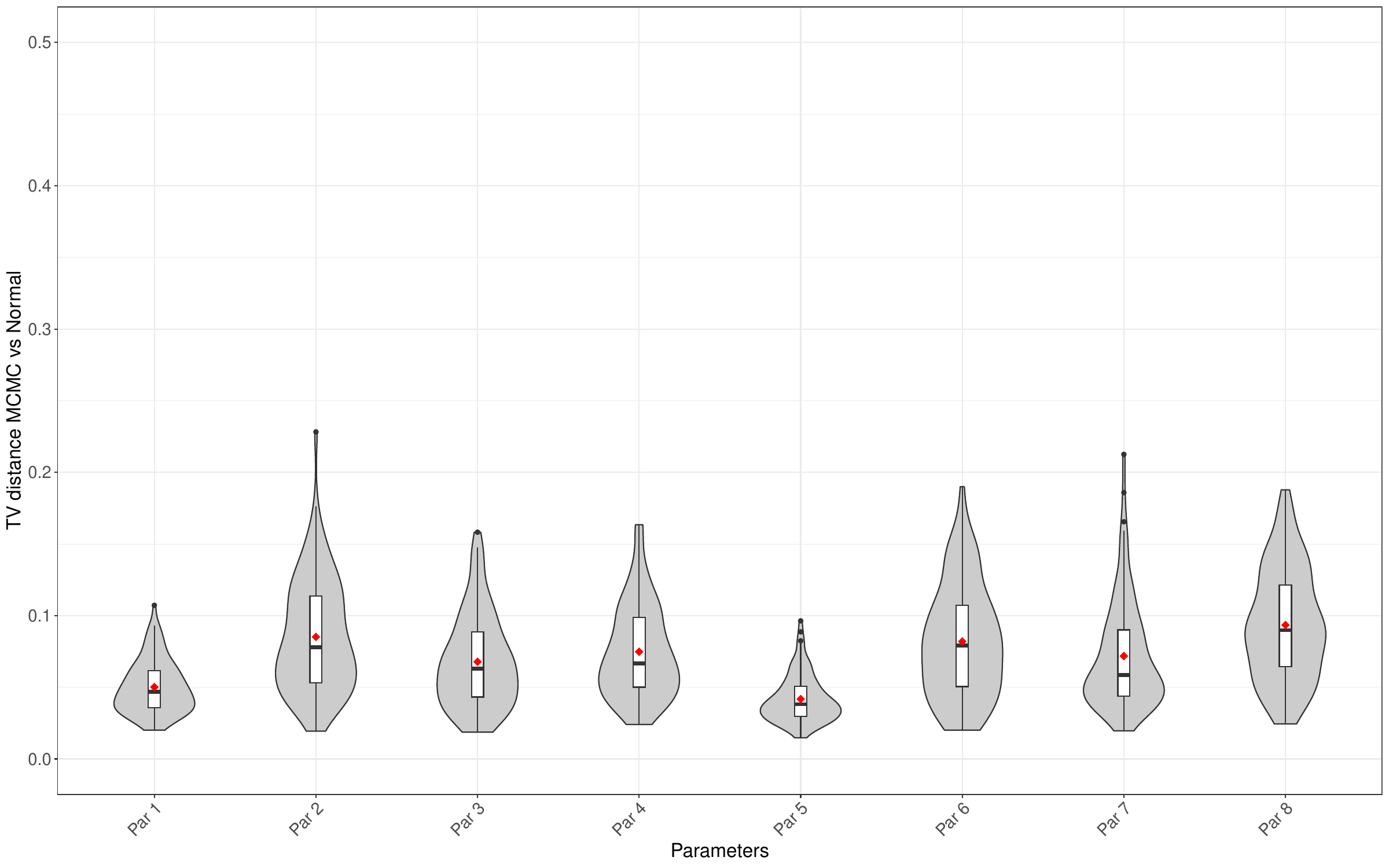} & \includegraphics[width=0.3\textwidth]{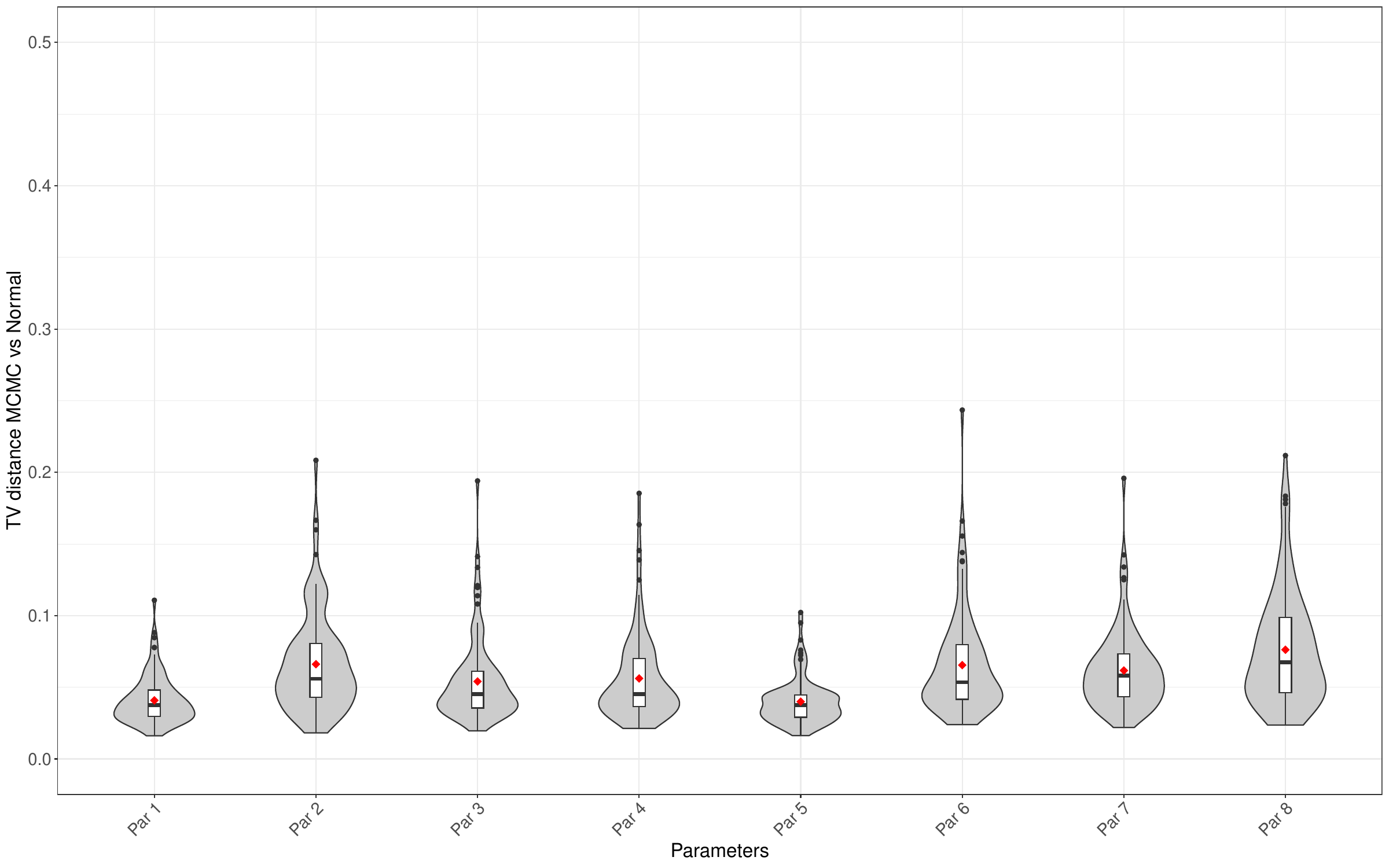} & \includegraphics[width=0.3\textwidth]{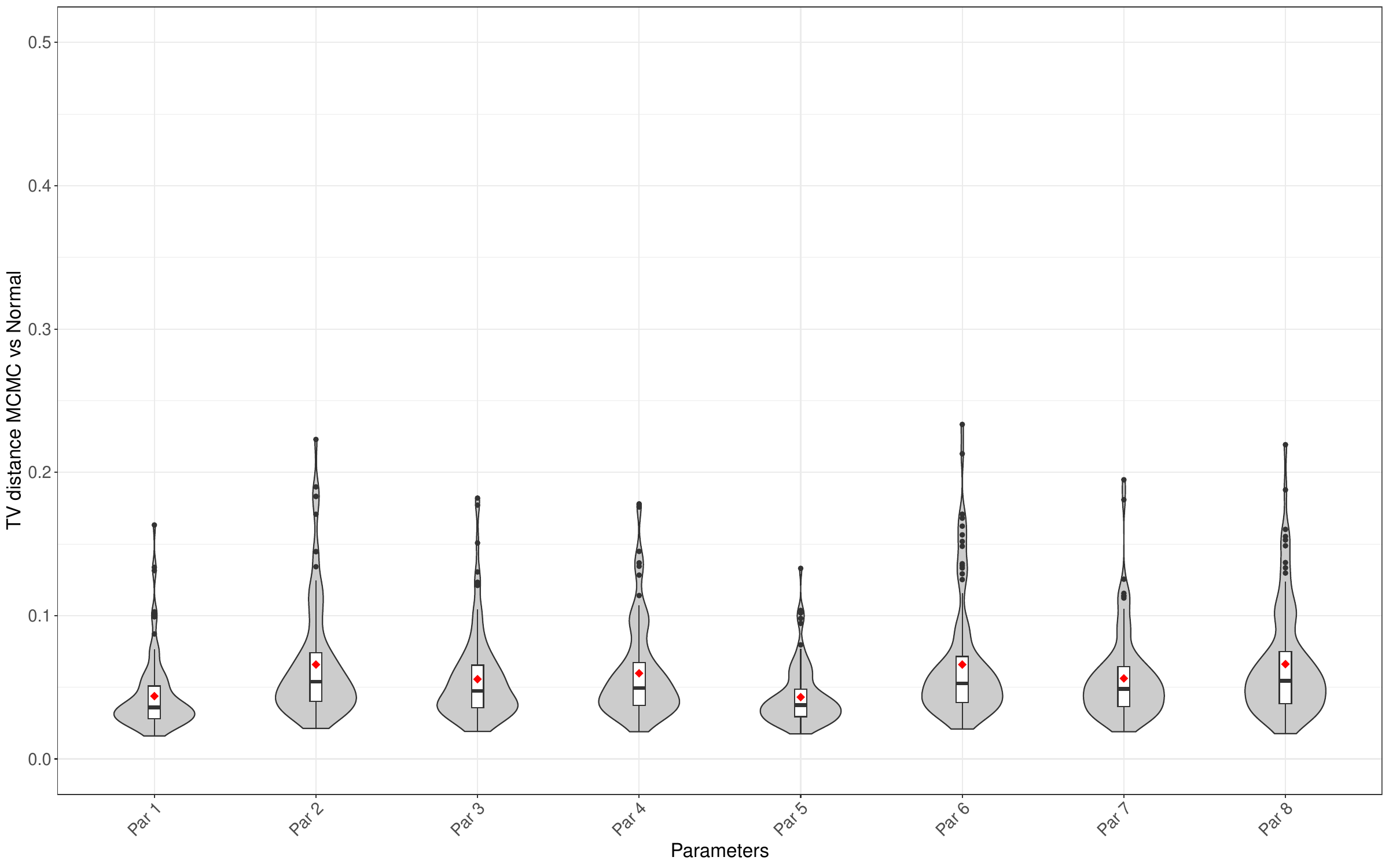} \\
 (d) & (e) & (f) \\
 \end{tabular}
    \caption{TV distance for MCMC \emph{vs.} Normal approximation marginal posterior samples: (a) $n=500$, $40\%$ censoring, (a) $n=1000$, $40\%$ censoring,(a) $n=200$, $40\%$ censoring, (a) $n=500$, $20\%$ censoring, (a) $n=1000$, $20\%$ censoring,(a) $n=200$, $20\%$ censoring.}
    \label{fig:TVDs}
\end{figure}

{To further assess the accuracy of the normal approximation relative to a posterior sample obtained via adaptive MCMC, we focus on the first $\tilde{M} = 100$ iterations of the simulation scenarios described above. First, we generate a posterior sample of size $1,000$ for the parameters $\bbeta$ using an adaptive Metropolis MCMC sampler (\texttt{AdaptiveMCMC.jl}), following a burn-in of $5,000$ iterations and thinning every $50$ iterations. Next, we draw a sample of size $1,000$ from the asymptotic normal approximation. Using these posterior samples, we construct kernel density estimates of the marginal posterior distributions, $\widehat{\pi}_{\text{MC}}$ and $\widehat{\pi}_{\text{N}}$, and compute the total variation (TV) distances:
\begin{equation*}
d_\text{TV}(j)= \dfrac{1}{2}\int_{-\infty}^{\infty} \vert \widehat{\pi}_{\text{MC}}(\beta_j)- \widehat{\pi}_{\text{N}}(\beta_j) \vert d\beta_j.
\label{eq:GH}
\end{equation*}
The numerical integration is performed using the \texttt{QuadGK.jl} library in Julia. We take the MCMC samples as the reference, as the adaptive sampler provides a more accurate exploration of the posterior distribution without relying on asymptotic assumptions.
Figure \ref{fig:TVDs} presents violin plots for the $\tilde{M}$ TV distances in each scenario. As shown, increasing the sample size $n$ reduces the distance between the MCMC samples and the normal approximation, with $n = 1000$ yielding most TV distances close to $0.1$. The results also reveal variability in TV distances across parameters, reflecting the typical intuition that different shape parameters govern distinct features of the distribution and may require different sample sizes for accurate estimation.
Moreover, reducing censoring from $40\%$ to $20\%$ has a marked effect on the TV distances, particularly for certain parameters (\textit{e.g.}, parameters 6 and 8), indicating that shorter follow-up periods hinder the ability to learn the shape of individual hazard functions. Overall, $n = 1000$ appears sufficient for the MCMC and normal approximation to produce similar inferences, though the interplay between censoring and follow-up duration can have a substantial impact.
}

\subsection{Benchmarking}\label{subsec:benchmarking}
{Now, we compare the performance of the hazard–response model \eqref{eq:hazardresponse} with that of the General Hazard (GH) model \citep{etezadi:1987, chen:2001, rubio:2019},
\begin{equation*}
h(t;\bx_i, \balpha, \bbeta) = h_0\left(t \exp\left\{\tilde{\bx}_i^{\top}\balpha\right\} \right)\exp\left\{\bx_i^{\top}\bbeta\right\},
\label{eq:GH}
\end{equation*}
where the covariates are defined as $\tilde{\bx}_i = \bx_i = (x_{1i},x_{2i},x_{3i},x_{4i})^{\top}$, as in Section \ref{subsec:validation}. We adopt a three-parameter power generalised Weibull distribution for the baseline hazard. We refer the reader to Section \ref{sec:applications} for a more detailed discussion of this model. We simulate $M_h = 250$ datasets under each of the scenarios described in Section \ref{subsec:validation}. For the GH model, we assign uniform priors to all parameters over wide compact sets: $\text{Unif}(0, 1000)$ for positive parameters and $\text{Unif}(-1000, 1000)$ for real-valued parameters. We then estimate the maximum a posteriori (MAP) values for both models. The MAPs for the GH model are obtained using the \texttt{HazReg.jl} Julia package, noting that under uniform priors, with large enough support, the MAPs coincide with the maximum likelihood estimators (MLEs).

To assess model performance, we compare the fitted hazard functions (using plug-in estimators given by the MAPs) from the GH and hazard–response models for 20 individuals, to the true hazard function, using the restricted $L_1$ distance between two hazard functions $h$ and $\tilde{h}$ \citep{de:2009}:
\begin{equation*}
d_R(h,\tilde{h}) = \int_0^{t^*} \vert h(t) - \tilde{h}(t) \vert dt.
\end{equation*}
We take $t^*$ to be the maximum follow-up (administrative censoring) time for each case. Figure \ref{fig:hazdist} in the Supplementary Material presents the violin plots of the $M_h$ mean distances (averaged over the 20 individuals). The results clearly indicate that the hazard–response model provides more accurate individual hazard predictions than the GH model, which lacks the flexibility to adapt the hazard shape across individuals.}

\pagebreak

\begin{table}[h!]
\centering
\begin{tabular}{l | cccccc}
  \hline
  Parameter & mean (MAP) & median (MAP) & se (MAP) & RMSE (MAP) &  width & coverage \\
  \hline 
$\beta_{10}$ (1.5) & 1.536 & 1.536 & 0.029 & 0.046 & 0.170 & 0.950 \\ 
  $\beta_{1,1}$ (0.5) & 0.497 & 0.496 & 0.038 & 0.038 & 0.150 & 0.926 \\ 
  $\beta_{2,0}$ (0.5) & 0.479 & 0.480 & 0.038 & 0.043 & 0.210 & 0.984 \\ 
  $\beta_{2,1}$ (-0.5) & -0.497 & -0.496 & 0.050 & 0.050 & 0.192 & 0.948 \\ 
  $\beta_{3,0}$ (1) & 1.059 & 1.061 & 0.038 & 0.070 & 0.267 & 0.976 \\ 
  $\beta_{3,1}$ (0.5) & 0.496 & 0.495 & 0.036 & 0.037 & 0.150 & 0.950 \\ 
  $\beta_{4,0}$ (3) & 3.007 & 3.005 & 0.051 & 0.052 & 0.202 & 0.958 \\ 
  $\beta_{4,1}$ (-0.5) & -0.485 & -0.481 & 0.048 & 0.050 & 0.183 & 0.922 \\ 
   \hline
\end{tabular}
\caption{Simulation results: censoring rate $20\%$, $n=1000$.}
\label{tab:C20n1000}
\end{table}

\begin{figure}[h!]
    \centering
\includegraphics[width=0.75\textwidth]{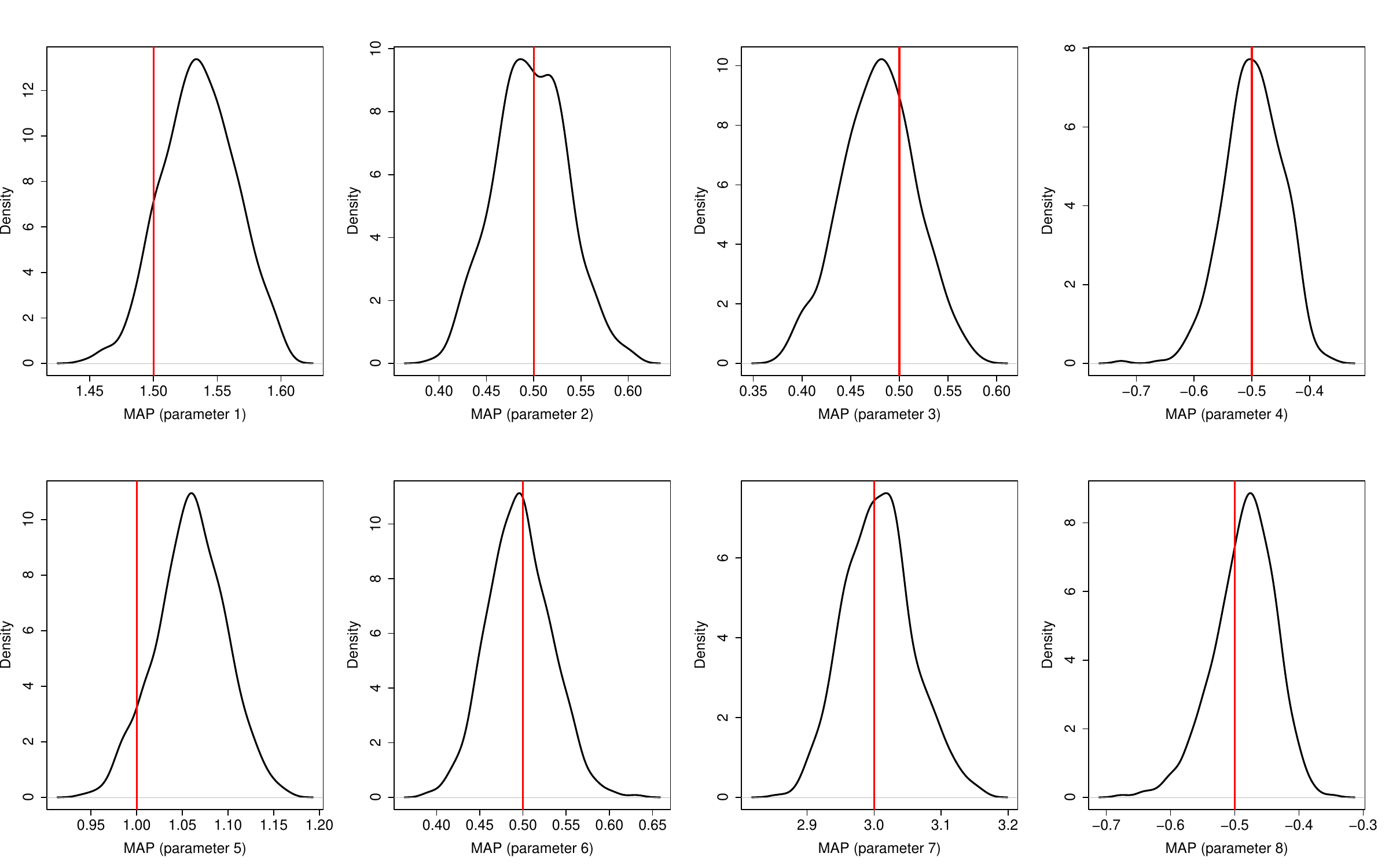}
\caption{Simulation results: censoring rate $20\%$, $n=1000$.}
\label{fig:C20n1000}
\end{figure}

\color{black}

\pagebreak
\section{Applications}\label{sec:applications}
In this section, we present two applications using real data where we illustrate the use of the regression models proposed in Section \ref{sec:model}. The first application addresses a common issue in clinical trials, where the survival and hazard curves for two treatments of interest (or treatment \textit{vs.}~placebo) may cross. This often arises when one treatment results in higher early mortality but offers better long-term outcomes. Such behaviour can be accounted for using model \eqref{eq:general_ode} with a logistic growth ODE, where treatment indicators are incorporated through the model parameters. The second application involves modelling cancer recurrence times following interventions (\textit{e.g.} treatments) aimed at improving survival. Treatment effectiveness may vary with individual characteristics, such as age and tumor stage, leading to different disease trajectories and corresponding individual hazard shapes.

\subsection{Ipilimumab immunotherapy trial}

In this example, we analyse data from the ipilimumab immunotherapy trial (NCT00861614) \cite{plana:2022}. This dataset includes information on $n=799$ individuals enrolled in a phase 3 trial investigating immunotherapy for advanced prostate cancer. The dataset contains time-to-event data, right-censoring indicators, and treatment arm assignments (ipilimumab \emph{vs.}~placebo). Previous studies have recommended fitting a two-parameter Weibull distribution separately to each trial arm \citep{plana:2022}, resulting in a double-arm Weibull model, as the corresponding survival functions cross at a certain time point.

We fit the logistic growth hazard model \eqref{eq:logisODE}, incorporating the treatment indicator (\texttt{trt}) into both parameters: $\lambda_i = \exp\left(\alpha_0 + \alpha_1 \texttt{trt}_i\right)$, and $\kappa_i = \exp\left(\beta_0 + \beta_1 \texttt{trt}_i\right)$. A single unknown initial condition $h_0$ is assumed. 
We specify weakly informative priors for all parameters: normal priors $N(0, 10)$ for the regression coefficients and a $\text{Gamma}(2, 1/2)$ prior for the initial condition $h_0$. Under this formulation, we obtain a posterior sample of size $N = 1{,}000$ after a burn-in of $5,000$ iterations and a thinning interval of $50$ (\textit{i.e.}~a total of 55,000 iterations), using the t-walk algorithm \citep{christen:2010} in R. Posterior summaries and plots of the posterior samples are presented in the Supplementary Material \ref{app:ipilimumab}. These summaries and the kernel density estimators of the posterior densities are concentrated around the posterior median, indicating that the data strongly inform the parameters (including the initial conditions).

\begin{figure}[h!]
    \centering
\begin{tabular}{c c}
\includegraphics[width=0.45\textwidth]{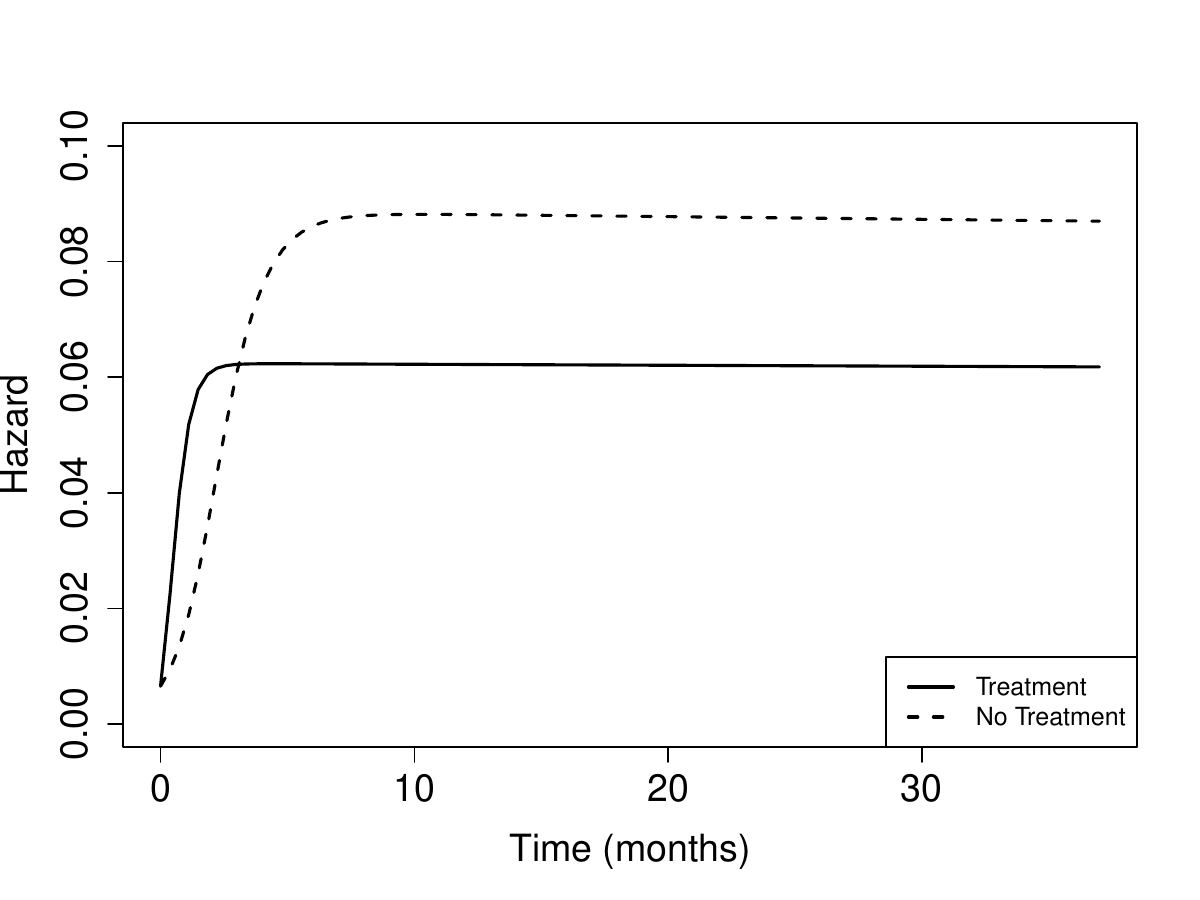} & \includegraphics[width=0.45\textwidth]{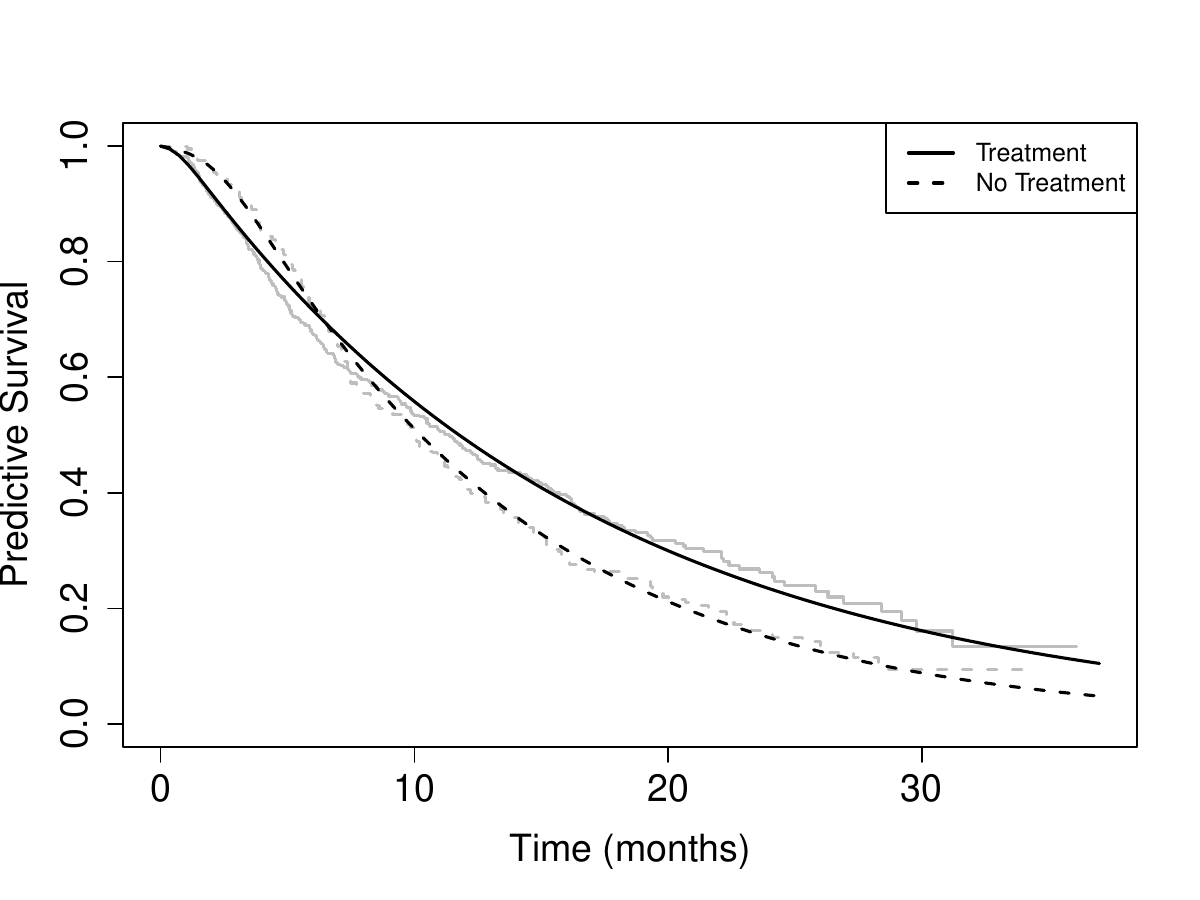} \\
 (a) & (b) \\
 \end{tabular}
    \caption{Ipilimumab data: (a) Predictive hazard functions for both arms, and (b) predictive survival functions for both arms (with Kaplan-Meier estimates in gray).}
    \label{fig:2A}
\end{figure}

Figure~\ref{fig:2A} presents the posterior predictive hazard and survival curves for each trial arm. The logistic growth model effectively captures the crossing of curves, which arises from differences in growth rates and carrying capacities between the arms. These plots highlight the importance of examining the crossing point of the survival functions, as it marks a change in the ordering of survival probabilities. They also show that analysing and interpreting hazard functions is informative: the crossing occurs earlier, indicating a shift in the ordering of mortality risk.
Finally, we compare the performance of this model using AIC and BIC. The AIC and BIC for the logistic growth regression model are $(4130.01,4153.43)$, while the corresponding quantities for the double-arm Weibull model are $(4163.77, 4182.51)$, clearly favouring the logistic growth regression model.

\pagebreak

\subsection{Breast cancer recurrence}

In this application, we model the recurrence (relapse) times of $n=2,939$ breast cancer patients from the \texttt{rotterdam} dataset in the R package \texttt{survival}. This is an observational dataset which includes information on recurrence times and censoring indicators (due to death or end of follow-up). We consider the following covariates: \texttt{age} (age at surgery, start of follow-up), \texttt{nodes} (number of positive lymph nodes), \texttt{size} (tumor size, dichotomised $0 =$ ``$\leq 50$'', $1 =$ ``$>50$''), and \texttt{trt} (chemotherapy, binary). The maximum follow-up duration is $19.3$ years, with $52\%$ of patients experiencing recurrence during the follow-up period. $580$ patients received chemotherapy, while $334$ patients also received hormonal treatment, with the majority ($306$) being on the non-chemotherapy group. 

Since some patients received treatment, it is reasonable to expect a reduction in the hazard associated with cancer recurrence. Consequently, we model this dataset using the hazard-response model \eqref{eq:hazardresponse}, incorporating covariates on the corresponding parameters as described in Section \ref{sec:model}. 
First, given that the minimum observed recurrence time is $36$ days, we do not expect the data to be informative about the initial conditions $h_0$ and $q_0$, which are assumed to be the same across all covariate values. Therefore, we fix these values using contextual information. Because the dataset pertains to breast cancer patients, who typically have a favourable prognosis, we adopt the strategy proposed in \cite{christen:2024} to reflect this prior knowledge. Specifically, we assume that the recurrence probability at one month $\Delta t = 1/12$ is approximately $S(\Delta t) \approx 0.999$. Using this, we estimate the initial hazard as $h_0  \approx  -\dfrac{S(\Delta t) - S(0)}{\Delta t S(\Delta t)} \approx 0.01$.
To specify the initial condition for $q$, we note that treatment typically does not begin immediately at the start of follow-up. As such, the initial response in reducing the hazard is expected to be minimal. We therefore fix this value at $q_0=10^{-6}$. 

To construct the regression model, we employ the Bayesian variable selection methodology introduced in Section \ref{sec:modelbuild} to identify the relevant variables for each linear predictor. The group $g$-prior \eqref{eq:gprior} is calibrated as the effective sample size, with $g_k = n - 0.5c$, where $c$ denotes the number of censored observations. This calibration is used for the parameters $\lambda$, $\kappa$, and $\mu$.  The factor $0.5$  is an empirical rule commonly used in practice \citep{liu:2012}, although alternative values may also be considered. For the competition parameter $\alpha$, we impose stronger shrinkage, reflecting the expectation of limited information about this parameter, by setting $g_k = (n - 0.5c)/100$.
For the model indicators $\boldsymbol{\gamma}$, we adopt the complexity prior of \citet{castillo:2015}, given by $\pi(\boldsymbol{\gamma}) \propto \tilde{d}^{-0.1 |\boldsymbol{\gamma}|}$. This prior strongly penalises models with more than 10 active variables, thereby allowing for the inclusion of a few covariates a priori while still enforcing sparsity (see plot in the Supplementary Material  \ref{app:breastcancer}). The Gibbs sampler (described in the Supplementary Material \ref{app:breastcancer}) is initialised at the intercept-only model and run for $1{,}100$ iterations, discarding the first $100$ as burn-in.
This step is computationally intensive: each iteration of the Gibbs algorithm stage evaluates one variable at a time (a total of $p=16$ variables). Additionally, each iteration of the Gibbs sampler requires a Laplace approximation, involving both an optimisation step and the computation of the Hessian matrix, as detailed in Section \ref{sec:modelbuild}. The visited models are saved in a dictionary to avoid repeated evaluations. Moreover, each evaluation of the log-posterior requires solving $n$ systems of ODEs \eqref{eq:hazardresponse}, which are parallelised using multi-threading in Julia (8 threads).
The model-building process takes approximately 36 hours on a Mac Studio (Apple M2 Ultra, 24-core CPU, 60-core GPU, 32-core Neural Engine, 64 GB RAM). We select the model with the highest posterior probability ($0.685$), which also coincides with the posterior median model (that is, the model obtained with variables with posterior inclusion probabilities larger than 0.5). This selected model includes: $(\text{intercept}, \texttt{nodes}, \texttt{trt})$ for $\lambda$, $(\text{intercept}, \texttt{nodes})$ for $\kappa$, $(\text{intercept}, \texttt{nodes}, \texttt{trt})$ for $\alpha$, and $(\text{intercept}, \texttt{age}, \texttt{size}, \texttt{trt})$ for $\mu$.
These results convey an important message: the growth rate ($\lambda$) of the hazard of recurrence depends on the number of nodes -- a key indicator of disease severity and prognosis -- as well as on treatment. In contrast, the carrying capacity (\textit{i.e.}~the bound of the hazard) is influenced solely by the number of nodes. The growth rate ($\mu$) of the response is associated with tumour size, treatment, and age, which are well-established clinical factors that affect both the response to chemotherapy and the decision to initiate treatment \citep{kroman:2000}. The parameter $\alpha$, which captures the efficacy of the response in its competition with the hazard, is influenced by both the number of nodes and treatment.

We now fit the selected model using both an adaptive Metropolis MCMC sampler (\texttt{AdaptiveMCMC.jl}) and a normal approximation based on the results presented in Section \ref{sec:postinf}. We employ weakly informative priors (normal distributions with zero mean and large variance (100)) for all parameters. First, we run an adaptive MCMC sampler (\texttt{AdaptiveMCMC.jl}) for $150,000$ iterations, discarding the first $50,000$ as burn-in and thinning every $100$ iterations, and initialised at the MAP. This sampling took approximately hour. On the other hand, obtaining an independent sample of size $10,000$ using the normal approximation takes approximately 5 minutes (including the optimisation step and the calculation of the Hessian matrix). Visual comparisons and summaries of the posterior samples under both approaches are presented in the Supplementary Material \ref{app:breastcancer}, which show a good normal approximation. 

\begin{figure}[h!]
    \centering
\begin{tabular}{c c c}
\includegraphics[width=0.3\textwidth]{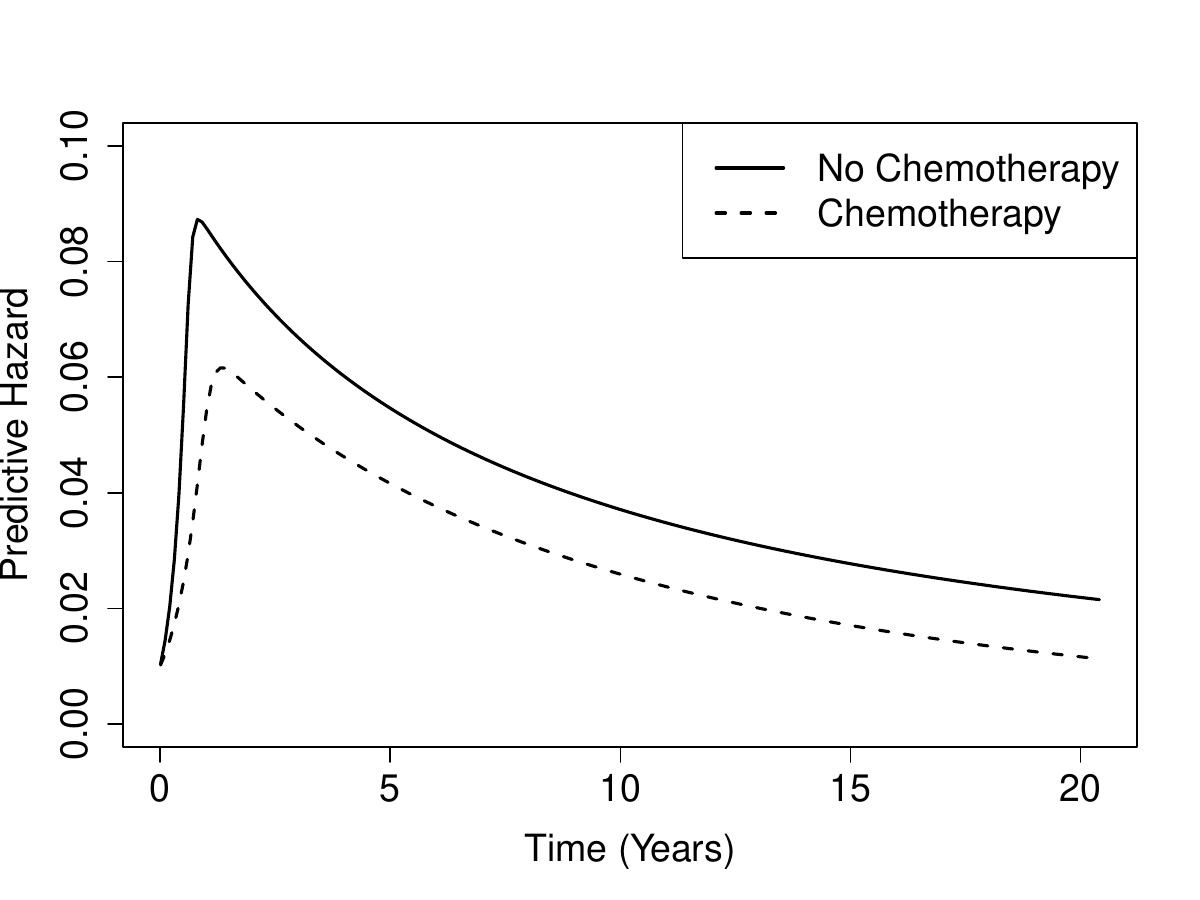} & \includegraphics[width=0.3\textwidth]{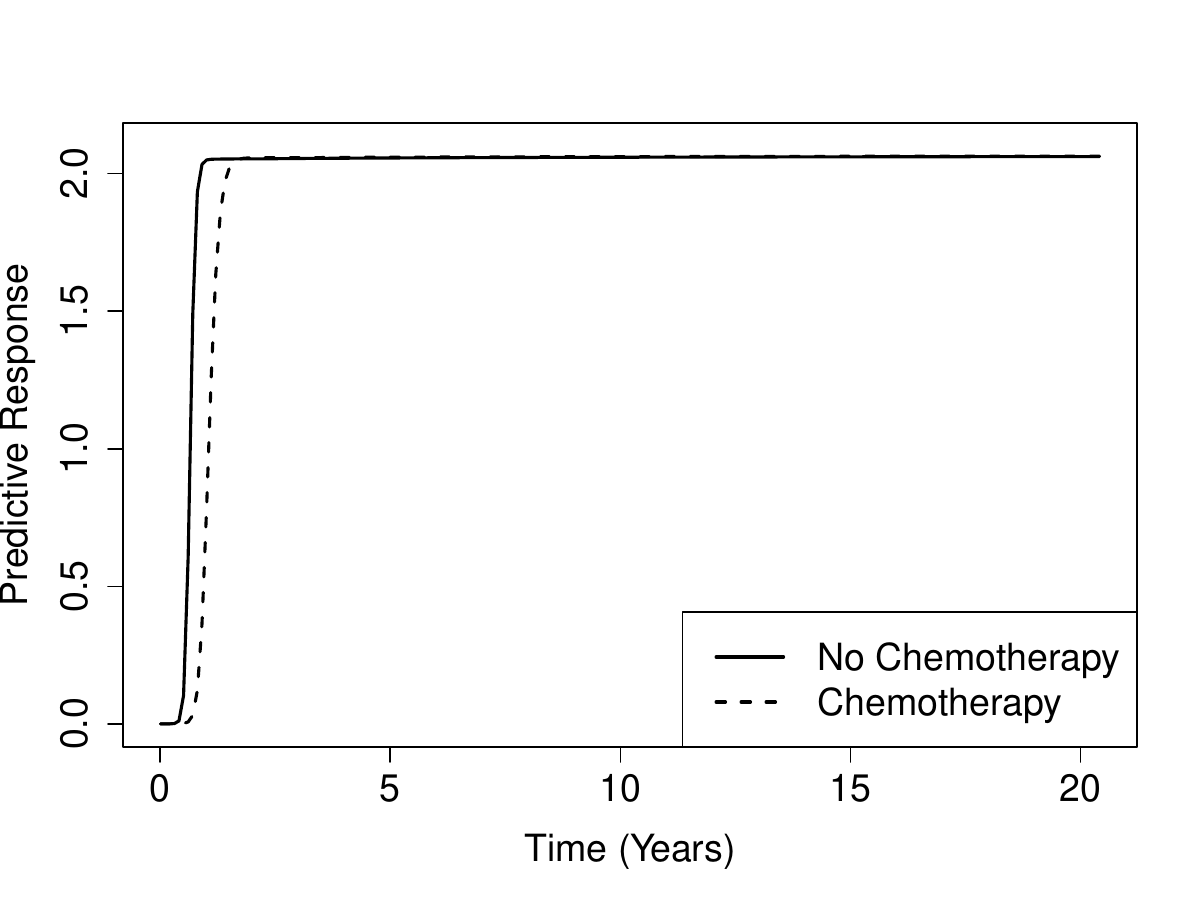} & \includegraphics[width=0.3\textwidth]{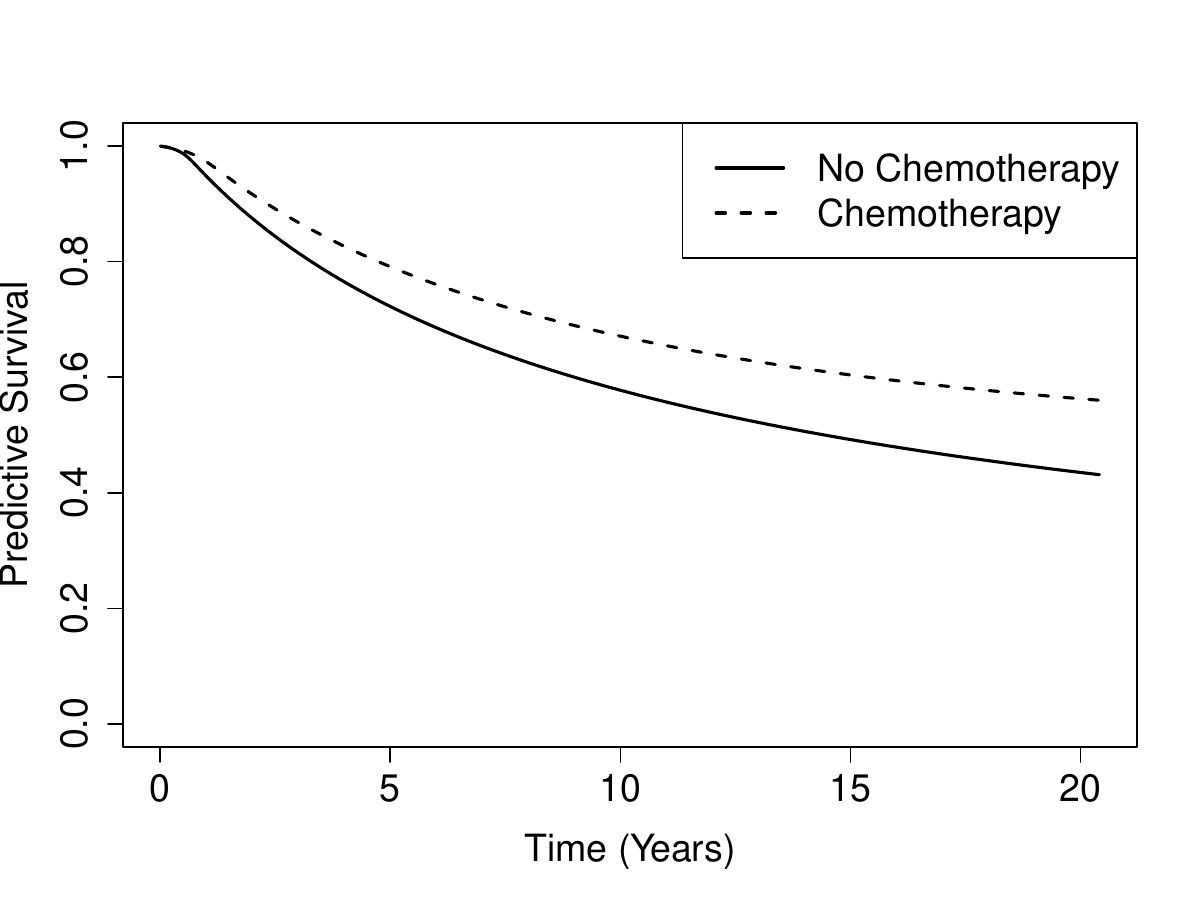} \\
 (a) & (b) & (c) \\
 \end{tabular}
    \caption{Breast cancer recurrence data (Early): (a) individual predictive hazard functions, (b) individual predictive response functions, and (c) individual predictive survival functions.}
    \label{fig:HRearly}
\end{figure}

\begin{figure}[h!]
    \centering
\begin{tabular}{c c c}
\includegraphics[width=0.3\textwidth]{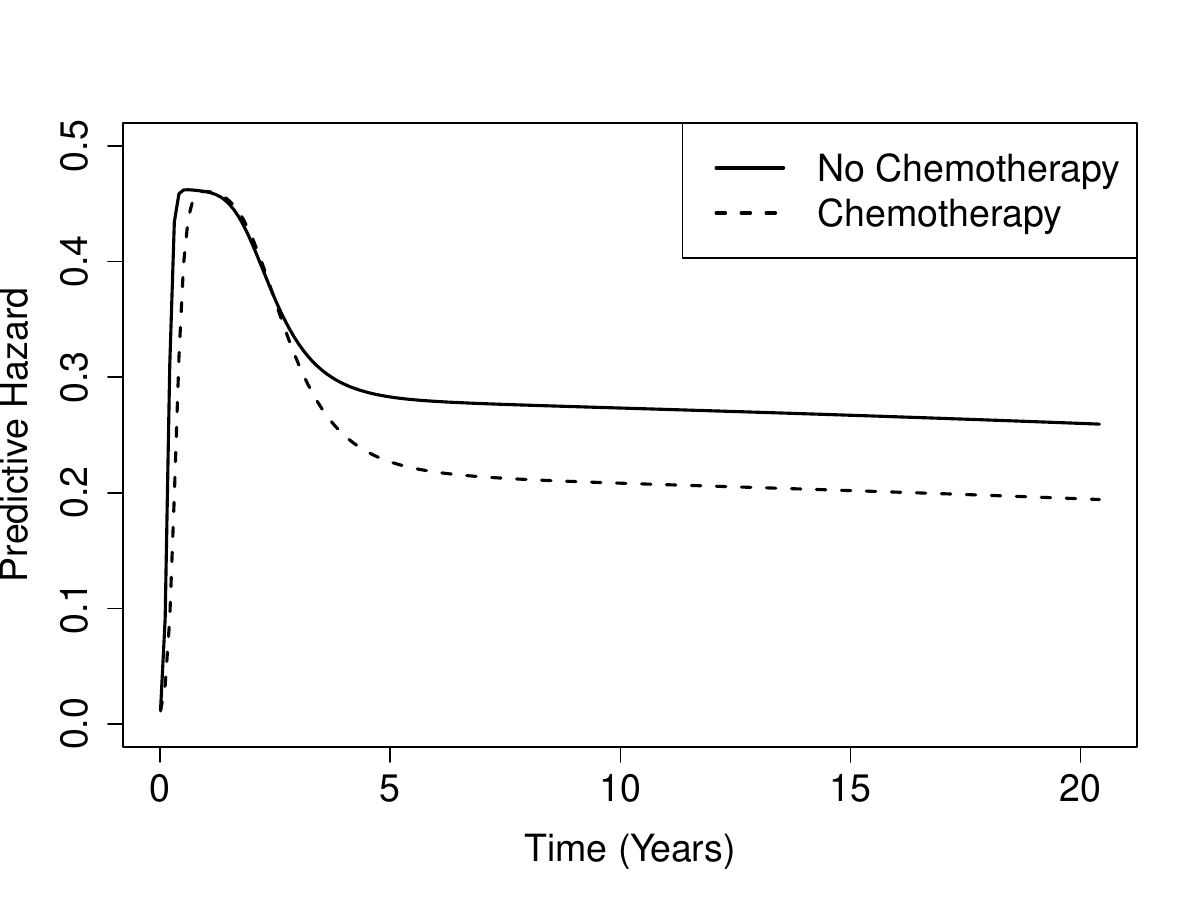} & \includegraphics[width=0.3\textwidth]{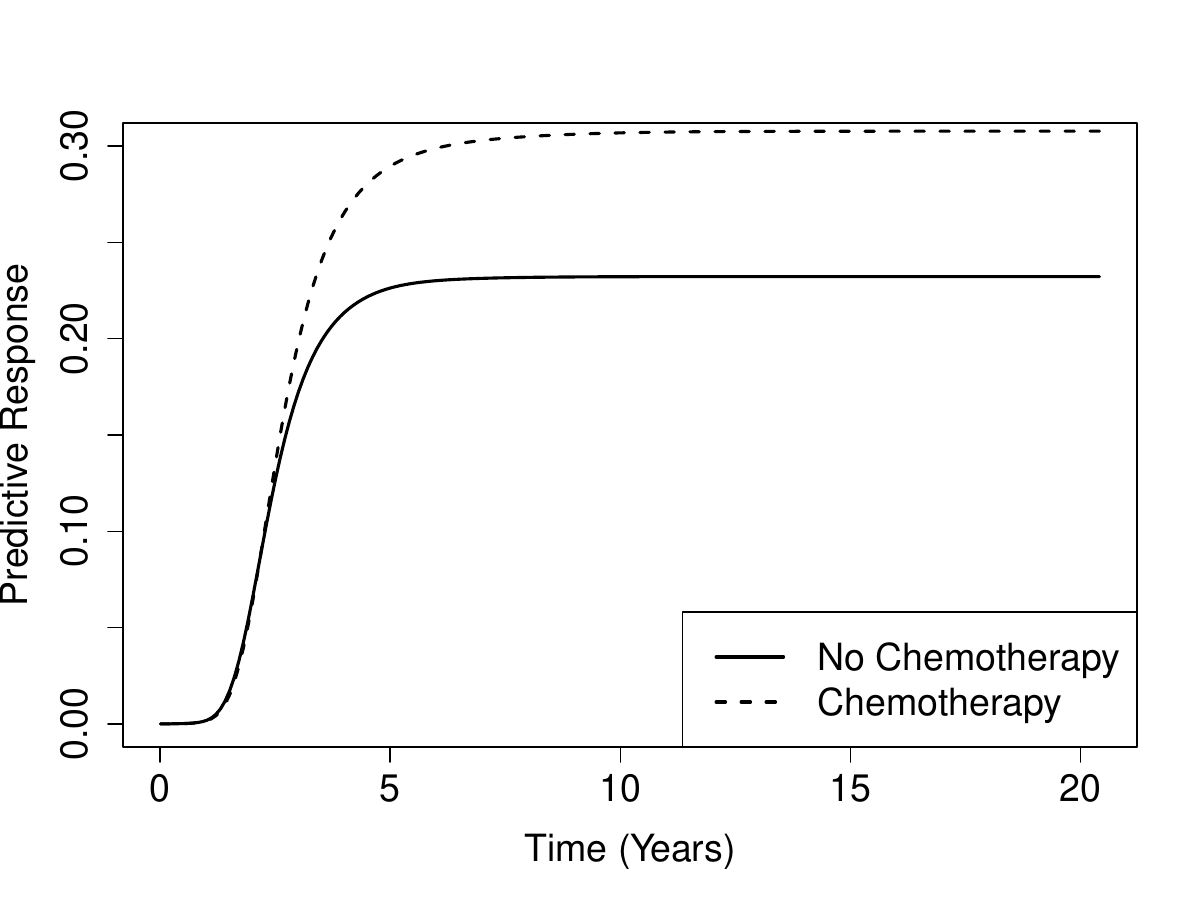} & \includegraphics[width=0.3\textwidth]{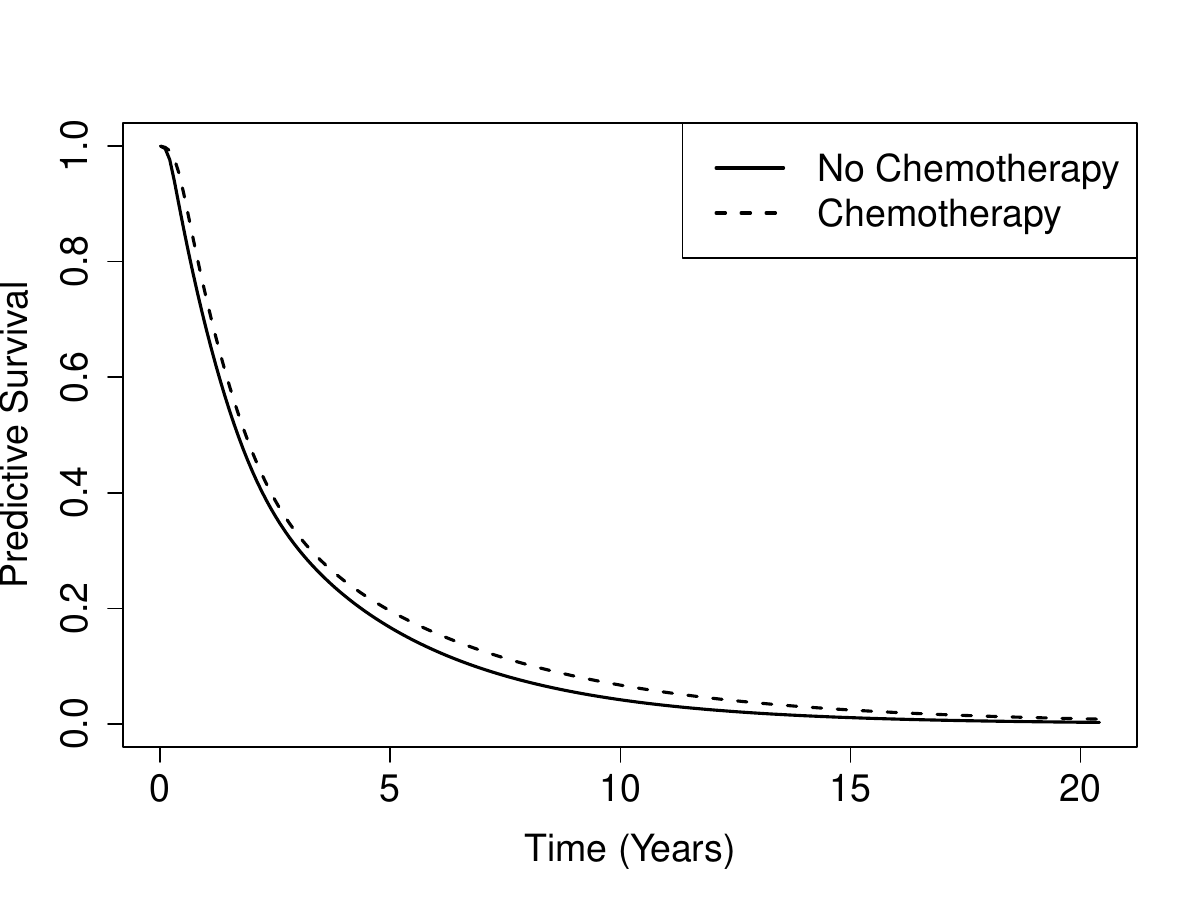} \\
 (a) & (b) & (c)\\
 \end{tabular}
    \caption{Breast cancer recurrence data (Late): (a) individual predictive hazard functions, (b) individual predictive response functions, and (c) individual predictive survival functions.}
    \label{fig:HRlate}
\end{figure}

Figure \ref{fig:HRpop} in the Supplementary Material \ref{app:breastcancer} displays the marginal predictive hazards, predictive responses, and predictive survival functions for the chemotherapy \textit{vs.} non-chemotherapy groups. The figure reveals crossing hazards and crossing survival functions. This figure shows that the survival function of recurrence is higher for the non-chemotherapy group.  In the same line, the predictive hazard function of recurrence for patients receiving chemotherapy remains higher than for those who did not. It is important to note that these are observational data, and the decision to treat is typically based on patients' clinical status. Since this conclusion applies only at the population level, we also perform a conditional analysis based on specific covariate values. Figure \ref{fig:HRearly} (Early) presents the conditional predictive functions for an individual with age (55), tumour size $\leq 50$, and a minimum number of nodes (0). Figure \ref{fig:HRlate} (Late) shows the corresponding functions for an individual with older age (74.5), tumour size $> 50$, and a larger number of nodes (15).
{For the first case (Early), we have the posterior probability $\Pr(h^* < 0) \approx 0.99$ for the treatment group, implying that these individuals belong to the attractor where the hazard ``loses'' the competition and the response ``wins'' ($h(t) \to 0$ as $t \to \infty$). Interestingly, $\Pr(h^* < 0) \approx 0.65$ and $\Pr(h^* > 0, q^* > 0) \approx 0.35$ for the non-treatment group, suggesting uncertainty about which attractor these individuals belong to. In other words, the predictive hazard is a mixture of the attractor where the hazard ``loses'' the competition and the response ``wins'' ($h(t) \to 0$ as $t \to \infty$), and an equilibrium attractor with the response ($h(t) \to h^* < \kappa$\; and also $q(t)\to q^* < \kappa$), which implies a much slower decay to zero. This behaviour is illustrated in Figure~\ref{fig:HRearly} and may indicate a good response to treatment.
In contrast, for the second case (Late), both groups fall into the attractor where the hazard and response functions are in equilibrium (none ``wins'', with $\Pr(q^* > 0) \approx 1$ and $\Pr(h^* > 0) \approx 1$), suggesting that treatment is less effective for this group.}

{For comparison, we also fit a General Hazard (GH) model \citep{etezadi:1987,chen:2001,rubio:2019}. The GH model incorporates effects on both the time and hazard scales as follows:
\begin{equation}
h(t \mid \bx_i, \balpha, \bbeta) = h_0\left(t \exp\left\{\tilde{\bx}_i^{\top}\balpha\right\} \right)\exp\left\{\bx_i^{\top}\bbeta\right\},
\label{eq:GH}
\end{equation}
where $\tilde{\bx}_i\in{\mathbb R}^q$ are the time-level effects, $\bx_i\in{\mathbb R}^p$ are the hazard-level effects, $\balpha\in{\mathbb R}^q$ and $\bbeta\in{\mathbb R}^p$ are the corresponding regression coefficients, and $h_0(\cdot)$ denotes the baseline hazard, which may be specified parametrically \citep{rubio:2019} or estimated nonparametrically \citep{chen:2001}.
The GH model in \eqref{eq:GH} encompasses several well-known survival models as special cases: the Proportional Hazards (PH) model ($\balpha = 0$), the Accelerated Failure Time (AFT) model ($\balpha = \bbeta$ and $\tilde{\bx}_i = \bx_i$), and the Accelerated Hazards (AH) model ($\bbeta = 0$). However, the GH model assumes a single baseline hazard shared by all patients. These models have been widely applied in survival analysis, as they offer a straightforward way to include time-level effects (through $\balpha$) while remaining computationally tractable.
In our implementation, we define the time-level and hazard-level effects as $\bx_i = \tilde{\bx}_i$, comprising the variables age, size, nodes, and treatment. We couple these with a power generalised Weibull (PGW) baseline hazard function, a flexible three-parameter distribution with positive support that can accommodate increasing, decreasing, unimodal, and bathtub-shaped hazards. The model is fitted using the R package \texttt{HazReg}.
The GH model serves as a natural general competitor, given that it includes the PH, AFT, and AH models as special cases. The AIC and BIC values for the GH model are $9677.11$ and $9742.96$, respectively, whereas for the hazard-response model they are $9582.26$ and $9654.09$. Both information criteria clearly favour the hazard-response model. Although the GH model is flexible and features a rich hazard structure, one likely explanation is that it imposes a common baseline hazard across the population, therefore assuming the same hazard shape for all subgroups.
}

\pagebreak
\section{Discussion}\label{sec:discusssion}
We introduce a new class of survival regression models that incorporates covariate information into the parameters of hazard functions modelled through systems of autonomous ordinary differential equations (ODEs). This class retains the interpretability of the ODE system parameters while enabling a clear connection between covariates and the dynamic features governed by those parameters. 
The proposed methodology can be coupled with any system of ODEs that produces a continuous, positive solution for the hazard function, enabling the use of population growth models, species competition systems, or more general families of dynamical systems \citep{hirsch:2013}. Since covariates are incorporated through suitably transformed linear predictors, this approach is compatible with any chosen dynamical system for modelling the hazard function. 
{The implementation cost of these models is equivalent to that of any other parametric regression model \citep{rubio:2019} in cases where the solution is available in closed form. In cases where an analytical solution is not available, the cost of evaluating the log-likelihood function corresponds to solving $n$ systems of ODEs (where $n$ is the sample size), which results in increased computational costs. Nevertheless, solving a single small system of ODEs is computationally inexpensive with modern computing power. For example, solving one Lotka–Volterra system in Julia (v1.11) takes approximately $0.000025$ seconds on a Mac Studio (Apple M2 Ultra, 24-core CPU, 60-core GPU, 32-core Neural Engine, 64 GB RAM), which is comparable to the computational cost of other parametric methods. We have proposed two efficient strategies to scale the methodology to larger samples and to make the use of MCMC methods feasible. The first is computational in nature and involves parallelisation (multi-threading in Julia), since ODE solvers are typically fast and efficient, and solving $n$ independent ODE systems is an embarrassingly parallelisable task. Further speed-ups can be achieved through GPU-accelerated solvers for ODE ensembles \citep{rackauckas:2017,utkarsh:2024}. However, the computational cost may still be substantial for large sample sizes. As an alternative, we propose a multivariate normal approximation to the posterior distribution, supported by a general Bernstein–von Mises theorem that justifies its use. Finally, we have also provided tools for building the regression model, that is, for selecting the variables included in the predictors used to model the parameters of the ODE system.}

The simulation study demonstrates good performance of the normal approximation to the posterior distribution in terms of recovering the true parameter values and achieving good coverage. It also illustrates the impact of sample size and censoring rate, particularly in the presence of administrative censoring, which presents a challenge by truncating the time span over which survival times are observed. {The simulation study also includes a comparison of individual hazard predictions with the proposed methodology (hazard-response model) against those obtained from a general (or extended) hazard regression model. This comparison illustrates a scenario in which a single baseline hazard fails to accurately capture the individual hazard shapes for certain subgroups, resulting in poorer predictions for those individuals.} 
Two applications using real data were presented. The first involves a case where an analytical solution is available and demonstrates strong performance in the context of clinical trials with crossing hazards and survival functions. This application also highlights the importance of examining the hazard function: crossing points in the hazard function occur earlier than those in the survival function, providing insight into when the risks associated with the two trial arms reverse.
The second application considers a survival regression model for which no analytical solution is available. The computational cost of implementing this model is justified by its interpretability, as it enables understanding of the covariate values that lead patients to fall into different attractors. This, in turn, provides information about the effects and limitations of an intervention across different population subgroups.
{In the second application, we also compared the results obtained using an adaptive Metropolis sampler with those based on the asymptotic normal approximation of the posterior proposed in Section \ref{sec:postinf}. The asymptotic normal approximation demonstrates competitive performance, producing posterior regions with similar shapes. In general, a well-tuned or adaptive MCMC sampler is preferable in practice as it does not rely on asymptotic arguments, although it can be orders of magnitude slower than the normal approximation. Nevertheless, the normal approximation represents a practical tool for rapid exploration of the posterior distribution.}

Extensions of the proposed methodology include replacing linear predictors with additive ones \citep{kneib:2021,thielmann:2024}. This increases the number of model parameters and requires the use of shrinkage or selection methods \citep{rossell:2023} to prevent overfitting. Despite this added complexity, the computational cost remains comparable regarding the need for solving $n$ systems of ODEs. 
{Alternatively, one could use approximations that target either the posterior distribution or the ODE solver. Examples of the first type include amortised variational inference, which learns the posterior distribution using neural networks and variational methods, making the approximation scalable to large samples. Examples of the second type include kernel and spline methods that avoid the intensive use of numerical ODE solvers, such as those proposed by \cite{ramsay:2007}, \cite{liang:2008ode}, \cite{wu:2012},  and \cite{huang:2022}. Although these methods have typically been applied in contexts where systems of ODEs are used to model observable quantities, in contrast to our framework, which models the hazard function via ODEs, they represent promising research directions for improving scalability to larger samples and higher dimensions.
However, note that these methods tackle the problem of ODE parameter estimation in one particular system.  Our problem is that we have in fact $n$ systems, one for each patient.  Extrapolating solutions of nonlinear ODEs from one parameter setting to another is highly challenging, and adapting the cited methods to our broader problem would require substantial methodological development.}

{The Bayesian variable selection methodology presented here was introduced as a formal, yet preliminary, approach to constructing the distributional regression model, that is, to identify the variables that enter each linear predictor. This development opens the door to further research. Natural follow-up questions include characterising the asymptotic properties of such variable selection methodology, exploring alternative priors for the covariates and model space \citep{rossell:2023}, designing more efficient MCMC samplers \citep{liang:2023}, and developing methods for high-dimensional variable selection. The development of scalable variable selection methods within the framework of distributional regression remains an open question. In particular, such advances would make it possible to conduct simulation studies to empirically evaluate properties such as sensitivity, specificity, and the ability to recover the true model, among other characteristics that lie beyond the scope of the present work.}

{Other possible extensions include the incorporation of random effects, spatial effects, and other complex components. 
Since we are already solving one ODE per patient to evaluate the likelihood, these extensions seem computationally feasible with the techniques we developed in this work.
Furthermore, the methodology can be adapted to alternative survival analysis frameworks, such as competing risks, cure models and relative survival, where the interpretability of the (excess) hazard function is particularly valuable for epidemiological research \citep{rubio:2019,eletti:2022,basak:2025}. Another promising application of these models lies in risk prediction modelling, especially when individual-level predictions are of interest.}

\section*{Acknolwedgments}\label{sec:ack}

We thank two reviewers, the Associate Editor, and the Editor for constructive comments. We thank Dr. Kolyan Ray for helpful discussions on the Bernstein–von Mises theorem.

\clearpage

\section*{Supplementary Material}

\section{Technical conditions}\label{app:technical}

Recall that the hazard function $h(t\mid \bmeta, \bx_i)$ and the cumulative hazard function $H(t\mid \bmeta, \bx_i)$ are retrieved as solutions to the system of ODEs \eqref{eq:general_ode_auto}. Note that the density function $f(t\mid \bmeta, \bx_i)$ can be obtained as $f(t \mid \bmeta, \bx_i) = h(t \mid \bmeta, \bx_i) \exp\{-H(t\mid \bmeta, \bx_i)\}$, the cumulative distribution function as $F(t \mid \bmeta, \bx_i) = 1 -  \exp\{-H(t\mid \bmeta, \bx_i)\}$, and the survival function $S(t \mid \bmeta, \bx_i) =  \exp\{-H(t\mid \bmeta, \bx_i)\}$. 


For the case where the initial conditions $\bY_0$ are fixed, the log-likelihood function is:
\begin{eqnarray*}
\ell_n(\bmeta) =  \sum_{i=1}^n \ell_i(\bmeta) 
&=& \sum_{i=1}^n   \delta_i \log h(t_i \mid \bmeta, \bx_i) - \sum_{i=1}^n H(t_i \mid \bmeta, \bx_i)\\
 &=& \sum_{i=1}^n   \delta_i \log f(t_i \mid \bmeta, \bx_i) + \sum_{i=1}^n (1-\delta_i)\log S(t_i \mid \bmeta, \bx_i),
\end{eqnarray*} 
From the below assumptions on the initial value problem (IVP) \eqref{eq:general_ode_auto}, the gradient of the log-likelihood $\nabla_{\bmeta} \ell_n(\bmeta)$ exist and is continuous.
Define $Q_n(\bmeta) = \left[\nabla_{\bmeta} \ell_n(\bmeta)\right] \left[\nabla_{\bmeta} \ell_n(\bmeta)\right]^{\top}$, then by the weak law of large numbers $\dfrac{1}{n}Q_n(\bmeta) \stackrel{\Pr}{\to} Q(\bmeta) = {\mathbb E}\left\{ \left[\nabla_{\bmeta} \ell_i(\bmeta)\right] \left[\nabla_{\bmeta} \ell_i(\bmeta)\right]^{\top} \right\}$, as $n\to\infty$, for each value of $\bmeta$, since each partial derivative of $\ell_n(\bmeta)$ is a summation of $n$ terms. 
The function $Q(\bmeta)$, which consists of the cross-products of all partial derivatives of the log-likelihood, can be seen as a generalised version of the standard Fisher information matrix \citep{hjort:1992}, incorporating the censoring process.


Consider the following regularity conditions.
\begin{itemize}
    \item[\textbf{C1.}] The parameter space $\tilde{\Theta}$ is a compact subset of ${\mathbb R}^{\td}$, and $\bmeta^*\in\tilde{\Theta}$ is the true value of the parameter. Additionally, we assume that $t \in [0,\tau]$, for $\tau>0$.
    
    \item[\textbf{C2.}] The vector field $\psi_{\bmeta}: D \to \mathbb{R}^{m+1}$, where $D \subseteq \mathbb{R}^{m+1}$ is a closed rectangle, is three-times continuously differentiable in $\bY$, three times continuously differentiable in $\bmeta$, and $\bY_0$ is in the interior of $D$.
    
    \item[\textbf{C3.}] The solution $h(t\mid \bmeta, \bx)$ is identifiable. That is, if $h(t\mid \bmeta_1, \bx) = h(t\mid \bmeta_2, \bx)$, for $t \in [0,\tau]$ and all $\bx$, implies that $\bmeta_1 = \bmeta_2$.   


    

    
\item[\textbf{C4.}] The matrix $Q(\bmeta^*)$ is non-singular.

\item[\textbf{C5.}] The covariates $\bx_i$ are independent and identically distributed. Let $\bX_k$, $k=1,\dots,d$, be the design matrix corresponding to the covariates included in the linear predictor for the $k$th parameter, $\theta_k$, in the ODE system. Assume that there exists $n_0$ such that, for $n > n_0$, the matrices $\bX_k^{\top}\bX_k$ are positive definite almost surely, and $\dfrac{1}{n}\bX_k^{\top}\bX_k \stackrel{\Pr}{\to} \Sigma_k$, for some $k\times k$ positive definite matrix $\Sigma_k$, as $n\to \infty$.

\item[\textbf{C6.}] $0 < {\mathbb P}(\delta_i = 1) \leq 1$, and $O_i \perp C_i \mid \bx_i$ (non-informative censoring conditional on covariates).

\end{itemize}

\section{Theoretical results}\label{app:theory}
Before presenting the proof of Propositon \ref{prop:BvM}, we present two preliminary results that guarantee the standard regularity conditions under conditions C1-C3.

\begin{lemma}\label{le:FIM}
Suppose that conditions C1-C3 are satisfied. Then, for each value of $\bx$, there exist functions $M_1(t, \bx)$ and $M_2(t, \bx)$ such that, for $i=1,\dots,\tilde{d}$ and $j=1,\dots,\tilde{d}$,
    \begin{equation*}
    \begin{split}
    \left \vert \dfrac{ \partial \log f(t \mid \bmeta, \bx)}{\partial \bmeta_{i}} \cdot \dfrac{ \partial \log f(t \mid \bmeta, \bx)}{\partial \bmeta_{j}}  \right\vert &\leq M_1(t, \bx), \\
    \left \vert \dfrac{ \partial^2 \log f(t \mid \bmeta, \bx)}{\partial \bmeta_{i} \partial\bmeta_{j}}  \right\vert &\leq M_2(t, \bx),  
    \end{split}
    \end{equation*}
    where 
    \begin{equation*}
        \begin{split}
        \int M_1 (t, \bx) dF(t\mid \bmeta, \bx) &<\infty,\\
        \int M_2 (t, \bx) dF(t\mid \bmeta, \bx) &<\infty.
            \end{split}
    \end{equation*}

\proof

Conditions C1-C3 $\psi_{\bmeta}( \bY, \bx_i)$ in \eqref{eq:general_ode_auto} has continuous third partial derivatives with respect to $\by$ and the parameters $\bmeta$ (C3),  this guarantees existence and uniqueness of a solution of the initial value problem (IVP)  \eqref{eq:general_ode_auto}. Moreover, the second partial derivatives of $\bY( t \mid \bmeta, \bx_i)$ and
$H(t \mid \bmeta, \bx_i)$ with respect to the parameters $\bmeta$ exist, and therefore the first derivatives are continuous.  These results can be found, for example, in \cite{jackiewicz:2009}, Theorem 1.5.1, and can be extended by reapplying the theorem to the derivatives themselves. Thus, conditions C1-C3 guarantee that the solution to the IVP are twice continuously differentiable with respect to $\bY$ and the parameters $\bmeta$. Note also that 
\begin{eqnarray*}
   \dfrac{ \partial \log f(t \mid \bmeta, \bx)}{\partial \bmeta_{i}} &=&  \dfrac{ \partial \log h(t \mid \bmeta, \bx)}{\partial \bmeta_{i}}  -  \dfrac{ \partial H(t \mid \bmeta, \bx)}{\partial \bmeta_{i}} \\
    &=&  \dfrac{\dfrac{ \partial h(t \mid \bmeta, \bx)}{\partial \bmeta_{i}}}{ h(t \mid \bmeta, \bx)} -  \dfrac{ \partial H(t \mid \bmeta, \bx)}{\partial \bmeta_{i}},
\end{eqnarray*}
and
\begin{eqnarray*}
  \dfrac{ \partial^2 \log f(t \mid \bmeta, \bx)}{\partial \bmeta_{i} \bmeta_{j}}  = \dfrac{\dfrac{ \partial^2 h(t \mid \bmeta, \bx)}{\partial \bmeta_{i}\partial \bmeta_{j}} -  \dfrac{ \partial h(t \mid \bmeta, \bx)}{\partial \bmeta_{i}}\dfrac{ \partial h(t \mid \bmeta, \bx)}{\partial \bmeta_{j}} }{ h(t \mid \bmeta, \bx)^2} -  \dfrac{ \partial H(t \mid \bmeta, \bx)}{\partial \bmeta_{i}}.
\end{eqnarray*}
By condition C3, and since we are focusing on autonomous systems \eqref{eq:general_ode_auto}, it follows that the terms in the previous equations are continuous in $t$. Recalling that $t\in [0,\tau]$, it follows that these functions are Lebesgue integrable for each value of $\bx$. Equivalently, there exist functions $M_1(t, \bx)$ and $M_2(t, \bx)$ such that
    \begin{equation*}
    \begin{split}
    \left \vert  \dfrac{\dfrac{ \partial h(t \mid \bmeta, \bx)}{\partial \bmeta_{i}}}{ h(t \mid \bmeta, \bx)} -  \dfrac{ \partial H(t \mid \bmeta, \bx)}{\partial \bmeta_{i}}  \right\vert &\leq M_1(t, \bx), \\
    \left \vert \dfrac{\dfrac{ \partial^2 h(t \mid \bmeta, \bx)}{\partial \bmeta_{i}\partial \bmeta_{j}} -  \dfrac{ \partial h(t \mid \bmeta, \bx)}{\partial \bmeta_{i}}\dfrac{ \partial h(t \mid \bmeta, \bx)}{\partial \bmeta_{j}} }{ h(t \mid \bmeta, \bx)^2} -  \dfrac{ \partial H(t \mid \bmeta, \bx)}{\partial \bmeta_{i}}  \right\vert &\leq M_2(t, \bx),  
    \end{split}
    \end{equation*}
    where 
    \begin{equation*}
        \begin{split}
        \int M_1 (t, \bx) dF(t\mid \bmeta, \bx) &<\infty,\\
        \int M_2 (t, \bx) dF(t\mid \bmeta, \bx) &<\infty.
            \end{split}
    \end{equation*}
Using the equivalence of the derivatives of $\log f(t \mid \bmeta, \bx)$ and those of $h(t \mid \bmeta, \bx)$ and $H(t \mid \bmeta, \bx)$, the result follows. 
\end{lemma}

Now, we prove a lemma that shows that under some regularity conditions on the right-hand side, the survival regression models defined by the system of ODEs  \eqref{eq:general_ode_auto} is differentiable in quadratic mean. 
 
\begin{lemma}\label{le:DQM}
Let $h(t\mid \bmeta, \bx)$ be the hazard function obtained as the solution to the system of ODEs  \eqref{eq:general_ode_auto} for a given covariate $\bx$. Suppose that conditions C1--C3 hold. Then, the model $f(t\mid \bmeta, \bx)$ is differentiable in quadratic mean.

\proof

First, note that condition C2 implies that the right-hand side $\Psi_{\bmeta}(\bY(t \mid \bmeta, \bx), \bx)$ is Lipschitz continuous in $\bY$, and consequently there exists a unique solution to the autonomous system of ODEs \eqref{eq:general_ode_auto} (see Chapter 4 from \cite{hirsch:2013}). Such solution is identifiable by assumption C3. The differentiability condition C2 guarantees that $h(t\mid \bmeta, \bx)$ is twice continuously differentiable with respect to $\bmeta$ (see Chapter 1 from \cite{jackiewicz:2009}). These points, together with the above discussion of the differentiability of the log-likelihood, imply that the map $\bmeta \mapsto \sqrt{f(t\mid \bmeta, \bx)}$ is continuously differentiable for each $t>0$.

Lemma \ref{le:FIM} implies that the entries of the matrix $Q(\bmeta)$ are continuous in $\bmeta$ (see Chapter 5 from \cite{lehmann:2006}). These results together with Lemma 7.6 from \cite{van:2000} imply differentiability of the root density $\bmeta \mapsto \sqrt{f(t\mid \bmeta, \bx)}$ in quadratic mean. That is, for each $\bx$
\begin{equation*}
    \int_{0}^{\infty} \left[ \sqrt{f(t\mid \bmeta + \bh, \bx)} - \sqrt{f(t\mid \bmeta, \bx)} - \dfrac{1}{2} \left(\bh^{\top}\nabla_{\bmeta} \log f(t\mid \bmeta, \bx)\right) \sqrt{f(t\mid \bmeta, \bx)} \right] dt = o\left(\lVert \bh \lVert^2 \right), 
\end{equation*} 
as $\lVert \bh \lVert \to 0$.

\textbf{Proof of Proposition \ref{prop:BvM}}.
First, note that, by Theorem 7.2 in \cite{van:2000}, we have that differentiability in square mean implies local asymptotic normality (LAN) when the sample is uncensored. Using the results (Example 7) in \cite{le:1988} and condition C6, it follows that LAN is preserved under non-informative right censoring. 

Now, the restriction of the parameter space and $t$ to compact sets in C1 implies the existence of uniformly consistent hypothesis tests \citep{van:2000,nickl:2013}. That is, there exists a sequence of tests $W_n$ for testing 
\begin{equation*}
H_0: \bmeta = \bmeta^* \quad \text{vs.} \quad H_1: \lVert \bmeta - \bmeta^* \lVert > \epsilon,
\end{equation*}
for every $\epsilon>0$, which satisfy
\begin{eqnarray}\label{eq:testcond}
    {\mathbb E}_{\bmeta^*}\left(W_n\right) &\to& 0, \nonumber\\
   \sup_{\lVert \bmeta - \bmeta^*\lVert \geq \epsilon} {\mathbb E}_{\bmeta}\left( 1- W_n \right) &\to& 0, \quad \text{as} \quad n \to \infty.
\end{eqnarray}


The proof of Proposition \ref{prop:BvM} follows from Lemma \ref{le:DQM}, which establishes differentiability in quadratic mean for the survival regression models; the argument above, which proves local asymptotic normality under censoring; the testing condition \eqref{eq:testcond}; and Theorem 10.3 from \cite{van:2000}.$\blacksquare$

\end{lemma}

\clearpage
\section{MCMC sampler for Bayesian variable selection}\label{app:gibbs}
In this section, we present the Gibbs sampler algorithm proposed for selecting the variables that enter each predictor of the distributional regression model. More advanced methods exist and could be seamlessly integrated with our methodology \citep{liang:2023}.

\begin{algorithm}
\caption{Gibbs sampler for Bayesian variable selection}\label{alg:gibbs}
\begin{algorithmic}[1]
    \STATE Initialise.
    \begin{itemize}
        \item $\bgamma^{(0)} = \left(\gamma^{(0)}_{k,j}\right)$, $j = 2,\dots, p_k$, $k=1,\dots, d$.
        \item Compute $\widehat{p}\left(\bt, \bdelta, \bX \mid \bgamma^{(0)}\right)$.
        \item Set $\text{iter} = 0$.
    \end{itemize}
    
    \STATE Iterate for $\text{iter} = 1,\dots,M$.
    
    For each predictor $k$ and each covariate $j$, update the inclusion matrix $\bgamma$ updating each $\gamma_{k,j}$, using its full conditional:
    \begin{itemize}
        \item 
        Let $\gamma_{k,j}' = 1 - \gamma^{(t-1)}_{k,j}$,
        and define the new model matrix $\bgamma{'}$, where only the $(k,j)$-th entry has changed.
        \item Compute the probability 

        \medskip

        $$
        P = \dfrac{\widehat{p}\left(\bt, \bdelta, \bX \mid \bgamma{'}\right)\pi(\bgamma{'}) }{\widehat{p}\left(\bt, \bdelta, \bX \mid \bgamma{'}\right)\pi(\bgamma{'}) + \widehat{p}\left(\bt, \bdelta, \bX \mid \bgamma^{(t-1))}\right)\pi(\bgamma^{(t-1))}) }.
    $$

    \STATE Simulate from the full conditional, that is change to $\gamma_{k,j}'$ with probability $P$ or remain with $\gamma^{(t-1)}_{k,j}$ with probability $1-P$: 

    \begin{itemize}
        \item Draw $U\sim \text{Unif}(0,1)$,
        \item If $U < P$, set $\gamma^{(t)}_{k,j} = \gamma_{k,j}'$,
        \item otherwise, keep $\gamma^{(t)}_{k,j} = \gamma^{(t-1)}_{k,j}$.
    \end{itemize}
    \end{itemize}

    \end{algorithmic}
\end{algorithm}

\section{Simulation from the hazard-response regression model}\label{app:hrsim}
The following algorithm presents a method to produce approximate simulations from the hazard-response regression model \eqref{eq:hazardresponse} -- \eqref{eq:reghr}. This algorithm represents an extension to that proposed in \citep{christen:2024} in the context without covariates.
\begin{algorithm}[ht]
\caption{Approximate Simulation from the Hazard-Response regression model \eqref{eq:hazardresponse} -- \eqref{eq:reghr}}
For a grid of $M$ time points $\tilde{\bt}_i = \{t_{1i},\dots,t_{Mi}\}$ in the interval $[0,t_{Mi}]$, and for each $i=1,\dots,n$:
\begin{itemize}
\item[1.] Obtain a numerical solution of the ODE system \eqref{eq:hazardresponse}, with parameters $\bmeta$ and covariate value $\bx_i$ at $\tilde{\bt}_i$, using an ODE Solver.
\item[2.]  Using the grid $\tilde{\bt}_i$ and the corresponding values of solution for the cumulative hazard function $H$ evaluated at $\tilde{\bt}_i$, construct an approximation $\tilde{H}^{-1}$, of $H^{-1}(\cdot \mid \bmeta, \bx_i)$.
\item[3.] Generate $u_i\sim U(0,1)$.
\item[4.] Calculate the simulated value $t_i^* = \tilde{H}^{-1}\left( -\log(u_i) \right)$.
\end{itemize} 
\label{alg:simHT}
\end{algorithm}

In step 1, the solution to the system of ODEs can be obtained using the Julia library\\ \texttt{DifferentialEquations.jl} and the \texttt{solve()} function.
Similarly, step 2 can be implemented automatically using the output from \texttt{solve()}. When simulating data with right-censoring, the upper bound $t_{Mi}$ can be set to the maximum follow-up time.

\clearpage

\section{Simulation results}\label{app:simresults}

\begin{table}[h!]
\centering
\begin{tabular}{l  | cccccc}
\hline
Parameter & mean (MAP) & median (MAP) & se (MAP) & RMSE (MAP) &  width & coverage \\
\hline 
$\beta_{1,0}$  (1.5) & 1.528 & 1.526 & 0.042 & 0.051 & 0.240 & 0.960 \\ 
$\beta_{1,1}$  (0.5) & 0.499 & 0.499 & 0.054 & 0.054 & 0.215 & 0.922 \\ 
$\beta_{2,0}$  (0.5) & 0.453 & 0.451 & 0.055 & 0.073 & 0.301 & 0.972 \\ 
$\beta_{2,1}$  (-0.5) & -0.497 & -0.490 & 0.073 & 0.073 & 0.278 & 0.910 \\ 
$\beta_{3,0}$  (1.0) & 1.079 & 1.081 & 0.053 & 0.095 & 0.374 & 0.992 \\ 
$\beta_{3,1}$  (0.5) & 0.500 & 0.497 & 0.056 & 0.056 & 0.219 & 0.930 \\ 
$\beta_{4,0}$  (3.0) & 3.001 & 2.994 & 0.076 & 0.076 & 0.291 & 0.956 \\ 
$\beta_{4,1}$  (-0.5) & -0.487 & -0.478 & 0.070 & 0.072 & 0.271 & 0.908 \\  
   \hline
\end{tabular}
\caption{Simulation results: censoring rate $20\%$, $n=500$.}
\label{tab:C20n500}
\end{table}

\begin{figure}[h!]
    \centering
\includegraphics[width=0.75\textwidth]{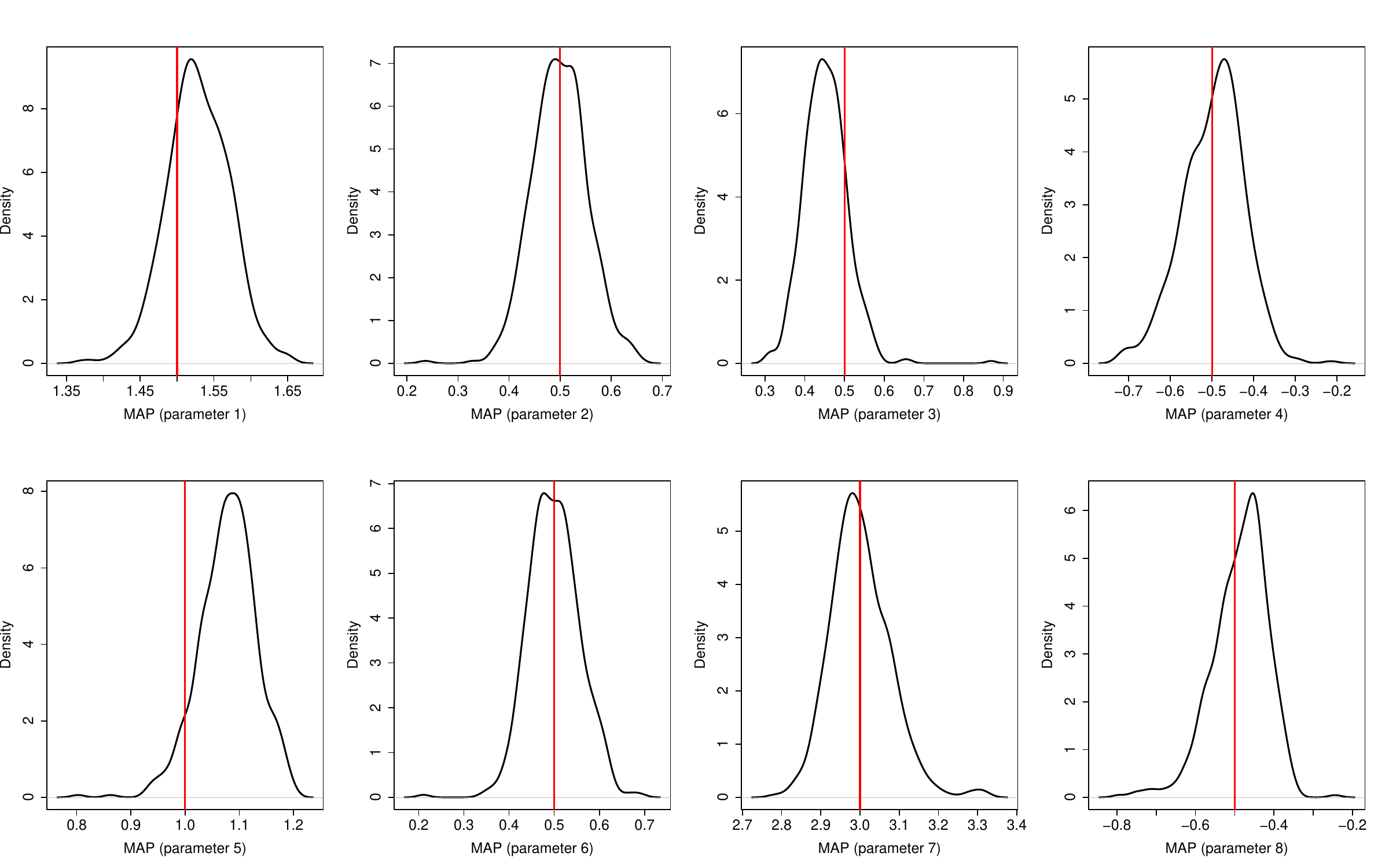}
\caption{Simulation results: censoring rate $20\%$, $n=500$.}
\label{fig:C20n500}
\end{figure}




\pagebreak

\begin{table}[h!]
\centering
\begin{tabular}{l | cccccc}
  \hline
  Parameter & mean (MAP) & median (MAP) & se (MAP) & RMSE (MAP) &  width & coverage \\
  \hline
$\beta_{1,0}$ (1.5) & 1.527 & 1.526 & 0.021 & 0.034 & 0.116 & 0.924 \\ 
  $\beta_{1,1}$ (0.5) & 0.496 & 0.495 & 0.025 & 0.026 & 0.103 & 0.950 \\ 
  $\beta_{2,0}$ (0.5) & 0.522 & 0.522 & 0.025 & 0.033 & 0.147 & 0.984 \\ 
  $\beta_{2,1}$ (-0.5) & -0.497 & -0.496 & 0.033 & 0.033 & 0.133 & 0.950 \\ 
  $\beta_{3,0}$ (1) & 1.005 & 1.005 & 0.026 & 0.027 & 0.185 & 0.998 \\ 
  $\beta_{3,1}$ (0.5) & 0.498 & 0.495 & 0.026 & 0.026 & 0.104 & 0.950 \\ 
  $\beta_{4,0}$ (3) & 2.999 & 2.996 & 0.037 & 0.037 & 0.139 & 0.940 \\ 
  $\beta_{4,1}$ (-0.5) & -0.485 & -0.484 & 0.029 & 0.033 & 0.125 & 0.934 \\  
   \hline
\end{tabular}
\caption{Simulation results: censoring rate $20\%$, $n=2000$.}
\label{tab:C20n2000}
\end{table}

\begin{figure}[h!]
    \centering
\includegraphics[width=0.75\textwidth]{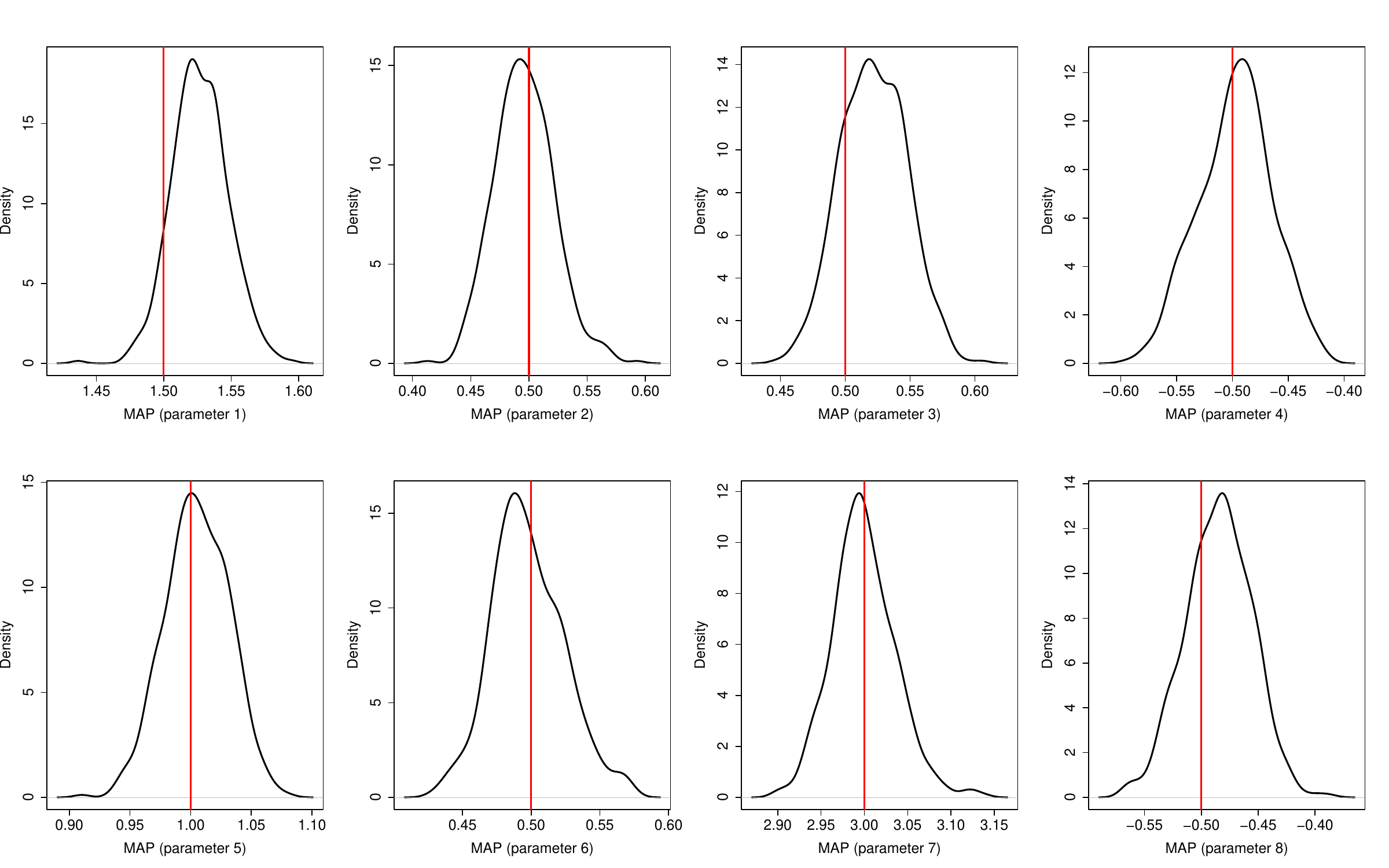}
\caption{Simulation results: censoring rate $20\%$, $n=2000$.}
\label{fig:C20n2000}
\end{figure}

\pagebreak

\begin{table}[h!]
\centering
\begin{tabular}{l | cccccc}
  \hline
  Parameter & mean (MAP) & median (MAP) & se (MAP) & RMSE (MAP) &  width & coverage \\
  \hline
$\beta_{1,0}$ (1.5) & 1.529 & 1.528 & 0.046 & 0.054 & 0.249 & 0.972 \\ 
  $\beta_{1,1}$ (0.5) & 0.499 & 0.499 & 0.065 & 0.065 & 0.252 & 0.934 \\ 
  $\beta_{2,0}$ (0.5) & 0.458 & 0.457 & 0.094 & 0.103 & 0.423 & 0.956 \\ 
  $\beta_{2,1}$ (-0.5) & -0.498 & -0.490 & 0.128 & 0.128 & 0.460 & 0.918 \\ 
  $\beta_{3,0}$ (1) & 1.081 & 1.083 & 0.077 & 0.112 & 0.453 & 0.972 \\ 
  $\beta_{3,1}$ (0.5) & 0.501 & 0.492 & 0.081 & 0.081 & 0.290 & 0.912 \\ 
  $\beta_{4,0}$ (3) & 3.001 & 2.990 & 0.090 & 0.090 & 0.326 & 0.940 \\ 
  $\beta_{4,1}$ (-0.5) & -0.483 & -0.469 & 0.092 & 0.094 & 0.331 & 0.886 \\ 
   \hline
\end{tabular}
\caption{Simulation results: censoring rate $40\%$, $n=500$.}
\label{tab:C40n500}
\end{table}

\begin{figure}[h!]
    \centering
\includegraphics[width=0.75\textwidth]{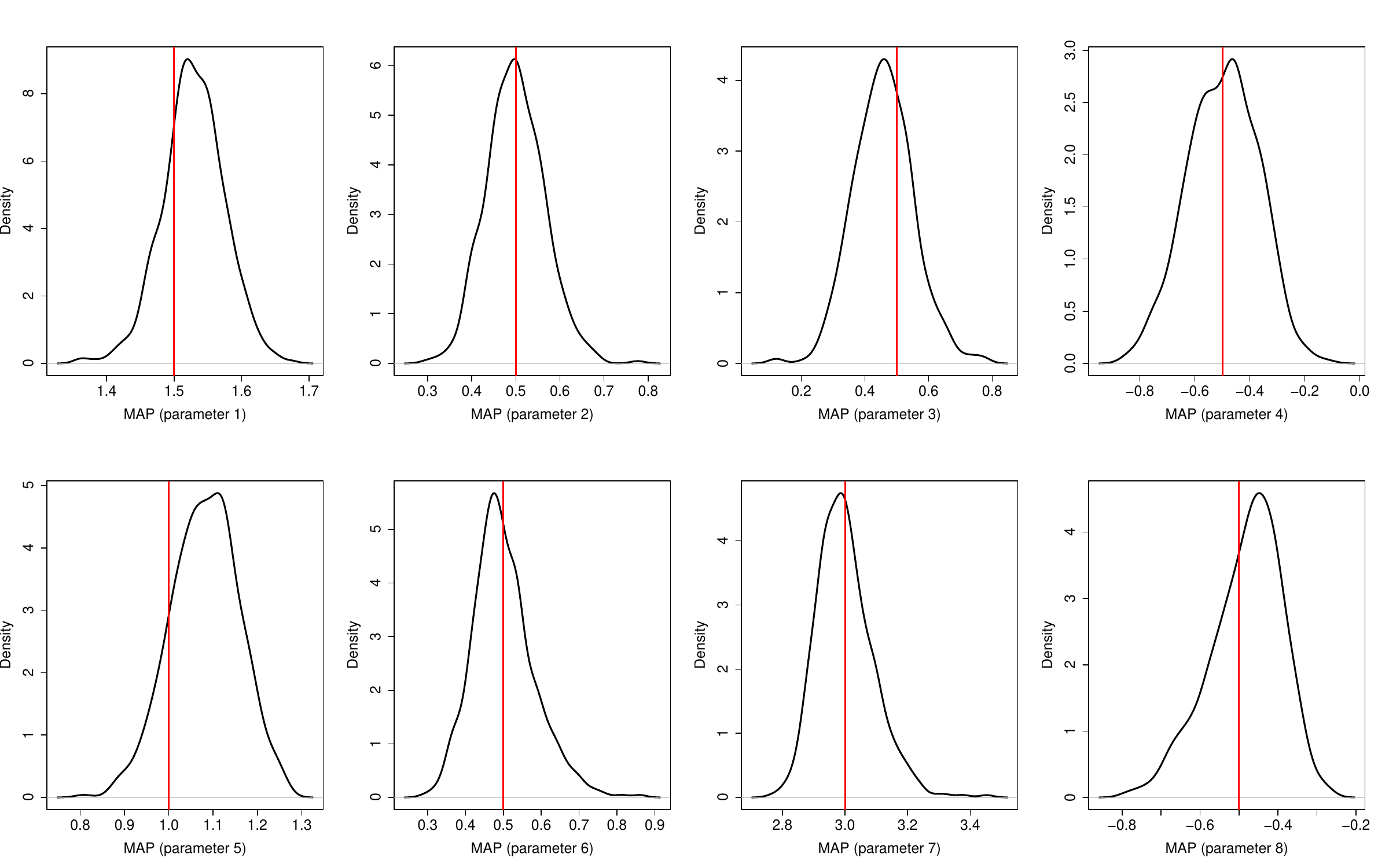}
\caption{Simulation results: censoring rate $40\%$, $n=500$.}
\label{fig:C40n500}
\end{figure}

\pagebreak

\begin{table}[h!]
\centering
\begin{tabular}{l | cccccc}
  \hline
  Parameter & mean (MAP) & median (MAP) & se (MAP) & RMSE (MAP) &  width & coverage \\
  \hline
$\beta_{1,0}$ (1.5) & 1.538 & 1.539 & 0.031 & 0.050 & 0.177 & 0.938 \\ 
  $\beta_{1,1}$ (0.5) & 0.495 & 0.493 & 0.045 & 0.045 & 0.175 & 0.938 \\ 
  $\beta_{2,0}$ (0.5) & 0.476 & 0.474 & 0.065 & 0.069 & 0.296 & 0.970 \\ 
  $\beta_{2,1}$ (-0.5) & -0.494 & -0.488 & 0.088 & 0.088 & 0.320 & 0.908 \\ 
  $\beta_{3,0}$ (1) & 1.069 & 1.073 & 0.053 & 0.087 & 0.321 & 0.968 \\ 
  $\beta_{3,1}$ (0.5) & 0.492 & 0.489 & 0.049 & 0.049 & 0.197 & 0.952 \\ 
  $\beta_{4,0}$ (3) & 3.006 & 3.002 & 0.056 & 0.057 & 0.222 & 0.960 \\ 
  $\beta_{4,1}$ (-0.5) & -0.479 & -0.473 & 0.061 & 0.064 & 0.223 & 0.882 \\ 
   \hline
\end{tabular}
\caption{Simulation results: censoring rate $40\%$, $n=1000$.}
\label{tab:C40n1000}
\end{table}

\begin{figure}[h!]
    \centering
\includegraphics[width=0.75\textwidth]{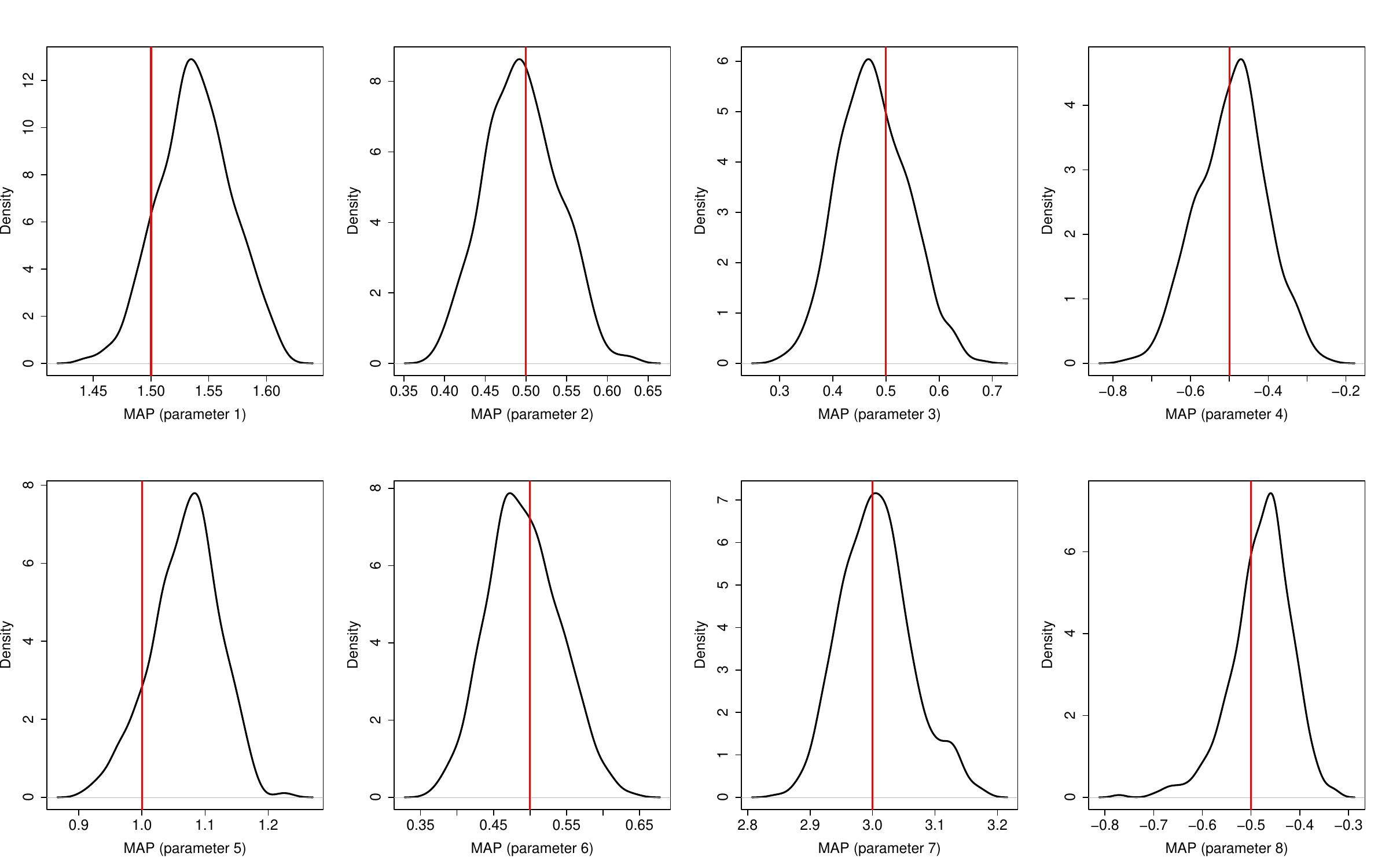}
\caption{Simulation results: censoring rate $40\%$, $n=1000$.}
\label{fig:C40n1000}
\end{figure}

\pagebreak

\begin{table}[h!]
\centering
\begin{tabular}{l | ccccc}
  \hline
  Parameter & mean (MAP) & median (MAP) & se (MAP) & RMSE (MAP) & coverage \\
  \hline
$\beta_{1,0}$ (1.5) & 1.525 & 1.524 & 0.022 & 0.033 & 0.934 \\ 
  $\beta_{1,1}$ (0.5) & 0.498 & 0.497 & 0.031 & 0.031 & 0.940 \\ 
  $\beta_{2,0}$ (0.5) & 0.530 & 0.532 & 0.043 & 0.053 & 0.934 \\ 
  $\beta_{2,1}$ (-0.5) & -0.497 & -0.495 & 0.055 & 0.055 & 0.948 \\ 
  $\beta_{3,0}$ (1) & 0.997 & 0.996 & 0.039 & 0.039 & 0.996 \\ 
  $\beta_{3,1}$ (0.5) & 0.498 & 0.495 & 0.036 & 0.036 & 0.954 \\ 
  $\beta_{4,0}$ (3) & 3.001 & 2.998 & 0.040 & 0.040 & 0.946 \\ 
  $\beta_{4,1}$ (-0.5) & -0.487 & -0.484 & 0.039 & 0.041 & 0.916 \\ 
   \hline
\end{tabular}
\caption{Simulation results: censoring rate $40\%$, $n=2000$.}
\label{tab:C40n2000}
\end{table}

\begin{figure}[h!]
    \centering
\includegraphics[width=0.75\textwidth]{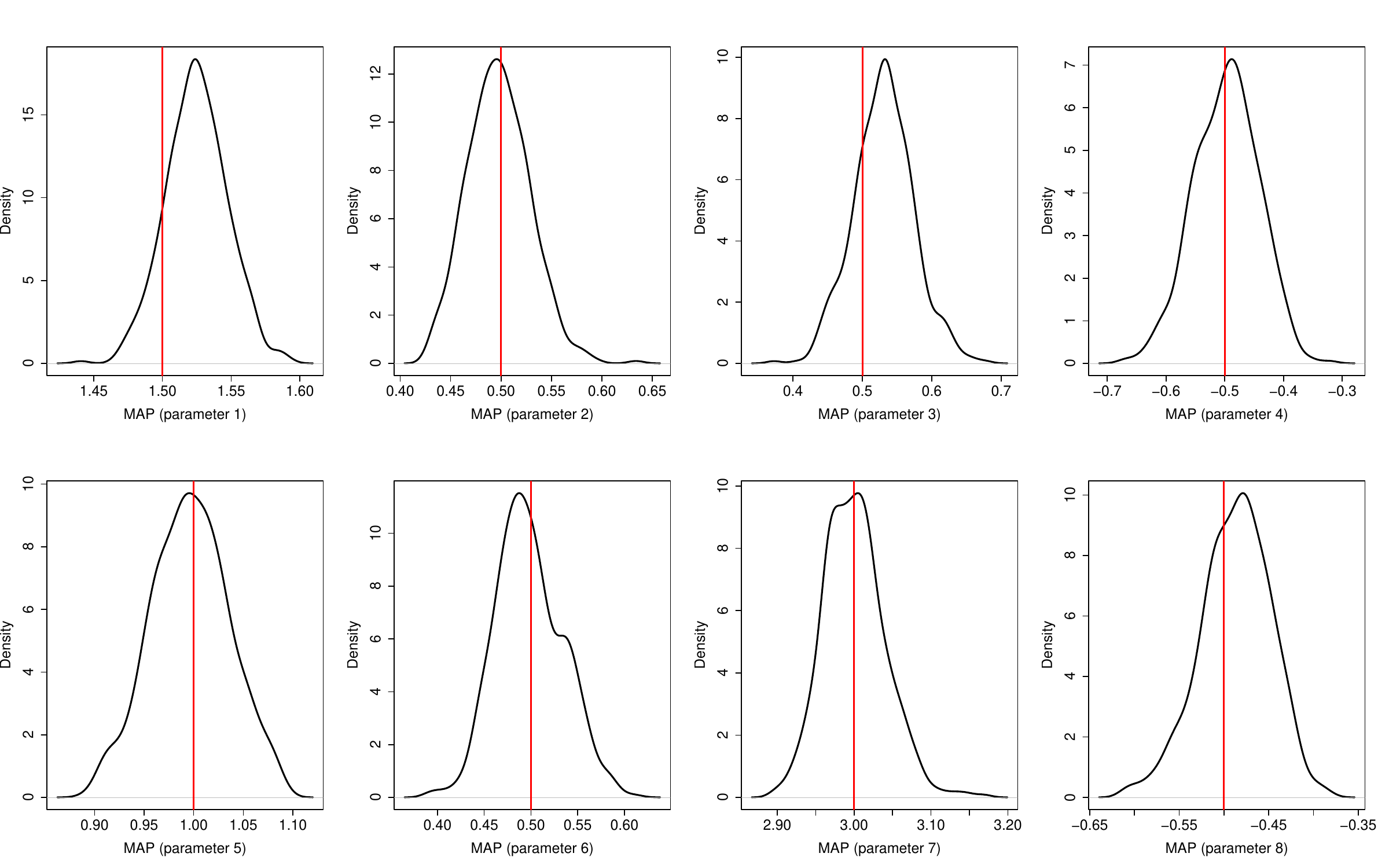}
\caption{Simulation results: censoring rate $40\%$, $n=2000$.}
\label{fig:C40n2000}
\end{figure}

\pagebreak

\begin{figure}[h!]
    \centering
\begin{tabular}{c c c}
\includegraphics[width=0.3\textwidth]{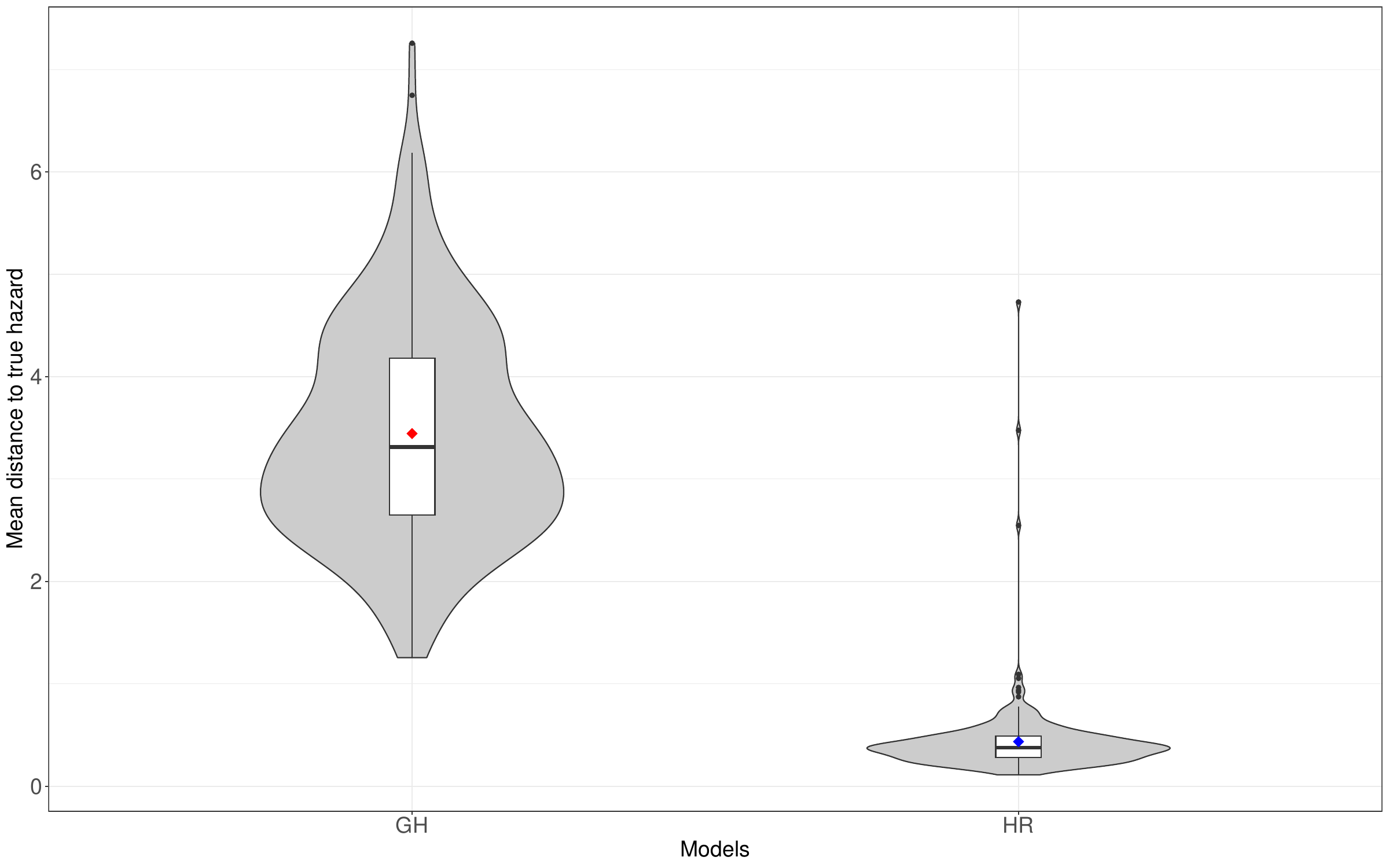} & \includegraphics[width=0.3\textwidth]{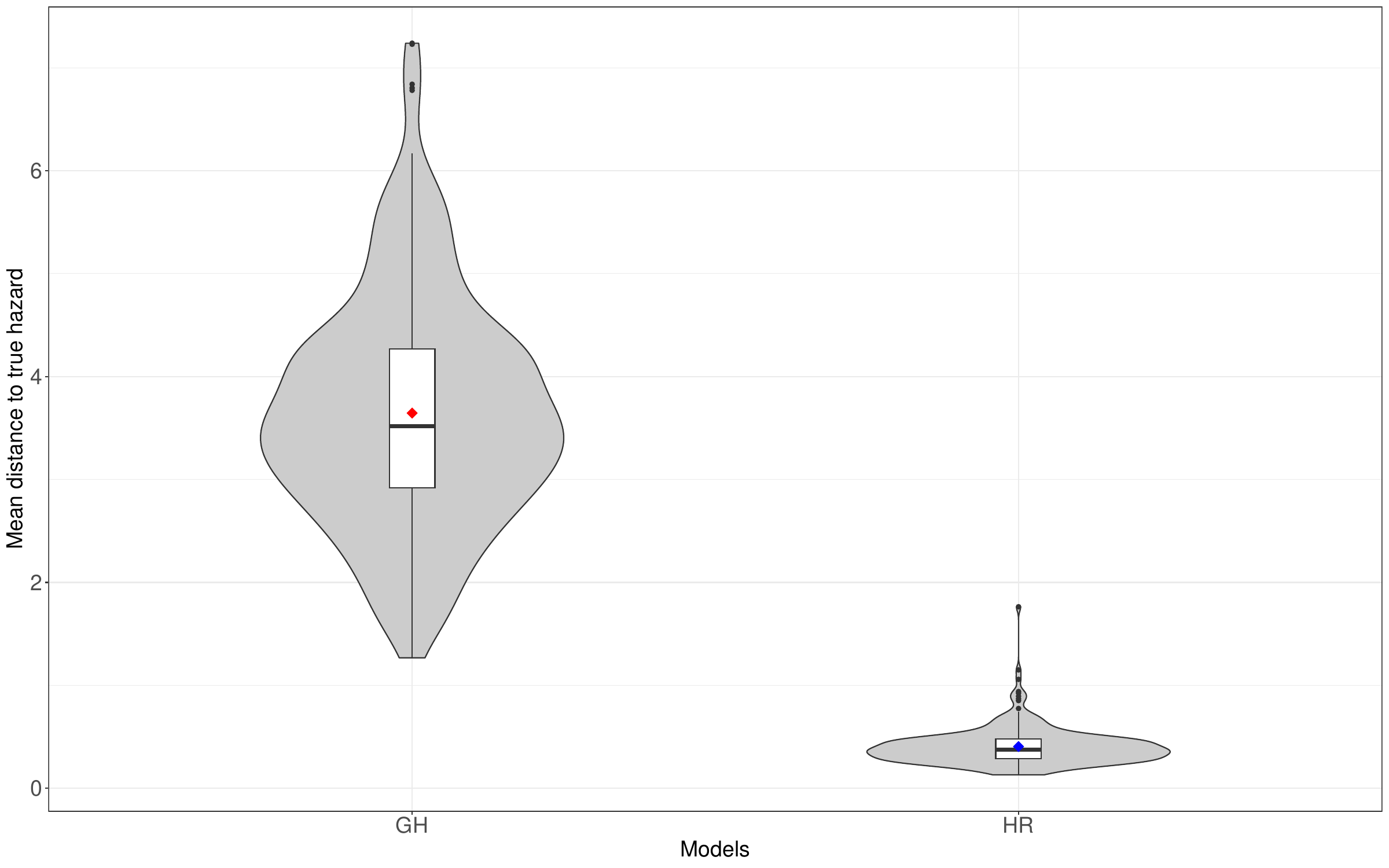} & \includegraphics[width=0.3\textwidth]{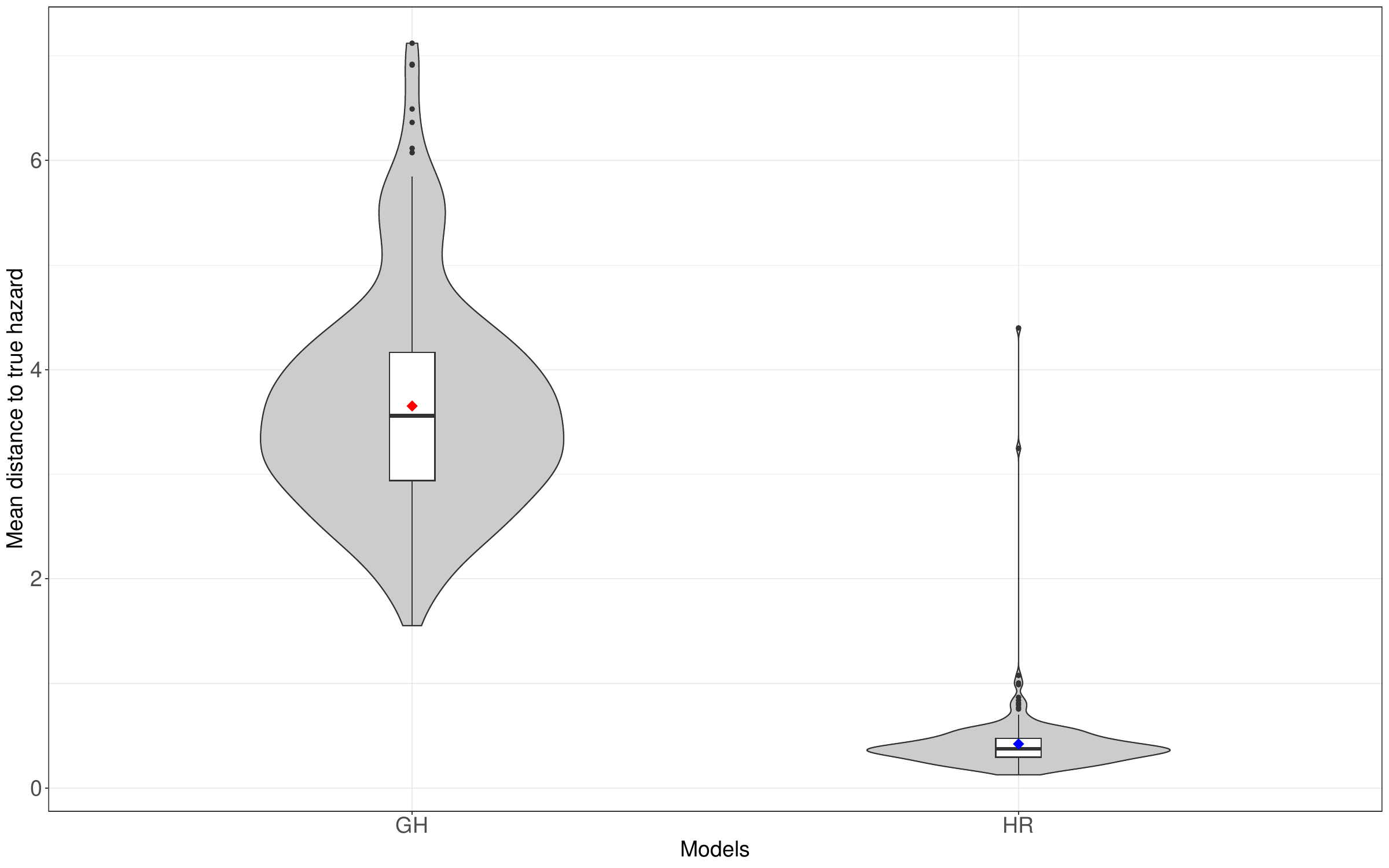} \\
 (a) & (b) & (c) \\
 \includegraphics[width=0.3\textwidth]{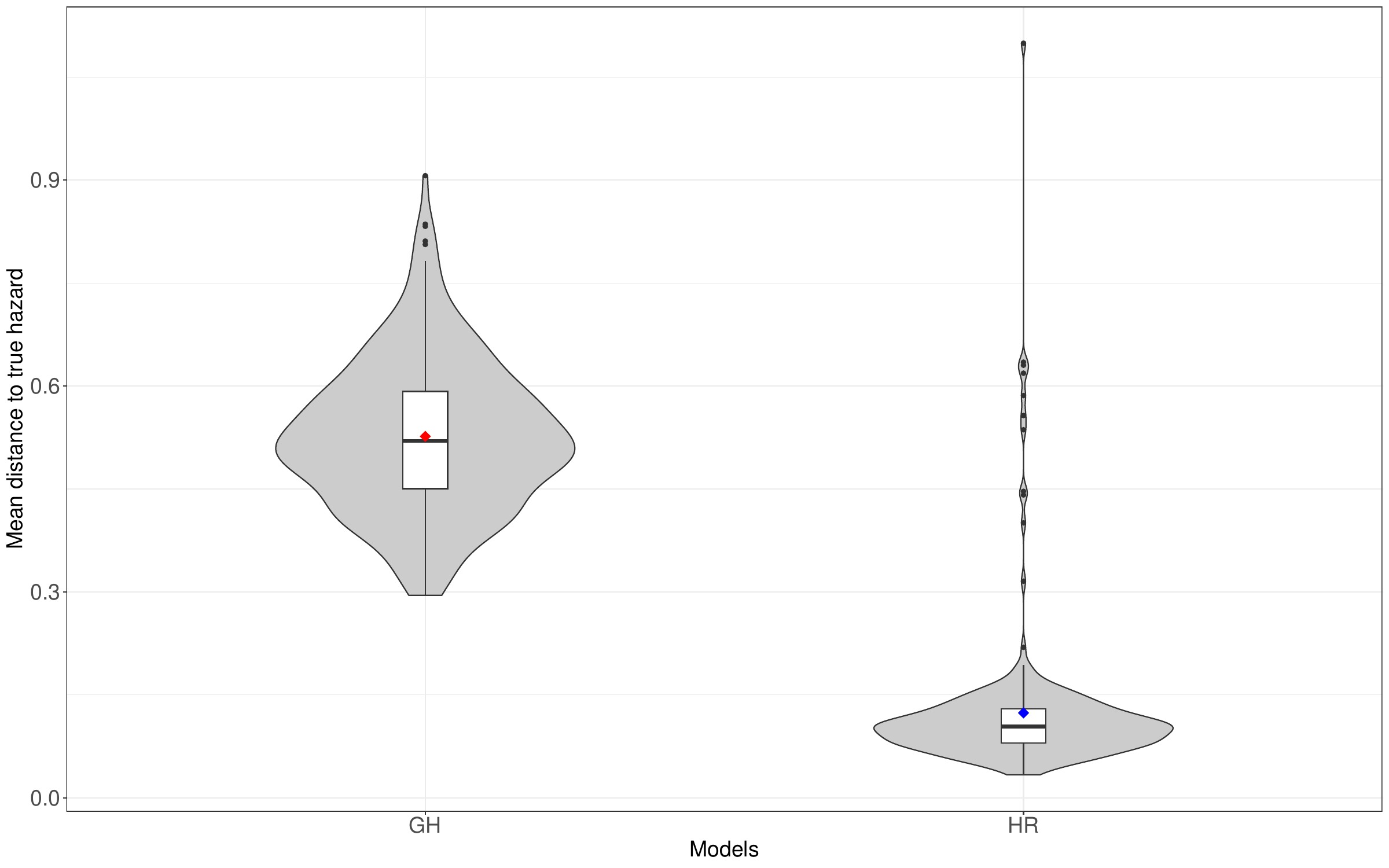} & \includegraphics[width=0.3\textwidth]{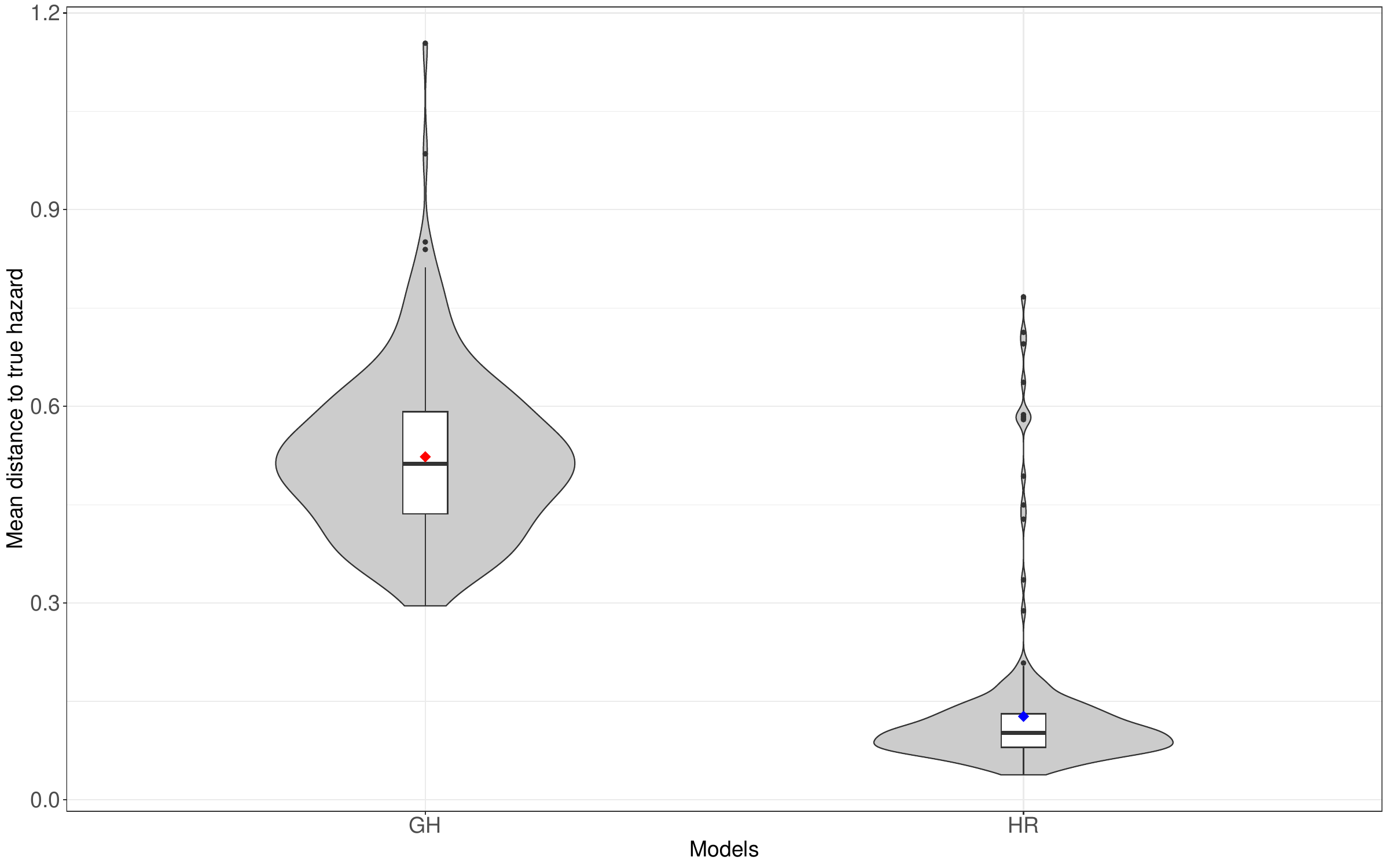} & \includegraphics[width=0.3\textwidth]{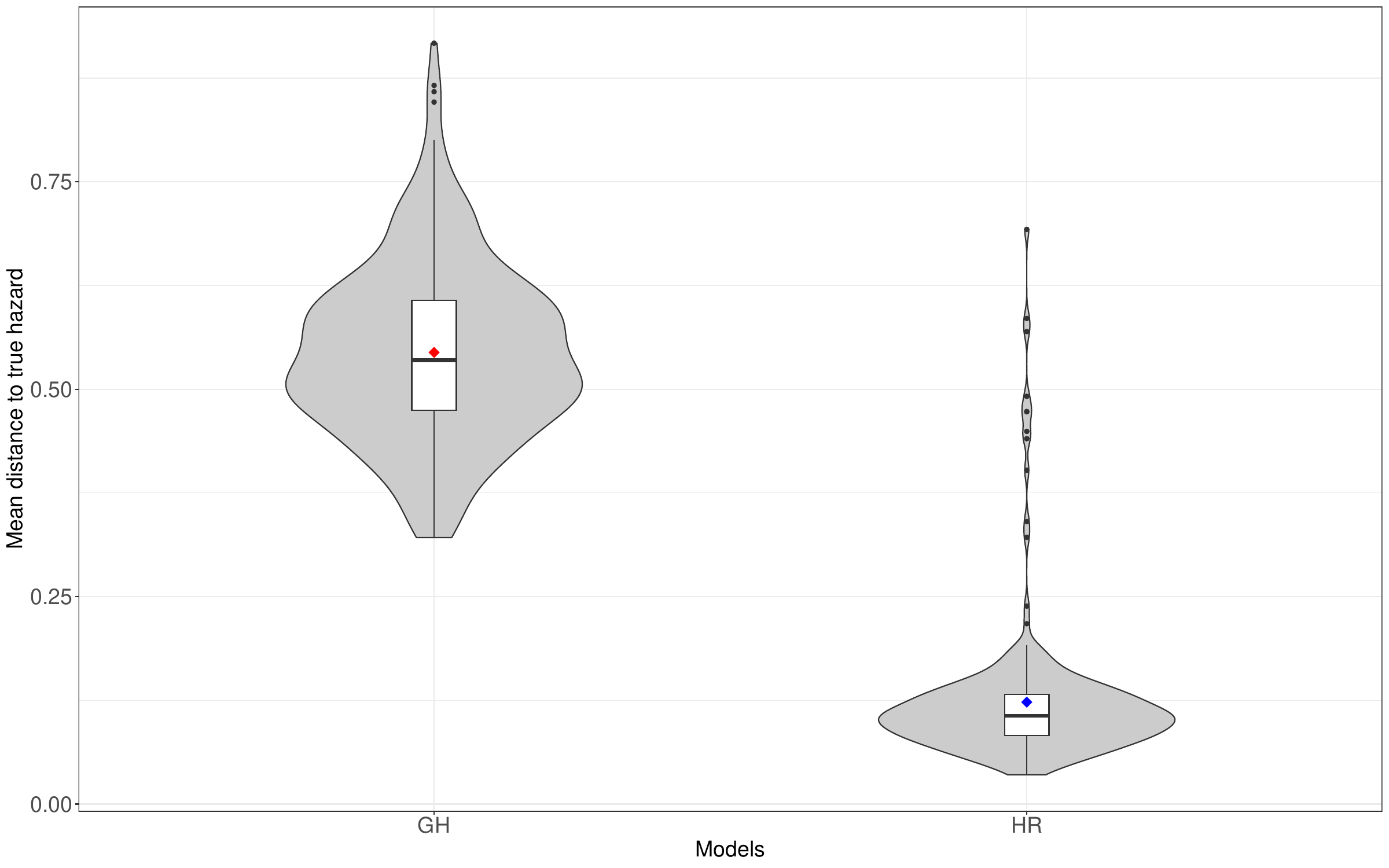} \\
 (d) & (e) & (f) \\
 \end{tabular}
    \caption{Mean hazard distance to the true hazard function: (a) $n=500$, $20\%$ censoring, (b) $n=1000$, $20\%$ censoring, (c) $n=2000$, $20\%$ censoring, (d) $n=500$, $40\%$ censoring, (e) $n=1000$, $40\%$ censoring, (f) $n=2000$, $40\%$ censoring.}
    \label{fig:hazdist}
\end{figure}

\clearpage

\section{Posterior summaries: Ipilimumab immunotherapy trial}\label{app:ipilimumab}

This section presents posterior summaries for the Ipilimumab immunotherapy trial example presented in the main paper.

\begin{table}[ht]
\centering
\begin{tabular}{|cccccc|}
  \hline
 & $\alpha_0$ & $\alpha_1$ & $\beta_0$ & $\beta_1$ & $h_0$ \\ 
  \hline
Min. & -1.20 & 0.18 & -2.68 & -0.70 & -7.22 \\ 
  1st Qu. & -0.04 & 0.93 & -2.48 & -0.41 & -5.38 \\ 
  Median & 0.11 & 1.17 & -2.43 & -0.35 & -5.08 \\ 
  Mean & 0.13 & 1.26 & -2.43 & -0.35 & -5.13 \\ 
  3rd Qu. & 0.29 & 1.45 & -2.38 & -0.28 & -4.81 \\ 
  Max. & 1.20 & 5.60 & -2.13 & -0.06 & -3.36 \\ 
   \hline
\end{tabular}
\caption{Ipilimumab immunotherapy trial data: Posterior summaries for the logistic growth model.}
\end{table}

\begin{figure}[h!]
    \centering
\includegraphics[width=0.75\textwidth]{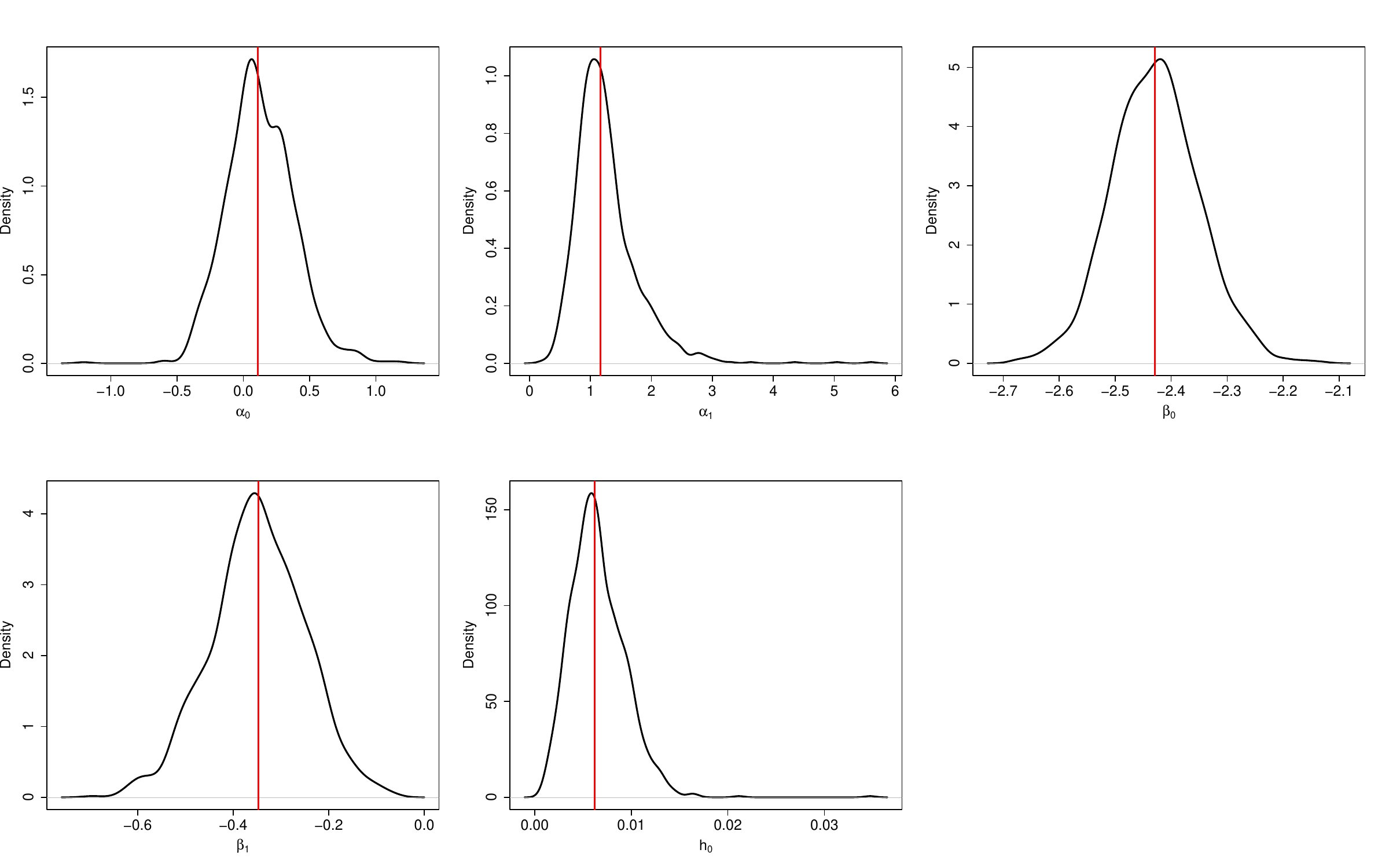}
\caption{Ipilimumab immunotherapy trial data: Posterior samples. The posterior median in shown in the vertical red line.}
\end{figure}

\clearpage
\section{Posterior summaries: Breast cancer recurrence}\label{app:breastcancer}

This section presents additional results for the Breast cancer recurrence application presented in the main paper. 

\begin{figure}[h!]
    \centering
\includegraphics[width=0.5\textwidth]{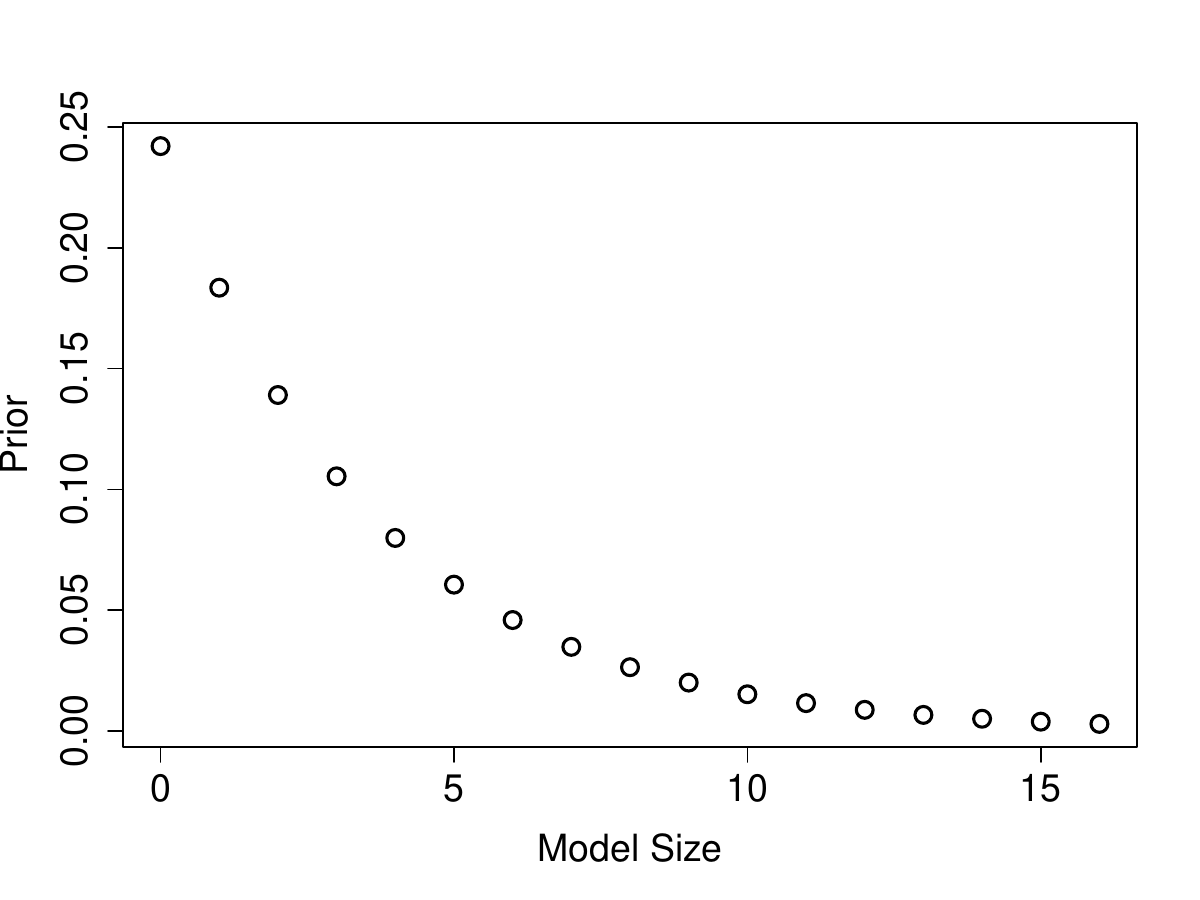}
\caption{Complexity prior (normalised) $\pi(\bgamma) \propto {\tilde{d}}^{-0.1 \vert \bgamma \vert}$ .}
\end{figure}

\begin{table}[ht]
\centering
\begin{tabular}{|ccccccccccccc|}
  \hline
 & $\beta_{1,0}$ & $\beta_{1,1}$ & $\beta_{1,2}$ & $\beta_{2,0}$ & $\beta_{2,1}$ & $\beta_{3,0}$ & $\beta_{3,1}$ & $\beta_{3,2}$ & $\beta_{4,0}$ & $\beta_{4,1}$ & $\beta_{4,2}$ & $\beta_{4,3}$ \\ 
  \hline
Min. & 1.34 & 0.40 & -1.11 & -0.85 & -0.59 & 0.53 & 0.48 & -1.01 & 2.58 & 0.04 & -0.43 & -0.94 \\ 
  1st Qu. & 1.49 & 0.50 & -0.75 & -0.01 & -0.39 & 1.10 & 0.71 & -0.69 & 2.98 & 0.10 & -0.27 & -0.57 \\ 
  Median & 1.53 & 0.53 & -0.66 & 0.19 & -0.34 & 1.29 & 0.77 & -0.62 & 3.05 & 0.12 & -0.23 & -0.48 \\ 
  Mean & 1.53 & 0.53 & -0.66 & 0.18 & -0.34 & 1.29 & 0.77 & -0.62 & 3.04 & 0.12 & -0.23 & -0.48 \\ 
  3rd Qu. & 1.58 & 0.56 & -0.58 & 0.38 & -0.28 & 1.48 & 0.84 & -0.54 & 3.12 & 0.13 & -0.19 & -0.39 \\ 
  Max. & 1.81 & 0.77 & -0.22 & 0.99 & -0.07 & 2.11 & 1.05 & -0.21 & 3.37 & 0.20 & -0.05 & -0.08 \\ 
   \hline
\end{tabular}
\caption{Breast cancer recurrence data: Posterior summaries from the hazard-response model fitted via MCMC.}
\end{table}

\begin{table}[ht]
\centering
\begin{tabular}{|ccccccccccccc|}
  \hline
 & $\beta_{1,0}$ & $\beta_{1,1}$ & $\beta_{1,2}$ & $\beta_{2,0}$ & $\beta_{2,1}$ & $\beta_{3,0}$ & $\beta_{3,1}$ & $\beta_{3,2}$ & $\beta_{4,0}$ & $\beta_{4,1}$ & $\beta_{4,2}$ & $\beta_{4,3}$ \\ 
  \hline
Min. & 1.23 & 0.42 & -1.05 & -0.95 & -0.62 & 0.41 & 0.45 & -1.01 & 2.71 & 0.02 & -0.45 & -0.87 \\ 
  1st Qu. & 1.49 & 0.50 & -0.72 & 0.03 & -0.39 & 1.12 & 0.71 & -0.68 & 3.00 & 0.10 & -0.27 & -0.54 \\ 
  Median & 1.54 & 0.52 & -0.64 & 0.20 & -0.34 & 1.28 & 0.77 & -0.61 & 3.06 & 0.11 & -0.24 & -0.46 \\ 
  Mean & 1.54 & 0.52 & -0.64 & 0.20 & -0.34 & 1.28 & 0.77 & -0.60 & 3.06 & 0.11 & -0.24 & -0.46 \\ 
  3rd Qu. & 1.59 & 0.54 & -0.57 & 0.37 & -0.29 & 1.44 & 0.83 & -0.53 & 3.11 & 0.13 & -0.20 & -0.38 \\ 
  Max. & 1.80 & 0.64 & -0.19 & 1.10 & -0.05 & 2.34 & 1.12 & -0.19 & 3.37 & 0.19 & -0.04 & -0.05 \\ 
  MAP & 1.54 & 0.52 & -0.64 & 0.20 & -0.34 & 1.28 & 0.77 & -0.61 & 3.06 & 0.11 & -0.24 & -0.46 \\ 
   \hline
\end{tabular}
\caption{Breast cancer recurrence data: Posterior summaries from the hazard-response model using the normal approximation.}
\end{table}

\begin{figure}[h!]
    \centering
\includegraphics[width=0.75\textwidth]{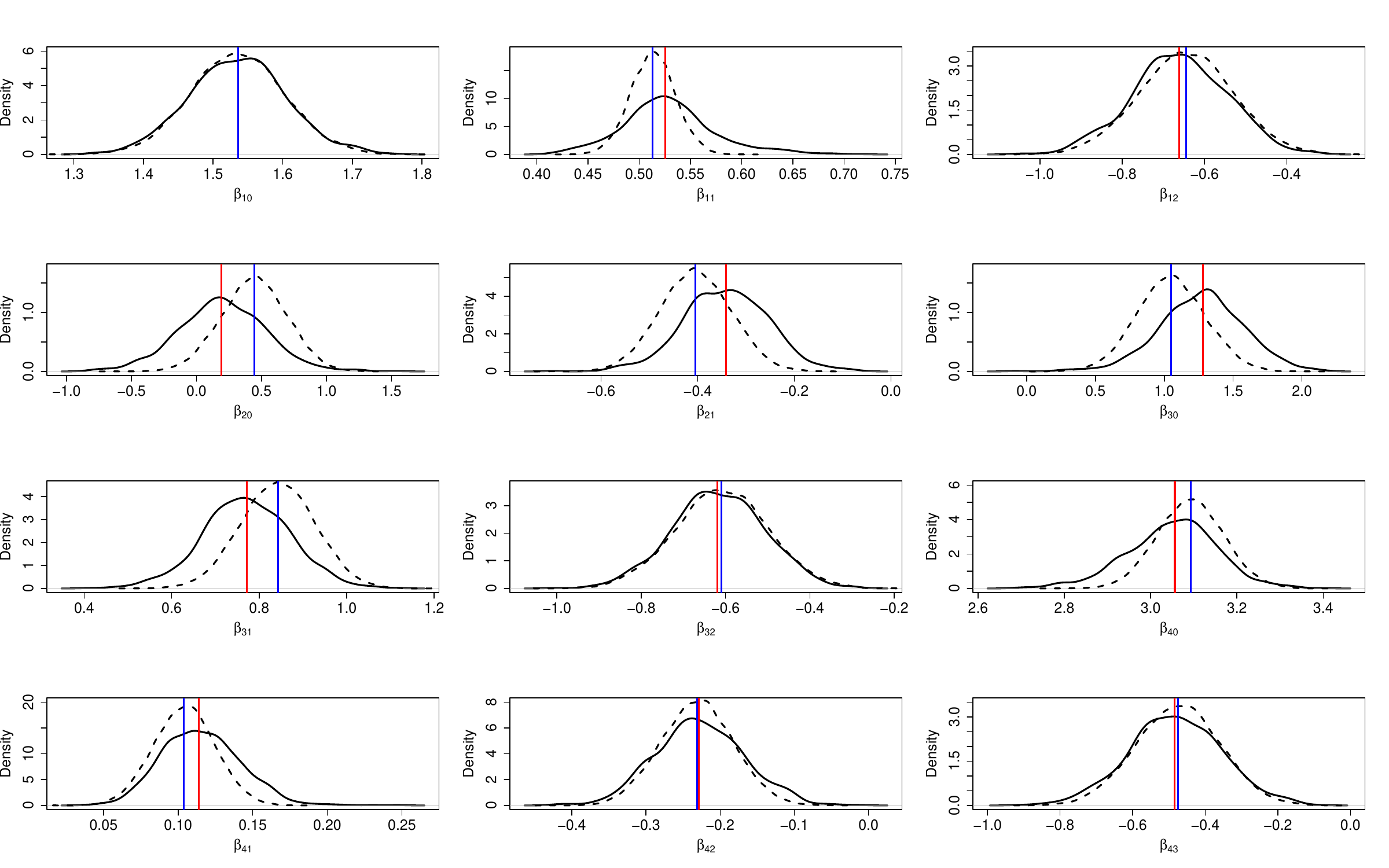}
\caption{Breast cancer recurrence data: marginal kernel density estimators of the posterior samples using adaptive MCMC (solid line) and the normal approximation (dashed line). The posterior median from the MCMC sample is shown in the vertical red line and the MAP in blue line.}
\end{figure}

\begin{figure}[h!]
    \centering
\includegraphics[width=0.75\textwidth]{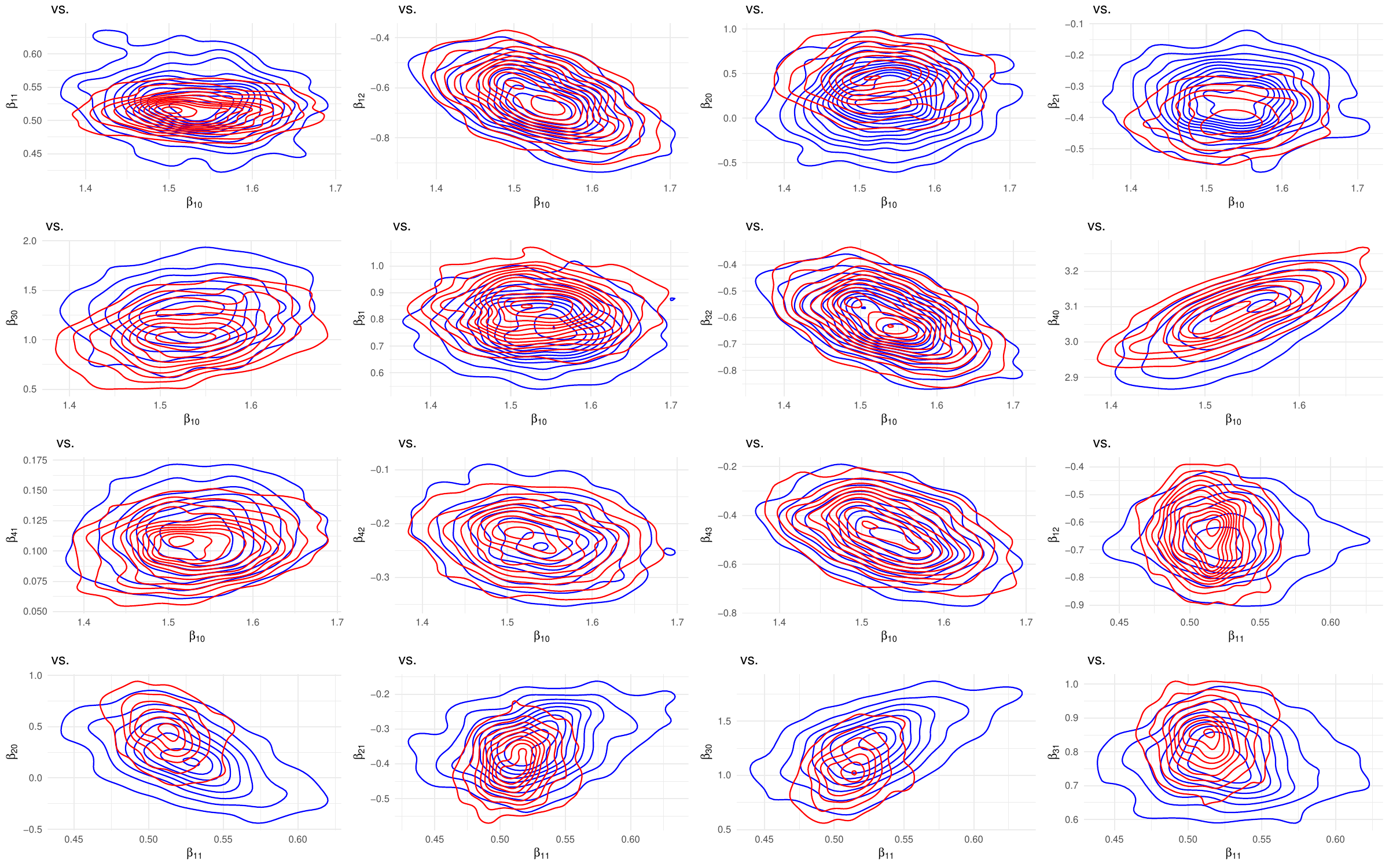}
\caption{Breast cancer recurrence data: Contour plots of the 2D kernel density estimates for the MCMC samples (blue) and the normal approximation (red).}
\end{figure}

\begin{figure}[h!]
    \centering
\includegraphics[width=0.75\textwidth]{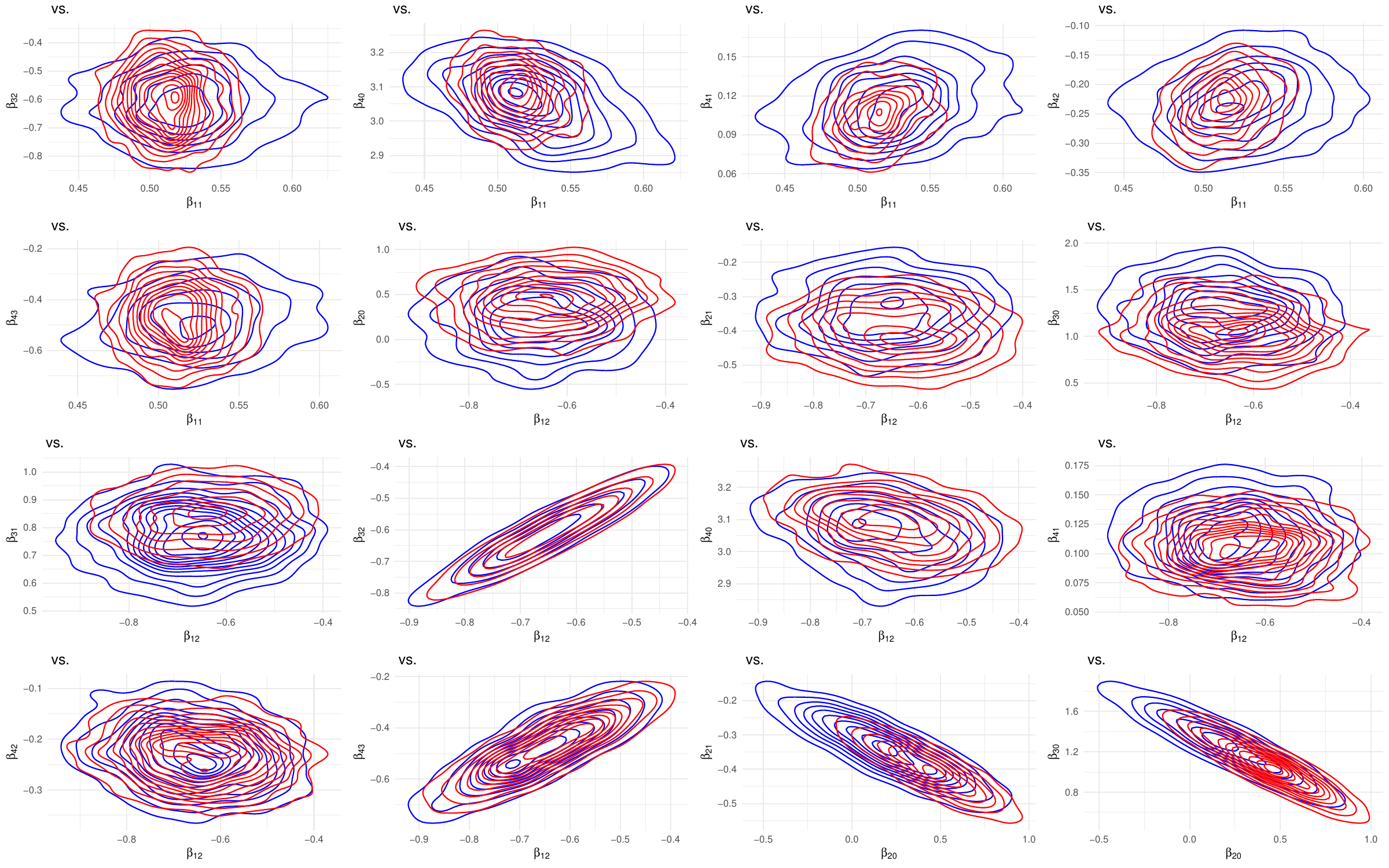}
\caption{Breast cancer recurrence data: Contour plots of the 2D kernel density estimates for the MCMC samples (blue) and the normal approximation (red).}
\end{figure}

\begin{figure}[h!]
    \centering
\includegraphics[width=0.75\textwidth]{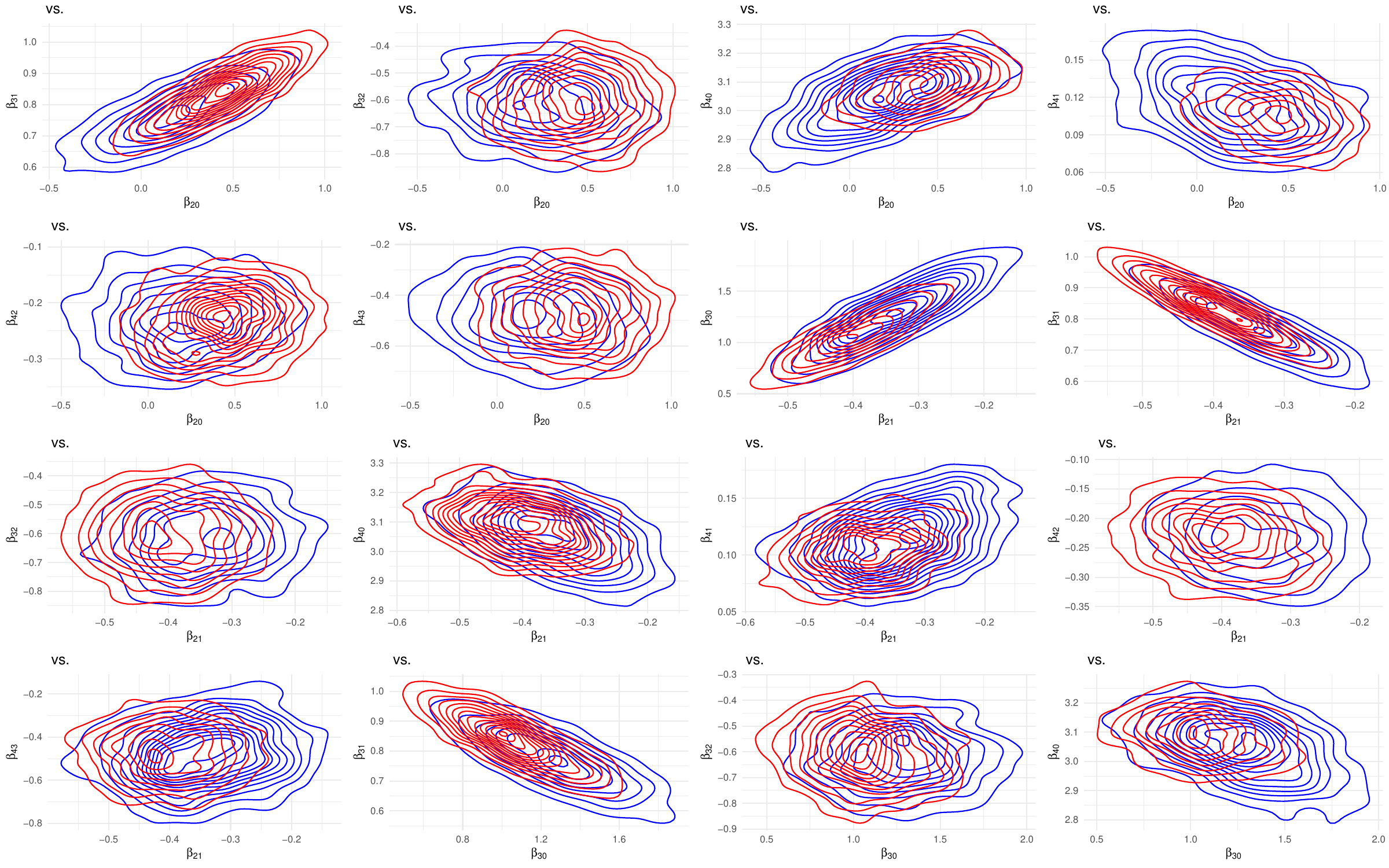}
\caption{Breast cancer recurrence data: Contour plots of the 2D kernel density estimates for the MCMC samples (blue) and the normal approximation (red).}
\end{figure}

\begin{figure}[h!]
    \centering
\includegraphics[width=0.75\textwidth]{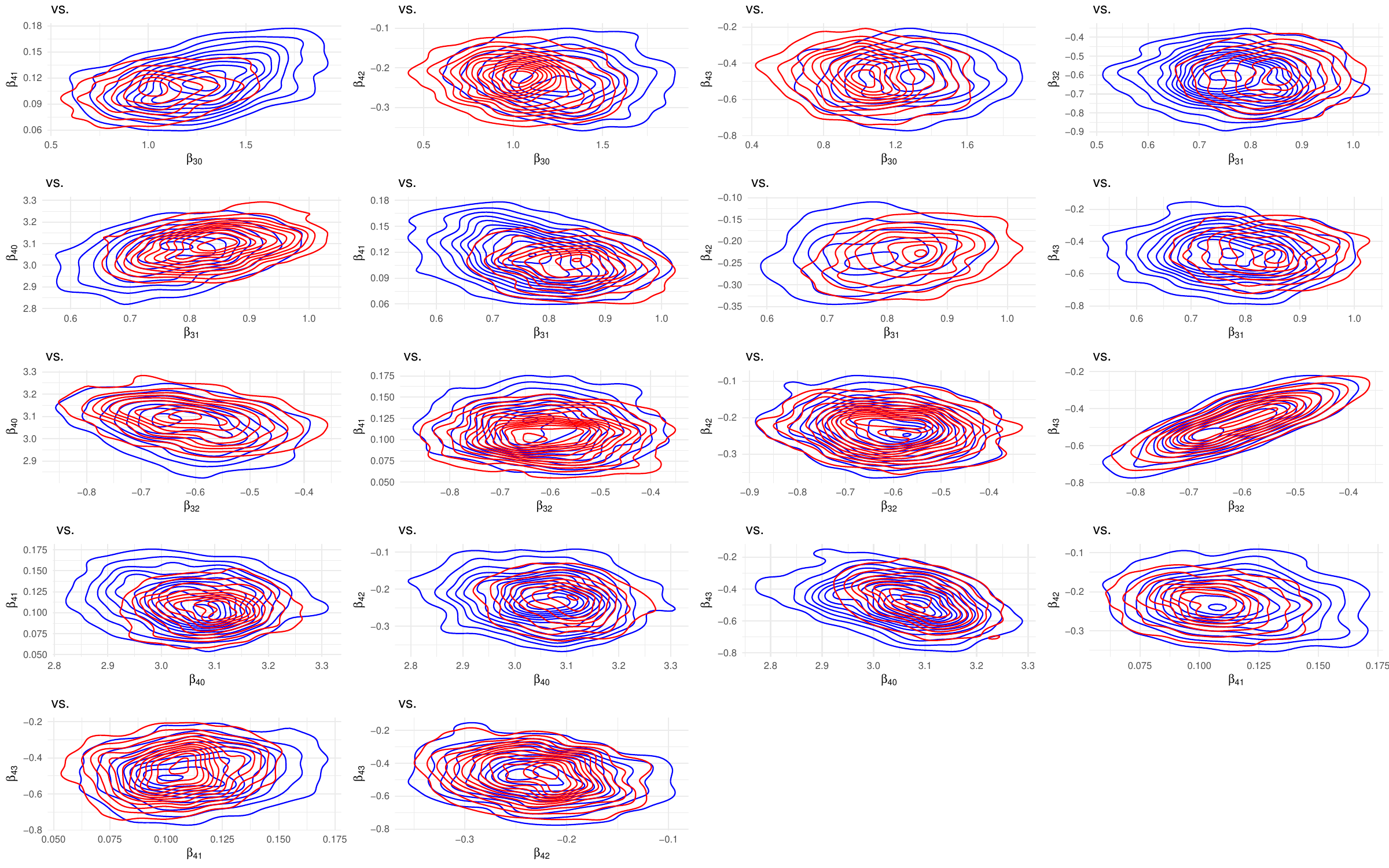}
\caption{Breast cancer recurrence data: Contour plots of the 2D kernel density estimates for the MCMC samples (blue) and the normal approximation (red).}
\end{figure}

\begin{figure}[h!]
    \centering
\begin{tabular}{c c c}
\includegraphics[width=0.3\textwidth]{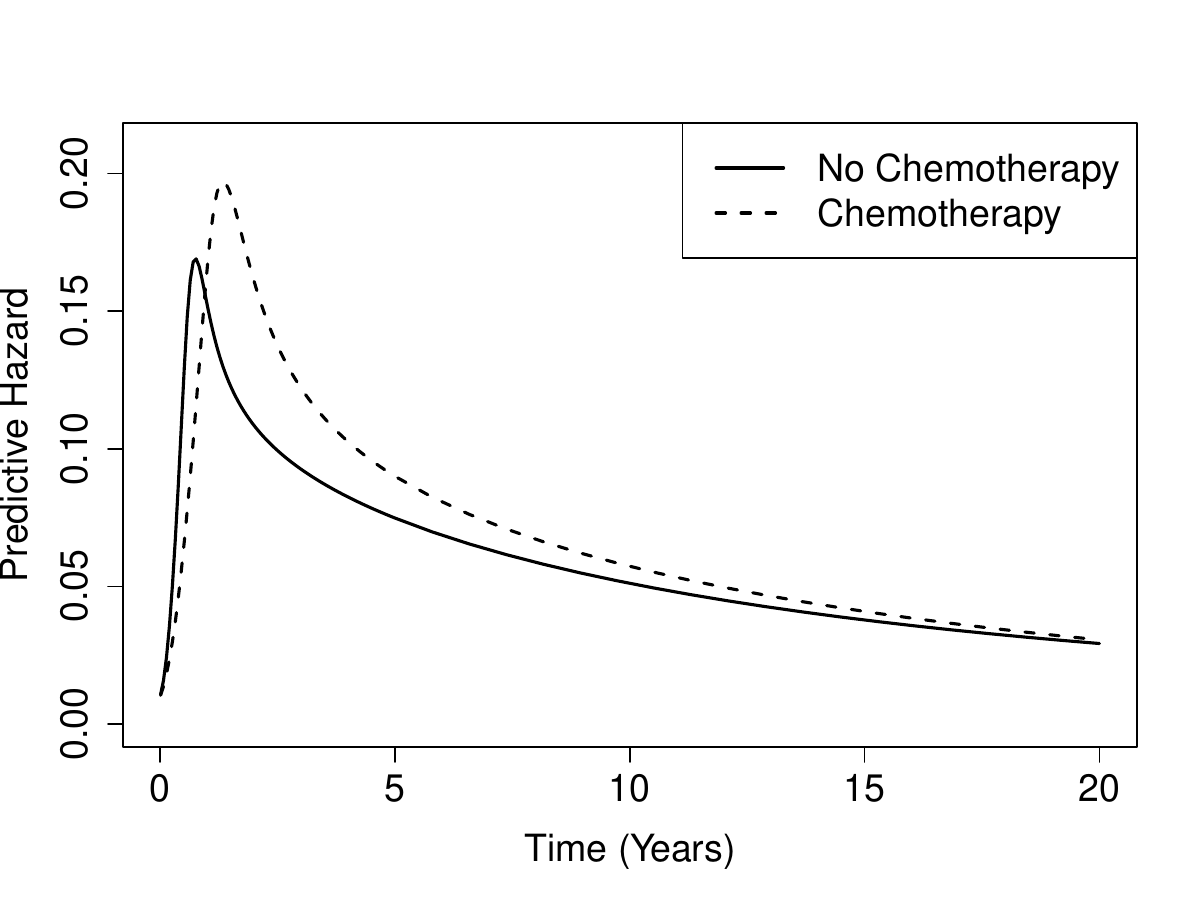} & \includegraphics[width=0.3\textwidth]{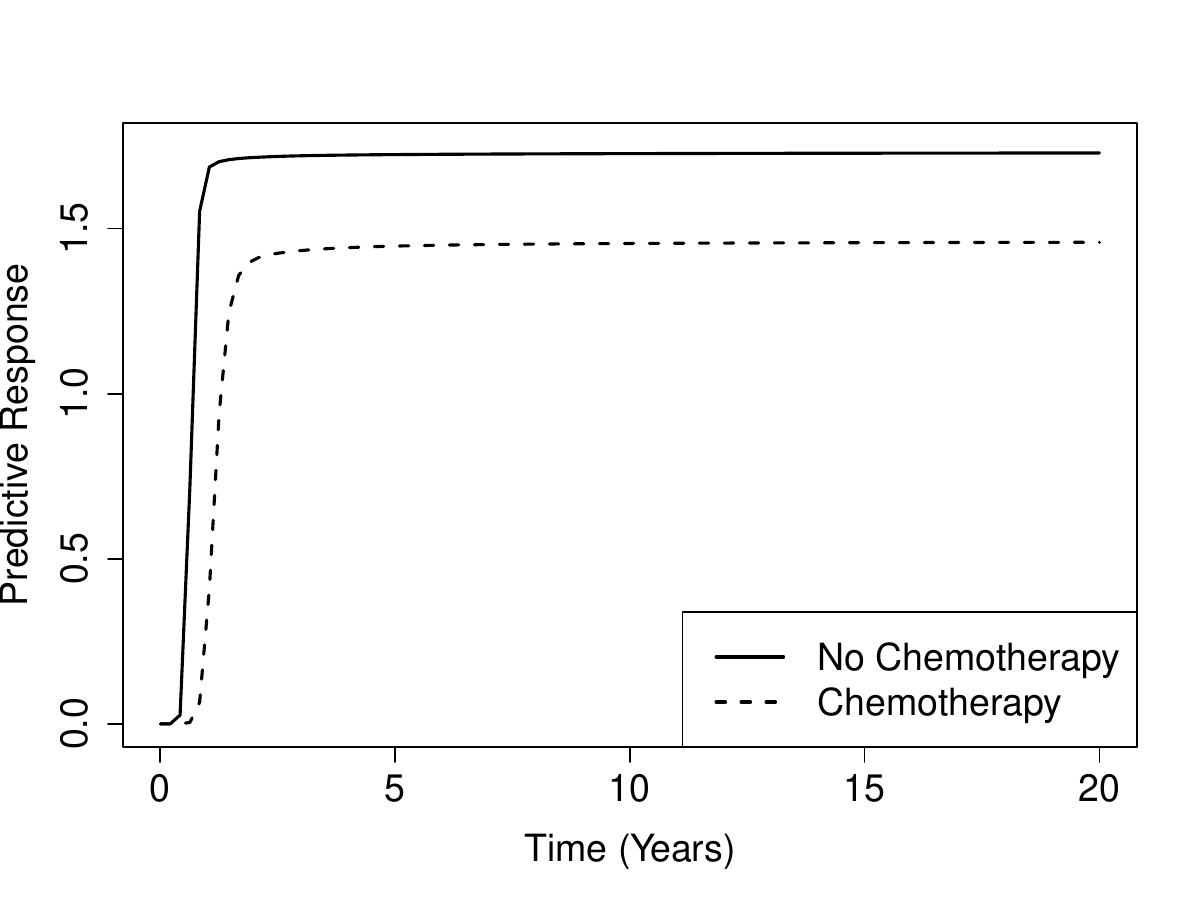} & \includegraphics[width=0.3\textwidth]{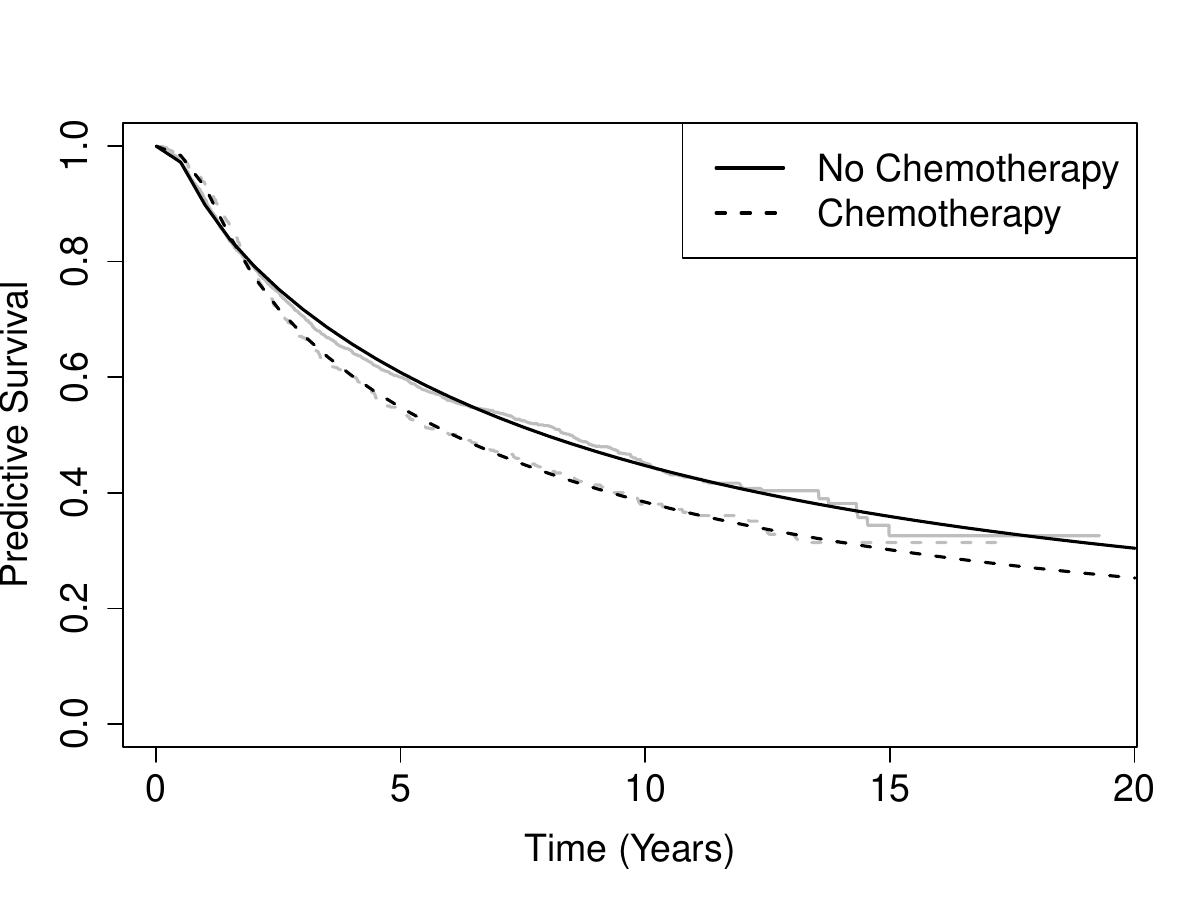} \\
 (a) & (b) & (c)\\
 \end{tabular}
    \caption{Breast cancer recurrence data (Population): (a) Predictive hazard functions, (b) predictive response functions, and (c) predictive survival functions and Kaplan-Meier estimates.}
    \label{fig:HRpop}
\end{figure}

\clearpage
    
\bibliographystyle{plainnat}
\bibliography{references}  

\end{document}